\documentclass[]{JHEP3}

\JHEPspecialurl{http://jhep.sissa.it/JOURNAL/JHEP3.tar.gz}

\usepackage{epsfig,multicol,bbm}
\usepackage{axodraw}
\newcommand\query[1]

\newcommand{\mrm}[1]{\mathrm{#1}}

\newcommand{\ttt}[1]{\texttt{#1}}
\newcommand{\br}[1]{\overline{#1}}

\newcommand{\brk}{\not \hspace{-3pt}U(1)}

\newcommand{\pT}{p_{\perp}}
\newcommand{\pTevol}{p_{\perp\mathrm{evol}}}
\renewcommand{\d}{\mathrm{d}}
\newcommand{\npT}{\not \hspace{-2pt}\pT}

\newcommand{\be}{\begin{equation}}
\newcommand{\ee}{\end{equation}}
\newcommand{\ba}{\begin{eqnarray}}
\newcommand{\ea}{\end{eqnarray}}

\title{Discerning Secluded Sector gauge structures}
\author{Lisa Carloni, Johan Rathsman  and  Torbj\"{o}rn Sj\"{o}strand\\ 
Theoretical High Energy Physics\\
Department of Astronomy and Theoretical Physics, Lund University,\\
S\"olvegatan 14A, SE 223-62 Lund, Sweden\\
E-mail: \email{lisa.carloni@thep.lu.se},\email{johan.rathsman@thep.lu.se},\email{torbjorn@thep.lu.se}
}

\preprint{
LU TP 11-09\\
MCnet 11-06
}

\abstract{

\vfill

New fundamental particles, charged under new gauge groups and only weakly
coupled to the standard sector, could exist at fairly low energy scales. In this
article we study a selection of such models, where the secluded group either
contains a softly broken $U(1)$ or an unbroken $SU(N)$. In the Abelian case new 
$\gamma_v$ gauge bosons can be radiated off and decay back into visible particles. 
In the non-Abelian case there will not only be a cascade in the hidden sector,
but also hadronization into new $\pi_v$ and $\rho_v$ mesons that
can decay back. This framework is developed to be applicable both 
for $e^+ e^-$ and $pp$ collisions, but for these first studies we 
concentrate on the former process type. For each Abelian and 
non-Abelian group we study three different scenarios for the 
communication between the standard sector and the secluded one.
We illustrate how to distinguish the various characteristics of 
the models and especially study to what extent the underlying 
gauge structure can be determined experimentally.}

\vfill
\vfill


\begin{document}

\section{Introduction}
\label{sec:intro}

There are basically two ways in which one can envision new physics beyond the standard model that can be searched for at future colliders. One possibility is to have theories with new heavy particles coupling to the standard model with either the ordinary gauge couplings, as in supersymmetry, or with couplings of a similar magnitude. This implies heavy particle masses, in order to avoid collider constraints. The other possibility, which we want to explore in this paper, is that new light particles are ultra-weakly coupled to the standard model particles, because they are not charged under the standard model gauge groups. Instead they couple  to the ordinary matter through some heavy state which carries both SM charges and charges of a new unknown gauge group, also carried by the light states.

There have been several suggestions for theories with this type of secluded sectors (sometimes also called hidden valleys or dark sectors), proposing non-conventional 
new physics with unexpected and unexplored signals could show up at current colliders such as the Large Hadron Collider (LHC) or a future linear electron-positron collider.

One example is the so-called hidden valley scenarios by Strassler 
and collaborators \cite{Strassler:2006im,Strassler:2006ri,
Strassler:2006qa,Han:2007ae,Strassler:2008bv,Strassler:2008fv,Juknevich:2009ji}, 
where  the SM gauge group is extended by a new unspecified gauge group $G$. 
In the original paper \cite{Strassler:2006im} this group is a 
$\not \hspace{-3pt} U(1)^\prime\times SU(N)$.  
The new matter sector consists of $v-$particles (where $v$ stands for 
``valley''), which are charged under the new gauge group and neutral under the 
standard one. The two sectors communicate via higher dimensional operators, 
induced either by heavy particle loops or by a $Z^\prime$ which can couple to 
both sectors. 

%{\bf do we need the following sentence?}

%Other examples of hidden valleys are found in String Theory \cite{Blumenhagen:2005mu}, 
%Twin Higgs models \cite{Chacko:2005pe}

An interesting feature of models with secluded sectors is that they naturally give rise to dark matter candidates. Likewise, some of the recently proposed dark matter models may present hidden sector features.
Specific dark matter models developed in the last few years such as 
\cite{ArkaniHamed:2008qn,Zurek:2008qg,Baumgart:2009tn,Cheung:2009qd,Morrissey:2009ur} suggest the existence of a GeV scale mass dark photon or scalar that is introduced to
enhance the dark matter annihilation cross section, in order to fit the data from PAMELA 
\cite{Adriani:2008zq,Adriani:2008zr} and originally ATIC \cite{:2008zzr}, although the latter data have later been superseded by more precise measurements from  FERMI \cite{Abdo:2009zk}. Here we will mainly be interested in models with dark photons originating from a softly broken $\brk$, which couple to standard model particles through so called kinetic mixing with the ordinary photon \cite{Holdom:1985ag} through  heavy particle loops in a similar way to the hidden valley scenarios.

The hidden valley-like theories and the dark matter models mentioned above share two features: the enlarging of the standard model symmetries to include a new gauge group $G$ and the presence of new light particle sectors that are solely charged under this new gauge group. 
If the new light particles can decay into standard model particles, their existence could be inferred from their effect on standard model particle phenomenology.
In \cite{Carloni:2010tw} we studied the effects of the new gauge group radiation, specifically 
the kinematic effects of $SU(3)^\prime$ radiation from fermions charged under both 
the SM and the new gauge group on the kinematic  distributions of visible particles.
In this paper we address the issue of discerning between different
gauge structures. Specifically,  we 
want to outline the differences between signatures arising from a 
secluded sector broken $\not\hspace{-3pt} U(1)^\prime$ gauge group with a light $\gamma^\prime$   and those 
arising from a confining $SU(N)$. In both cases we assume there is a mechanism for the secluded particles to decay back into the SM.

\FIGURE[t]{
%production
\begin{minipage}[b]{0.3\linewidth}
\epsfig{file=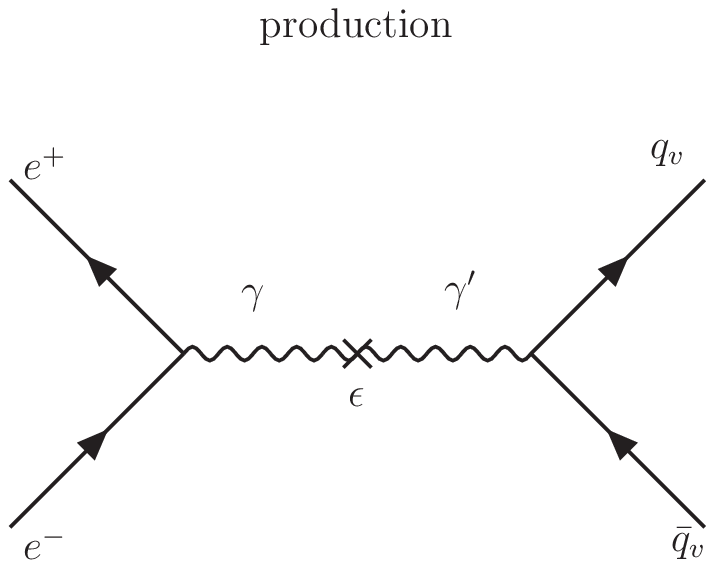,scale=0.5, angle=0}
\epsfig{file=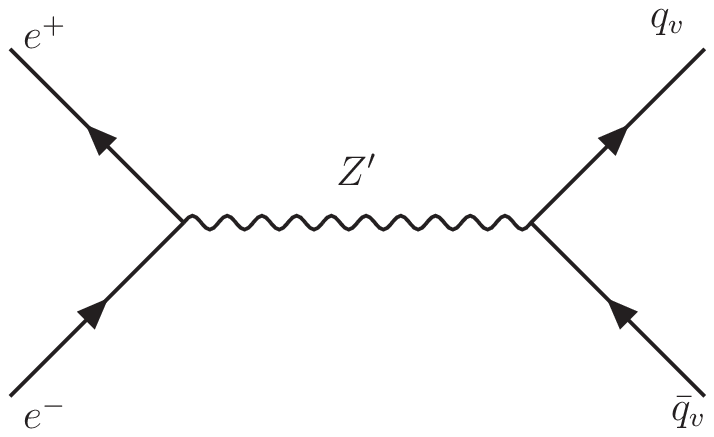,scale=0.5, angle=0}
\epsfig{file=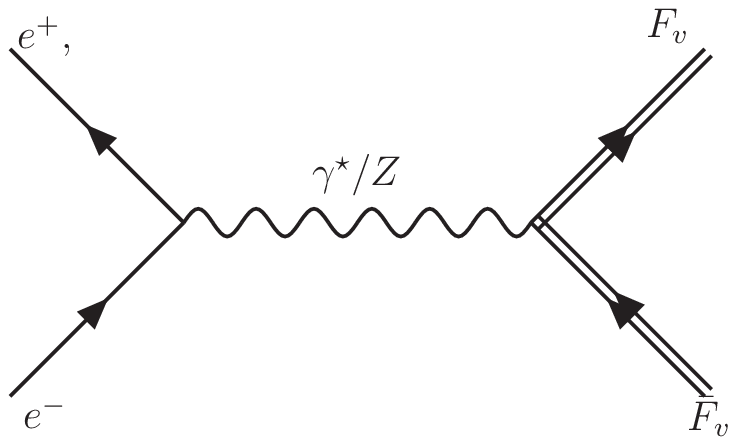,scale=0.5, angle=0}
\end{minipage}
%hadronization
\hspace{12pt}
\begin{minipage}[b]{0.3\linewidth}
\epsfig{file=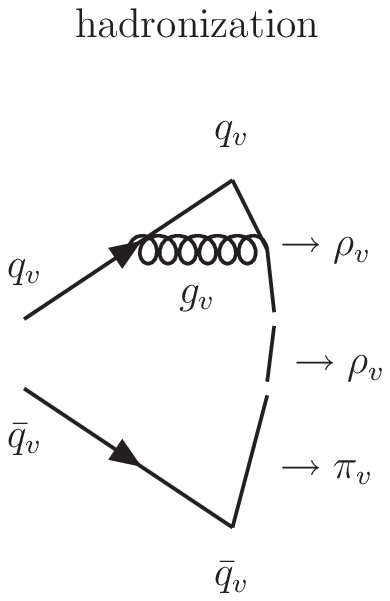,scale=0.6, angle=0}
\vspace{2.2cm}
\end{minipage}
%decay
\begin{minipage}[b]{0.3\linewidth}
\epsfig{file=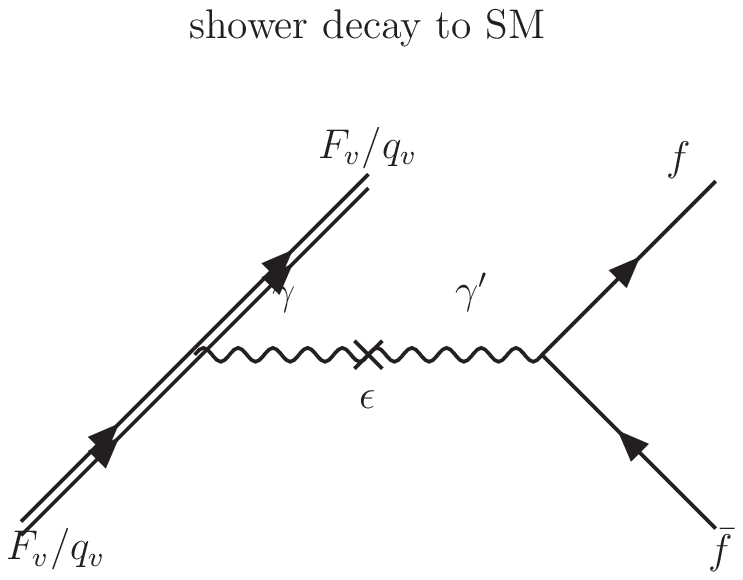,scale=0.5, angle=0}
\epsfig{file=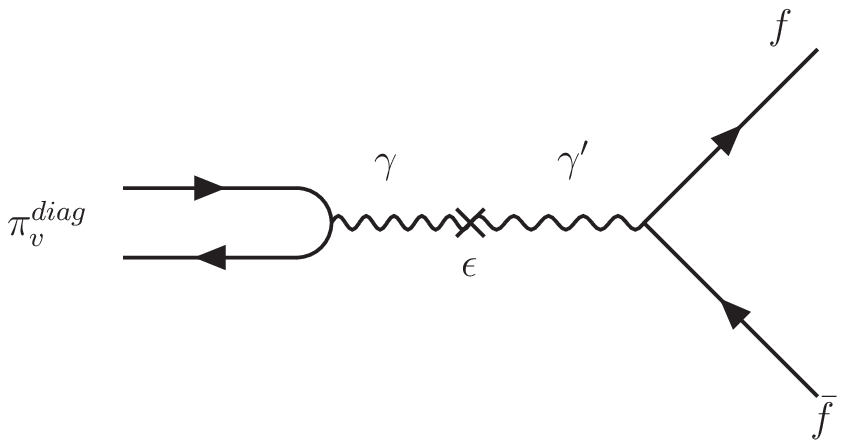,scale=0.5, angle=0}
\vspace{2cm}
\end{minipage}
\label{fig:scenarios}
\caption{The different mechanisms for production, hadronization  and decay that we consider as explained in the text.}
}

In order to distinguish which features of a given model are linked to the gauge group structure
 and which to the other details of the model, we consider different production processes and different mechanisms for the decay back into the SM. 
The various possibilities are summarized in Fig.~\ref{fig:scenarios}.  
For the production we consider three mechanisms. In the first case the portal to the hidden sector is through kinetic mixing between the SM photon and a light $\not\hspace{-3pt}U(1)$ gauge boson. In the second case we have production via a $Z^\prime$. 
In the kinetic mixing of $\gamma$-$\gamma_v$, the $\gamma_v$ is assumed to have a mass around 1-10 GeV and the mixing $\epsilon$ is assumed to be $\epsilon \sim 10^{-3}$, while in the $Z^\prime$  case, the mass of the $Z^\prime$ would be around 1-6 TeV \cite{Strassler:2006im}. The third case is where the production happens via SM gauge bosons as in \cite{Carloni:2010tw}. In this case the particles are assumed to have both SM charges and secluded sector ones. To distinguish them we will call them $F_v$ in the following, to separate them from particles that are only charged under the secluded gauge group, which we call $q_v$. We will also assume that the $F_v$ particles will decay into a standard model particle $f$ and a secluded sector particle $q_v$, {\it i.e.} $F_v \rightarrow f q_v$.

If the particles of the secluded sector are charged under a non-Abelian $SU(N)$
or a softly broken Abelian $\brk$ with a light $\gamma_v$, there will also be additional radiation of gauge bosons. In the former case, the $v$-gluons will be connected to the $q_v$s (produced directly or via the $F_v$s) and form a confined system which will then hadronize. 

Depending on the nature of
the secluded hadrons thus produced, they may then decay back into standard model
particles through kinetic mixing or a heavy $Z^\prime$. In both cases this decay 
can be very slow, so much so as to generate displaced vertices and other exotic
signatures as for example discussed in \cite{Strassler:2006ri}. In the case of
 $\gamma_v$ radiation instead, the gauge bosons may decay directly back into the SM
through kinetic mixing $\gamma_v$-$\gamma$, while the $q_v$s will not
be able to decay back into the SM since they carry the secluded gauge charge. 

Thus, in both the non-Abelian and Abelian cases, we can have models where some of the particles produced will decay back into SM particles and some of them will be invisible. The questions we want to address is thus how the production of   visible particles depends on the secluded gauge structure and whether it is possible to tell a non-Abelian and Abelian gauge group apart also when other features of the models are very similar.

The paper is structured as follows. Section \ref{sec:models}
gives a short overview of the general model considerations that underlie
our studies, with particular emphasis on the production mechanisms
that are relevant in various scenarios, and some comments on the
 decay mechanisms that lead to signals
in visible distributions. In section \ref{sec:pythia} we provide a
more in-depth overview of the new physics aspects that we have
implemented in \textsc{Pythia}~8: the particle content, the parton
showers, notably for the broken $U(1)$ case, the hadronization in
the secluded sector, and the decay back to the visible one. In
Section \ref{sec:analysis} we proceed to describe the phenomenology
of the various scenarios, in the context of an $e^+ e^-$ linear
collider. While less interesting than a corresponding LHC
phenomenology, it allows us to better highlight the relevant
features of the model as such. In section \ref{sec:compare} we
further study distributions that could offer a discrimination between
an Abelian and a non-Abelian scenario for the secluded sector.
In Section \ref{sec:conclusions} we summarize our findings, and give
an outlook. Finally in Appendix~\ref{appendix} we provide information
how the simulation of a wide range of scenarios can be set-up.

\section{Overview of hidden sector scenarios}
\label{sec:models}

As already mentioned in the introduction there are many different types of models that can display hidden sectors and the common feature is that they communicate with the standard model through some heavy states. This communication can occur in many different ways and we will distinguish three different types in the following: via kinetic mixing, via a heavy $Z^\prime$, and via heavy fermions that carry SM charges.  

The common feature of the models we consider is that the SM group 
$SU(3)_C\times SU(2)_L \times U(1)_Y$ is augmented by a new 
gauge group $G$. For each scenario we will consider two cases - one {\it Abelian} where $G$ 
contains a softly broken $\brk$ with a light gauge boson $\gamma_v$ and 
one {\it non-Abelian} where $G$ contains an unbroken $SU(N)$ factor mediated by a then massless $g_v$.

The particle content consists of  $q_v$ particles and/or $F_v$ particles.
With $q_v$ particles we indicate fermions or scalars (with spin = 1/2, 0, 1)
charged solely under the new gauge group. With $F_v$ we indicate particles 
(spin $s=0,1/2,1$) which may couple to both secluded sector and standard model sector.
Though in principle one could choose any spin assignment among the ones above, we have chosen to analyze the case in which $F_v$ and $q_v$ are fermions, except in the case when the $q_v$s are produced from a $F_v$ decay when we assume them to be scalars. 
In addition, in all the scenarios we consider both $F_v$ and $q_v$ belong to the fundamental representation of the group $G$. Finally, the $G$ sector charges are constrained by anomaly cancellation. For an example see \cite{Strassler:2006im}.

\subsection{Kinetic mixing scenarios}
As already alluded to, one way of producing the secluded sector particles is through kinetic mixing. In the scenarios we wish to investigate, the SM U(1) (effectively the photon) mixes kinetically with a new GeV mass $\gamma^\prime$ and produces a pair of secluded sector particles, see Fig.~\ref{fig:scenarios}. This mechanism is mostly relevant in the case when 
the secluded sector contains new fermions which are charged only 
under the new gauge group $G$. In addition we will in this scenario only consider those cases when the SM particles are not charged under the new U(1). Communication 
between the SM  and secluded sectors then only comes from kinetic mixing 
between the standard model U(1) gauge boson and the new gauge boson, as described by
\be
\mathcal{L}_\mrm {kin}=-\frac{1}{4}\epsilon_{1}\left(F^{\mu \nu}_1\right)^2
-\frac{1}{2}\epsilon F^{\mu \nu}_1F_{2,\mu \nu}
-\frac{1}{4}\epsilon_2 \left(F^{\mu \nu}_2\right)^2.
\ee
In the case of two $U(1)$ gauge symmetries ($U(1)_1\times U(1)_2$), the 
non-vanishing mixing $\epsilon$ arises naturally as one 
integrates out loops of heavy fermions coupling to both 
the associated gauge bosons \cite{Holdom:1985ag}
so long as there is a mass splitting among them. 
The relation between the size of the mixing 
and the mass splitting is given by
\be
\epsilon= \frac{e_1 e_2}{16 \pi^2} \ln\left(\frac{M^{(1)}_{12}}{M^{(2)}_{12}}\right),
\ee  
where $e_1$ and $e_2$ are the gauge couplings 
of the fermions in the loop to the two $U(1)$ gauge bosons, $A_1$ and the new $A_2$
respectively, and $M^{(1)}_{12}$ and $M^{(2)}_{12}$ are their masses. 
In general, the $U(1)_1$ and the $U(1)_2$ will not be orthogonal. One may however chose the $U(1)_1$ generator so that the fermions that are only charged under $U(1)_1$ do not have any charge shift, while those that couple to $U(1)_2$ do \cite{Holdom:1985ag}.

For the case of non-Abelian groups, $G_1\times G_2 \times G_3$, a mixing can come from the spontaneous breaking of the group down to $H\times U(1)_1\times U(1)_2$. Also in this case the $U(1)_1$ and the $U(1)_2$ will not be orthogonal, as long as the three couplings associated to the unbroken symmetries are different.

The kinetic mixing mechanism has been used in model that want to describe various recent cosmic ray  measurements in terms of dark matter models. The most important signal here is the 
positron excess observed by PAMELA \cite{Adriani:2008zr}. At the same time, any model wanting to explain this excess also has to explain the absence of an anti-proton excess observed by 
PAMELA \cite{Adriani:2008zq} and  finally the measurements of the total electron and positron flux observed by the Fermi LAT collaboration \cite{Abdo:2009zk}. 
The models are set up so that the dark matter particles will annihilate into a dark photon or scalar which couples to SM particles through kinetic mixing. The mass of the dark matter particle is then determined by the scale at which the positron excess is observed, to be of order 0.1--1 TeV.

In addition, the large positron excess observed also means that there must be some enhancement mechanism of the dark matter annihilation cross section. One way to do this is to invoke 
Sommerfeld enhancement\footnote{Resummation of t-channel exchanges of a new light particle.} by introducing a light dark photon or scalar. 
The mass of the dark photon (or scalar) in these models is typically in the GeV range, which means that decays into $\bar{p}$ and $\pi^0$ are kinematically suppressed relative to the lepton decays and thus also explain the non-observation of any anti-proton 
excess by PAMELA\cite{Adriani:2008zq}. 

A recent example of models that fits all these data is given by \cite{Bergstrom:2009fa}, but there are still large uncertainties due to cosmological assumptions such as the dark matter distribution and propagation of cosmic particles. 

The dark gauge group $G_\mrm {dark}$ is largely unspecified in these types of models except that it must contain a U(1) factor in order for the kinetic mixing with the SM photon. This means that there could also be additional Abelian or non-Abelian factors in $G_\mrm {dark}$. In the following we will consider the cases when $G_\mrm {dark}$ contains an additional $U(1)$, which is spontaneously broken giving a massive $Z^\prime$, or an additional $SU(N)$ factor giving a confining force for the secluded sector particles.

The phenomenology and constraints on these types of models at low energy $e^+ e^-$ colliders such as Belle, BaBar, DA$\Phi$NE, KLOE and CLEO  have been 
studied by \cite{McElrath:2005bp,Batell:2009yf,Essig:2009nc,Reece:2009un}. 
%In addition the Belle and CLEO collaborations have set an upper bound on the mixing $\epsilon < 10^{-2}$ in this type of set up. 

%{\bf what is the reference above?}

%Given the cosmological uncertainties in fitting these type of models to data we will concentrate on the lower mass range for the dark matter particles of the order 100 GeV in the following.

\subsection{${Z^\prime}$ mediated scenarios}

The second type of scenarios we want to consider are those that are similar to the
original hidden valley scenario \cite{Strassler:2006im} with a massive $Z^\prime$ coupling to both SM fermions and secluded sector ones. 
Thus, the processes we are interested in are when SM fermions annihilate into the secluded sector $Z^\prime$ which in turn gives a pair of secluded particles, as depicted in Fig.~\ref{fig:scenarios}.

In these types of models it is typically assumed that the  $Z^\prime$ acquires a mass by spontaneous symmetry breaking of a $U(1)$ symmetry by a $\langle \phi \rangle$ whereas the
origin of the secluded sector $\not \hspace{-3pt}U(1)$ is not discussed. 

The secluded sector particles that the $Z^\prime$ would decay to could be either charged solely under the valley gauge group $G$ or charged under $G$ and (parts of) the SM $SU(3)_C\times SU(2)_L\times U(1)_Y$. In the latter case, the particles would on the one hand have to be very massive (several hundreds of GeV) due to experimental constraints and on the other hand they would be more effectively produced through their SM couplings. Thus we will not consider this possibility more here. In contrast the particles charged solely under the secluded gauge group could be light with a mass in the  $1-50$ GeV range, thanks to the reduced coupling through the heavy $Z^\prime$.

As a consequence of the heavy mass of the $Z^\prime$, the $s$-channel pair production cross section will be peaked at $\sqrt{\hat{s}}\sim m_{Z^\prime}$ and be suppressed at an $e^+ e^-$ collider unless $\sqrt{s}\sim M_{Z^\prime}$. At a hadron collider the production of the $Z^\prime$ would be dominantly on-shell if the overall center of mass energy is large enough and there is enough support from the  parton density functions. 

In the original hidden valley model the secluded sector group also contains a confining 
$SU(N)$. Thus the produced secluded sector particles would have to hadronize into hadrons which are neutral under this $SU(N)$. Another possibility is that there is instead an additional $\brk$ which would instead give radiation of $\gamma^\prime$s. 

Finally it should be noted that also in this case there is kinetic mixing between the $Z$ and the $Z^\prime$, which primarily is important for setting limits on the mass and couplings of the $Z^\prime$ from LEP as discussed in \cite{Strassler:2006im}.  

\subsection{SM gauge boson mediated scenarios}
The final type of scenario that we consider are ones where the "communicator'' is charged under both the SM and new interactions. This scenario and its implementation into {\scshape pythia 8} has been described in \cite{Carloni:2010tw} so here we only briefly recapitulate the main features.

In this model the new heavy communicator particle $F_v$ would be pair produced with SM strength, which means that it would have to be quite heavy in order to not have been already seen at colliders. 
Another consequence is that the communicator would decay into a 
SM and pure hidden sector particle, dubbed $q_v$, so that quantum numbers are conserved. In the 
simple case in which neither $q_v$s nor $v$-gauge bosons leak back into the SM, as in the scenario in \cite{Carloni:2010tw}, this entails a missing energy signal. 

Also in this case, the secluded sector group can be either Abelian or non-Abelian. In both cases we will assume that the produced $\gamma^\prime$s or hadrons can decay back to SM particles through kinetic mixing via loops of the $F_v$ particles or via a $Z^\prime$.

\subsection{Decays back to the SM}
\label{sec:decay}

First of all we mention again the case of secluded particles which are charged both under the SM and secluded gauge groups, $F_v$, which we assume decay according to $F_v \to f q_v$. 
All other particles produced by either of the mechanisms described above may decay back to SM particles as long as they do not carry any charge under the secluded gauge group. Essentially these decays will be through kinetic mixing with SM gauge bosons or through a heavy $Z^\prime$ as detailed below. 

In the Abelian case, with a light secluded sector $\gamma^\prime$, the $q_v$s will be stable, but the $\gamma^\prime$s that are radiated in connection with the primary hard process will decay back to SM particles, ${\gamma^\prime} \rightarrow f\bar{f}$. The strength of the
kinetic mixing $\epsilon$, together with the available phase space, determines the decay width $\Gamma_{{\gamma^\prime}\rightarrow f\bar{f}}$. Since the  $\gamma^\prime$ is light, it will mainly mix with the standard model photon and thus the branching ratios for different channels will depend on the electric charge of the produced SM particles. In essence this means that the decays will be similar to a off-shell photon, $\gamma^*$ with the virtuality given by $m_{\gamma^\prime}$. We also note that if the kinetic  mixing is small, the life-time could be so large as to give displaced vertices.

In the non-Abelian case the secluded sector hadrons may also decay back into the SM via kinetic mixing of the $\gamma^\prime$ with the SM photon or via a heavy $Z^\prime$.  In this case the phenomenology will depend on the number of light flavours $N_{\rm flav}$ in the secluded sector.
In the following we will assume that $N_{\rm flav} \geq 2$ and only consider the case when the fundamental particles are fermions as in \cite{Strassler:2006im} although similar arguments can be made also in the case of scalar constituents. Thus, the bound states will be the secluded sector version of mesons, baryons and possibly also glueballs. For the decays back to SM particles, it is the meson states that are of primary interest and therefore we concentrate on them here. 

With $N_{\rm flav}$ light flavours, there will be of the order $N_{\rm flav}^2$ mesons with a given spin out of which approximately 
$N_{\rm flav}$ are flavour neutral and can decay back into the SM via kinetic mixing or a $Z^\prime$.
The SM decay products will depend on the spin of the secluded meson. For a spin zero meson, helicity suppression leads to dominance by the heaviest SM particle available whereas for a spin 1 meson it will depend on the couplings to the particle mediating the decay, {\it i.e.} either to the photon in the case of kinetic mixing or to the $Z^\prime$.

The phenomenology will thus depend on the relative production of spin-0 and spin-1 mesons and their masses. If the confinement scale $\Lambda_v$  in the secluded sector is large compared to masses of the lightest secluded sector fermions the situation will be similar to QCD. In other words there will be a light spin-0 $\pi_v$ with mass much smaller than the spin-1 $\rho_v$. Thus all $\rho_v$ will decay to  pairs of $\pi_v$s and the SM particles produced will be the heaviest one available.

If $\Lambda_v$ is of the order of the masses of the lightest secluded sector fermions then 
the mass splitting between the spin-0 and spin-1 mesons will be small and thus the spin-1 meson will be metastable and instead  decay back into the SM, again via either kinetic mixing or a $Z^\prime$, but in this case, there not being any helicity suppression, the decay will be similar to that of an off-shell photon. Thus in this case there will also be an abundance of leptons produced along with hadrons.

If all constituent masses are much larger than the confinement scale, the lowest lying $SU(N)$ neutral states would be glueballs as discussed in \cite{Juknevich:2009ji}. We do not discuss their phenomenology here. 
We will also not consider so called quirks \cite{Kang:2008ea} which are charged both under the SM $SU(3)_C$ and a secluded $SU(N)$ with the confinement scale $\Lambda$ being much smaller than the $F_v$ masses. 

Finally we note that similarly to the Abelian case some of the secluded sector hadrons could be metastable and decay back into the detectors with displaced vertices.

\section{Physics in the secluded sector}
\label{sec:pythia}

For the studies in this article we have developed a framework to 
simulate the physics of a secluded sector. It contains a flexible 
setup that can be used to study different production mechanisms, 
perturbative shower evolution scenarios, non-perturbative hadronization 
sequences and decays back into the visible sector. Parts of the 
framework were already in use for our previous study 
\cite{Carloni:2010tw} but significant new capabilities have been 
added. These are available starting with \textsc{Pythia}~8.150.
The physics content will be described in the following, while 
technical details on how to set up a variety of scenarios 
is outlined in Appendix A. The studies presented in this article 
only give a glimpse of the possibilities. 

\subsection {Particles and their properties}

The key aspect of a scenario is that of the valley gauge group $G$, 
which we allow to be either $U(1)$ or $SU(N)$. The gauge bosons of 
these groups are named $\gamma_v$ and $g_v$, respectively. The former 
can be broken or unbroken, i.e.\ $\gamma_v$ can have a mass, while 
the latter is always unbroken so that $g_v$ remains massless.

The rest of the particles, i.e.\ the ``matter'' content, fall into 
two main categories: those charged under both the SM and the $v$ sector, 
and those that are pure $v$-sector particles. 

For the doubly charged ones, dubbed $F_v$, 12 particles are introduced 
to mirror the Standard Model flavour structure, see Tab.~\ref{tab:Fv_codes} 
in the appendix. Each $F_v$ particles couples flavour-diagonally to the 
corresponding SM particle. In addition to its SM charges, it is also
put in the fundamental representation of $G$. For $U(1)$ the charge
is taken to be unity, while for $SU(N)$ the ``charge'' is
$C_F = (N^2 -1)/(2 N)$ while pair production cross 
sections obtain a factor of $N$ enhancement. Although the name 
suggests that the $F_v$ are fermions, they can be spin 0, 1/2 or 1 
particles. If the $F_v$ particles have spin 1 then their production
cross section depends also on the presence or not of an anomalous magnetic 
dipole moment.

The valley secluded sector further contains a purely $G$ interacting
sector. At the parton level this consists of $q_v$s, belonging to the
fundamental representation of $G$. The name is introduced to reflect
the similarities with the quark in QCD. The $q_v$ particle is stable 
and invisible to SM interactions. Its spin, 0 or 1/2, is adapted to 
the choice of spin made for $F_v$, in case the scenario allows for 
$F_v \rightarrow f q_v$ decay, where $f$ is a SM particle. The spin 
structure of the $F_v \to f q_v$ decay is currently not specified, 
so the decay is isotropic.

In the $G = U(1)$ scenarios only one $q_v$ is assumed to exist.
$F_v$ decays, if allowed kinematically, are flavour diagonal,
$F^i_v \rightarrow f^i q_v$, with a common (Yukawa) coupling
strength. Given that both the $F_v$s and the $q_v$s have a unit of 
$U(1)$ charge, they can radiate $\gamma_v$ gauge bosons. If $U(1)$
is unbroken the $\gamma_v$ is massless and stable. For a broken
symmetry, $G=\not\hspace{-3pt}U(1)$, the $\gamma_v$  can decay back 
to a SM fermion pair through the mechanisms discussed in the previous
Section~\ref{sec:models}. For kinetic mixing or decay via a $Z^\prime$,
branching ratios by default are assumed to be proportional to the
respective fermion coupling to the photon, whenever the production 
channel is allowed by kinematics. The $\gamma_v$ decay can be either 
prompt or displaced.

If instead $G = SU(N)$, the massless $g_v$ gauge bosons are 
self-interacting, such that the parton shower will also have to
allow for $g_v\rightarrow g_v g_v$ splittings, with no equivalence 
in the $U(1)$ case. The self-interactions also lead to confinement,
like in QCD. In Section~\ref{sec:fragmentation} below we will explain
how the resulting picture can be described in terms of ``strings'' 
stretched from a $q_v$ end via a number of intermediate $g_v$s to a
$\bar{q_v}$ end. The string can break, by the production of new
$q_v \bar{q_v}$ pairs, to produce a set of $v$-mesons formed by 
the $q_v$ of one break and the $\bar{q_v}$ from an adjacent one.
To first approximation these $v$-mesons would be stable, and so 
the whole $v$-hadronization process would be invisible. One would not
even have the kind of indirect recoil effects that the $v$-shower
can give. If kinetic mixing or decay via a $Z^\prime$ is assumed, 
it would again be possible to let the $v$-mesons decay back to a 
SM fermion pair. 

With only one $q_v$ species there would only be one kind of $v$-mesons,
and so the choice would be between two extremes: either all the
energy deposited in the hidden sector decays back to be visible, 
or none of it. The more interesting scenarios --- e.g.\ in terms
of offering a bigger challenge to sort out what is going on ---
are the ones where only part of the $v$-mesons can decay back.
Therefore a variable number $N_{\mathrm{flav}}$ of separate $q_v$ 
flavours are assumed to exist (at most 8 in the current 
implementation). This gives $N^2_{\mathrm{flav}}$ different possible
$v$-meson flavour combinations, out of which only $N_{\mathrm{flav}}$
are flavour-diagonal and thus able to decay back into the SM sector.
It would be possible to assign individual masses to the $q_v$s and
$v$-mesons, but for now we assume one common $q_v$ ``constituent''
mass and one common $v$-meson mass, twice as large as the former.   

By analogy with QCD two separate spin states are assumed, denoted
$\pi_v$ and $\rho_v$. For now mass splitting is taken to be small, 
such that $\rho_v \to \pi_v \pi_v$ is kinematically forbidden, as is
the case in QCD for the $s$ and heavier quarks. The decay of the 
flavour-diagonal mesons is different in the two cases: by helicity
(non)conservation the $\pi_v$ couplings to a pair of SM fermions $f$ 
provides an extra factor $m_f^2$, an addition to the squared charge
and phase space factors factors present for the $\rho_v$ mesons.

In the confining $SU(N)$ case also a  $v$-glueball is introduced.
It is only rarely used, to handle cases where the invariant mass 
of the invisible-sector fragmenting system is too large to produce 
one single on-shell $v$-meson and too small to give two of them.
Then it is assumed that an excited $v$-meson state is produced, that 
can de-excite by the emission of these invisible and stable $g_v g_v$ 
bound states.

In summary, by default the $v$-particles with no SM couplings are not
visible. Their presence can only be deduced by the observation of
missing (transverse) momentum in the event as a whole. On top of this 
we allow two different mechanisms by which activity can leak back from 
the hidden sector. The first is the $F_v \rightarrow f q_v$ decay and 
showers from the $F_v$ and $q_v$, in the scenario in which $F_v$ has 
both SM charges and $G$ charges, as discussed in our previous article
\cite{Carloni:2010tw}. The second is the decay of SM gauge bosons 
produced through mixing by the $G$ group gauge bosons in the kinetic 
mixing case, either the massive $\gamma_v$ for $\not\hspace{-3pt}U(1)$
or the diagonal $v$-mesons for $SU(N)$.

\subsection{Valley parton showers}
\label{sec:MC}

Parton showers (PS) offer a convenient approximation to higher-order
matrix elements, which by the use of Sudakov form factors contain
a resummation of virtual corrections to match the real emissions
\cite{Buckley:2011ms}. For the current studies, the \textsc{Pythia}
$\pT$-ordered parton showers \cite{Sjostrand:2004ef} are extended to 
the secluded sector, and the approach used to take into account 
massive radiating particles \cite{Norrbin:2000uu} must, for the 
$\not\hspace{-3pt}U(1)$ scenario, be extended to the case where also 
the radiated gauge boson is massive. This section gives a summary of 
the showering framework, with emphasis on aspects new to this study
(relative to \cite{Carloni:2010tw}). 
 
In the most general case, final-state QCD, QED and valley radiation 
are interleaved in one common sequence of decreasing emission $\pT$ 
scales. That is, emissions of a SM $g/ \gamma$ or a hidden 
$\gamma_v/g_v$ can alternate in the evolution of a $F_v$.
Of course any of the related charges can be zero in a specific
process, in which case the following expressions simplify accordingly.
For the $i$'th emission, the $\pT$ evolution starts from the maximum 
scale given by the previous emission, with an overall starting scale 
$p_{\perp 0}$ set by the scale of the hard process, or of the decay in
which the radiating particle was produced. Thus the probability to 
pick a given $\pT$ takes the form
\begin{equation}
  \frac{\d \mathcal{P}}{\d \pT} = \left( 
    \frac{\d \mathcal{P}_{\mrm{QCD}}}{\d \pT} + 
    \frac{\d \mathcal{P}_{\mrm{QED}}}{\d \pT} +
    \frac{\d \mathcal{P}_{\mrm{secl}}}{\d \pT}
  \right) \;
  \exp \left( - \int_{\pT}^{p_{\perp i-1}} \left(
      \frac{\d \mathcal{P}_{\mrm{QCD}}}{\d \pT'} + 
      \frac{\d \mathcal{P}_{\mrm{QED}}}{\d \pT'} + 
      \frac{\d \mathcal{P}_{\mrm{secl}}}{\d \pT'} 
    \right) \d \pT' \right)
\end{equation}
where the exponential corresponds to the Sudakov form factor.
Implicitly one must also sum over all partons that can radiate.

To be more precise, radiation is based on a dipole picture, where it
is a pair of partons that collectively radiates a new parton. The
dipole assignment is worked out in the limit of infinitely many
(hidden or ordinary) colours, so that only planar colour flows need be
considered.

Technically the total radiation of the dipole is split into two ends,
where one end acts as radiator and the other as recoiler.  The
recoiler ensures that total energy and momentum is conserved during
the emission, with partons on the mass shell before and after the
emission. Each radiation kind defines its set of dipoles.
To take an example, consider $q \br{q} \to F_v \br{F}_v$, which
proceeds via an intermediate $s$-channel gluon. Since this gluon
carries no QED or hidden charge it follows that the $F_v \br{F}_v$
pair forms a dipole with respect to these two emission kinds. The
gluon \textit{does} carry QCD octet charge, however, so $F_v \br{F}_v$
do \textit{not} form a QCD dipole.  Instead each of them is attached
to another parton, either the beam remnant that carries the
corresponding anticolour or some other parton emitted as part of the
initial-state shower.  This means that QCD radiation can change the
invariant mass of the $F_v \br{F}_v$ system, while QED and hidden
radiation could not.  When a $\gamma$ or $\gamma_v$ is emitted the
dipole assignments are not modified, since these bosons do not carry
away any charge.  A $g$ or $g_v$ would, and so a new dipole would be
formed.  For QCD the dipole between $F_v$ and one beam remnant, say,
would be split into one between the $F_v$ and the $g$, and one further
from the $g$ to the remnant. For the secluded sector the $F_v \br{F}_v$
dipole would be split into two, $F_v g_v$ and $g_v \br{F}_v$. As the
shower evolves, the three different kinds of dipoles will diverge
further.

Note that, in the full event-generation machinery, the final-state
radiation considered here is also interleaved in $\pT$ with the
initial-state showers and with multiple parton-parton interactions
\cite{Corke:2010yf}.

If the $F_v$ fermion is allowed to decay into a SM and a hidden
particle, one must also consider the hidden radiation from the hidden
particle.

There is a clean separation between radiation in the production stage
of the $F_v \br{F}_v$ pair and in their respective decay.  Strictly
speaking this would only be valid when the $F_v$ width is small, but
that is the case that interests us here.

In the decay $F_v \to f q_v$ the QCD and QED charges go with the $f$
and the valley one with $q_v$. For all three interactions the dipole
is formed between the $f$ and the $q_v$, so that radiation preserves
the $F_v$ system mass, but in each case only the relevant dipole end
is allowed to radiate the kind of gauge bosons that goes with its
charge.  (Strictly speaking dipoles are stretched between the $f$ or
$q_v$ and the ``hole'' left behind by the decaying $F_v$. The
situation is closely analogous to $t \to b W^+$ decays.)

The number of parameters of the hidden shower depends upon the
scenario.  In the case of the interleaved shower, there are only two,
the most important on being one the coupling strength $\alpha_v$,
i.e. the equivalent of $\alpha_s$.  This coupling is taken to be a
constant, i.e. no running is included.

From a practical point of view it is doubtful that such a running
could be pinned down anyway, and from a theory point of view it means
we do not have to specify the full flavour structure of the hidden
sector. The second parameter is the lower cutoff scale for shower
evolution, by default chosen the same as for the QCD shower,
$p_{\perp\mathrm{min}} = 0.4$~GeV.

\subsubsection{Shower kinematics with massive hidden photons}
\label{sub:timeshower}

Showers are expected to reproduce the soft and collinear behaviour
of (leading-order) matrix elements (MEs), but there is no guarantee how 
trustworthy they are for hard wide-angle emissions. Therefore various 
correction techniques have been developed \cite{Buckley:2011ms}.
The technique we will use here is to generate trial emissions 
according to the PS, but then use the weights ratio ME/PS to accept 
emissions, i.e.\ PS times ME/PS equals ME.  For this re-weighting 
recipe to work, obviously the ME weight has to be below the PS one, 
but the difference should not be too big or else the efficiency will 
suffer. It should also be noted that the ME/PS ratio is evaluated 
without including the Sudakov form factor of the shower, while the 
shower evolution itself does build up the Sudakov. By the veto 
algorithm it then follows that the ME expression is exponentiated 
to provide the kernel of the Sudakov \cite{Bengtsson:1986hr}, a 
technique nowadays used as a key ingredient of the POWHEG approach 
\cite{Frixione:2007vw}. The choice of shower evolution variable  
lives on in the integration range of the Sudakov, but for the rest
the PS expressions disappear in the final answer.

In the past, this approach has only been developed for the emission 
of a massless gluon, however, and we now need to generalize that to 
an arbitrary combination of masses. A technical task is to recast 
the ME and PS expressions to use the same phase space variables, 
such that the ratio is well-defined.

We follow the existing approach of mapping the PS variables onto the
ME ones. Below we therefore introduce the ME three-body phase space,
subsequently how the PS variables populate this phase space, and
finally how the presence of two shower histories can be taken into
account.

Consider a dipole of invariant mass $m_0$, consisting of two endpoint
partons 1 and 2, with nominal masses $m_1$ and $m_2$. Assume that a
shower emission occurs from the parton-1 dipole end, generating a new
particle 3 with mass $m_3$. This implies that there was an
intermediate off-shell state 13 with mass $m_{13}$. That is, the
kinematics to describe is $p_0 \to p_{13} + p_2 \to p_1 + p_3 +
p_2$. Averaging over the angular orientation of events, the MEs can be
written in terms of the $x_i = 2 p_i p_0 / m_0^2$ and the $r_i = m_i^2
/ m_0^2$ variables, where the $x_i$ reduce to energy fractions in the
dipole rest frame, with normalization $x_1 + x_2 + x_3 = 2$. This
means there are only two free independent variables, traditionally
$x_1$ and $x_2$.

The PS is instead described in terms of the $\pTevol^2$ and $z$
variables. In the soft and collinear emission limit these are
well defined, but away from these limits different possibilities 
could be contemplated. Our choice is such that
\be
m_{13}^2 = m_1^2 + \frac{\pTevol^2}{z(1-z)} ~,
\ee 
or 
\be 
\pTevol^2 = z (1 - z) (m_{13}^2 - m_1^2) ~.  
\ee 
By standard two-body kinematics
for $p_0 \to p_{13} + p_2$ it follows that 
\be 
x_2 = \frac{m_0^2 + m_2^2 - m_{13}^2}{m_0^2} = 1 + r_2 - r_{13} ~, 
\ee 
and thus $x_1 + x_3 = 2 - x_2 = 1 + r_{13} - r_2$. If $m_1 = m_3 = 0$ 
one would further require that $z = x_1 / (x_1 + x_3)$. Taken together, 
this is enough to specify the three four-vectors $p_2$, $p_1^{(0)}$ and
$p_3^{(0)}$, up to three angles. These are chosen at follows: in the
$p_0$ rest frame parton 2 is assumed to keep its direction of
motion when $m_1 \to m_{13}$, while 1 and 3 are selected to have an
flat distribution in the azimuthal angle around the 13 direction,
which is parallel with the 1 direction before the emission.

The kinematics for the case with massive partons 1 and 3 can then be
constructed from the massless four-vectors as 
\ba
p_1 & = & (1 - k_1) p_1^{(0)} + k_3 p_3^{(0)} \\
p_3 & = & (1 - k_3) p_3^{(0)} + k_1 p_1^{(0)} \\
k_{1,3} & = & \frac{m_{13}^2 - \lambda_{13}
  \pm (m_3^3 - m_1^2)}{2 m_{13}^2}  \\
\lambda_{13} & = & \sqrt{(m_{13}^2 - m_1^2 - m_3^2)^2 - 4 m_1^2 m_3^2}
\ea 
The physics content is that the directions of partons 1 and 3 in
the $p_{13}$ rest frame are retained, while their three-momenta are
scaled down by a common factor sufficient to put the two partons on
their mass shells. Since $m_{13}$ is not changed by the operation it
is necessary that $m_{13} > m_1 + m_3$ for the rescaling to work.

The rescalings imply that 
\be 
\frac{x_1}{x_1 + x_3} = \frac{x_1}{2 - x_2} = (1 - k_1) z + k_3 (1 - z) 
= (1 - k_1 - k_3) z + k_3 ~, 
\ee
and thus 
\be 
z = \frac{1}{1 - k_1 - k_3} \left( \frac{x_1}{2 - x_2} - k_3 \right) ~.  
\ee

Now we need to find the Jacobian to translate the shower emission rate
from the $(\pTevol^2, z)$ space to the $(x_1, x_2)$ one. Note that 
$m_{13}^2 = m_0^2 (1 - x_2) + m_2^2$ is independent of $x_1$, and thus 
so are $k_1$ and $k_3$. Therefore only the ``diagonal'' terms 
$\partial\pTevol^2/\partial x_2$ and $\partial z /\partial x_1$ are 
needed. 

The shower emission rate itself is 
\be 
\frac{\d\pTevol^2}{\pTevol^2} \, \frac{2 \, \d z}{1 - z} ~.
\label{eq:psrate}
\ee 
Here an overall coupling factor $C_F \ \alpha_v / 2 \pi$ is
omitted for simplicity. Also the Sudakov form factor is omitted, as
already motivated. The $z$-dependent part may seem unfamiliar, but is
an upper approximation to the more familiar $q \to q g$ splitting
kernel $(1 + z^2)/(1 - z)$, where the difference between the two is
absorbed into the ME/PS weighting.

Put together, the shower emission rate translates into 
\ba 
\frac{\d \pTevol^2}{\pTevol^2} & = & 
\frac{\d (m_{13}^2 - m_1^2)}{m_{13}^2 - m_1^2}
= \frac{\d x_2}{ 1 - x_2 + r_2 - r_1} ~,\\
\frac{2 \, \d z}{1 - z} & = & 
\frac{2 \, \d x_1}{(1 - k_1 - k_3)(2 - x_2)} \, 
\frac{1}{1 - \frac{1}{1 - k_1 - k_3} \left( \frac{x_1}{2 - x_2}
- k_3 \right)} \nonumber \\
& = & \frac{2 \, \d x_1}{x_3 - k_1 (x_1 + x_3)} ~.  
\ea
When $m_3 \to 0$, and hence $k_1 \to 0$, this simplifies to the 
familiar expression \cite{Norrbin:2000uu} 
\be 
W_{PS,1} = \frac{\d \pTevol^2}{\pTevol^2} \, \frac{2 \, \d z}{1 - z} = 
\frac{2 \, \d x_1 \, \d x_2}{(1 - x_2 + r_2 - r_1) x_3}.  
\ee

If only parton 1 can radiate, as in 
$F_v \to q_v + f \to q_v + \gamma_v + f$, we are done. The fact that 
the MEs also contain a contribution from $\gamma_v$ emission off the 
$F_v$ does not change the picture, since that does not introduce any 
new singularities, and empirically the PS expression provides a valid 
upper limit.

For the radiation $F_v \overline{F}_v \to F_v \overline{F}_v \gamma_v$
the sum of the two possible shower emissions are needed to match to
the full MEs.  Alternatively, and more conveniently, the ME expression
can be split into two parts, each to be compared with only one shower
history. This split is done in proportion to the respective
propagator, i.e. assumed emission off parton $i$ is proportional to
$1/(m_{i3}^2 - m_i^2)$. The relative probability for parton 1 to
radiate thus is 
\be 
P_1 = \frac{m_{23}^2 - m_2^2}{(m_{13}^2 - m_1^2) +
  (m_{23}^2 - m_2^2)} = \frac{1 - x_1 + r_1 - r_2}{x_3} ~, 
\ee 
so that the ME weight to be associated with this dipole end is 
\be 
W_{ME.1} = P_1 \, \frac{1}{\sigma_0} \, \frac{\d \sigma}{\d x_1 \,\d x_2} 
\, \d x_1 \, \d x_2 ~.  
\ee
Thus we arrive at the ME/PS correction factor 
\ba 
R_1 = \frac{W_{ME,1}}{W_{PS,1}} & = & 
\frac{(1 - x_1 + r_1 - r_2)(1 - x_2 + r_2 - r_1)}{2} \,
\, \frac{1}{\sigma_0} \, \frac{\d \sigma}{\d x_1 \,\d x_2} \nonumber \\
& \times & \frac{x_3 - k_1 (x_1 + x_3)}{x_3} ~.  
\ea 
All the explicit dependence on $m_3$ is located in $k_1$ in the 
last term, but obviously implicitly the whole kinematics setup 
is affected by the value of $m_3$.

The matrix elements for the radiation off $F_v \overline{F}_v$ are
calculated with them as stable final-state particles. This means that,
to preserve gauge invariance, they must be assigned the same mass. On
the other hand, since they are supposed to decay, we allow them to
have a Breit-Wigner mass distribution. To resolve this discrepancy,
the real kinematics with two different masses is shifted to a
fictitious one where $F_v$ and $\overline{F}_v$ have the same mass,
and it is this fictitious one that is used in the three-parton
matrix-element evaluation. As a guiding principle, the $F_v$ and
$\overline{F}_v$ three-momenta are kept unchanged in the 
$F_v \overline{F}_v$ rest frame, and only energy is shuffled so as to
equalize the masses. Denoting the average mass $\overline{m}$, the
conservation of three-momentum implies that 
\be
\sqrt{\frac{m_{12}^2}{4} - \overline{m}^2} = \sqrt{\frac{(m_{12}^2 -
    m_1^2 - m_2^2)^2 - 4 m_1^2 m_2^2}{4 m_{12}^2}} ~, \ee which gives
\be \overline{m}^2 = \frac{m_1^2 + m_2^2}{2} - \frac{(m_1^2 -
  m_2^2)^2}{4 m_{12}^2} ~.  
\ee 
As above, the modified four-vectors $\overline{p}_1$ and 
$\overline{p}_2$ can be written as linear combinations of the 
original ones, with the constraints $\overline{p}_1^2 = 
\overline{p}_2^2 = \overline{m}^2$ giving the solution 
\ba
\overline{p}_1 & = & p_1 + 
  \frac{m_2^2 - m_1^2}{2 m_{12}^2} (p_1 + p_2) ~, \\
\overline{p}_2 & = & p_2 - 
  \frac{m_2^2 - m_1^2}{2 m_{12}^2} (p_1 + p_2) ~.  
\ea 
This translates into identical relationships for the
modified matrix-element variables $\overline{x}_1$ and
$\overline{x}_2$ in terms of the original $x_1$ and $x_2$ ones.

\subsubsection{Matrix element for radiation in production}

The implementation of the $\not \hspace{-3 pt} U(1)$ has required the
calculation of matrix element corrections 
$|M|^2_{f\bar{f}\rightarrow F_v  \bar{F_v}\gamma_v}$ for the pair production
 process $f \bar{f} \rightarrow F_v \bar{F_v} \gamma_v$ described in
Fig.~\ref{fig:production}.  
\FIGURE[ht]{
  \epsfig{file=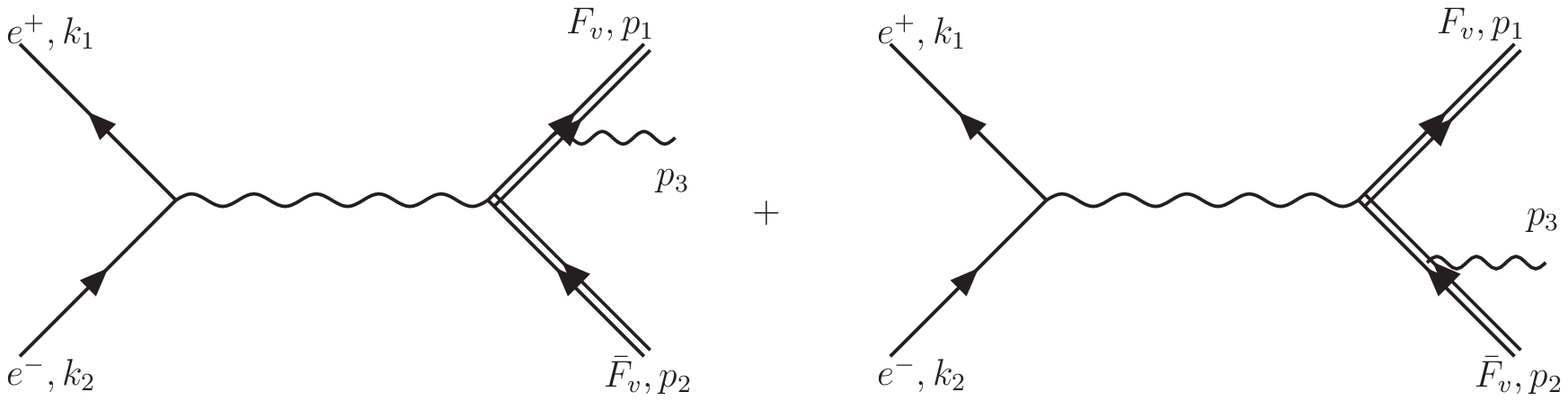,width=0.6\textwidth}
  \caption{The Feynman diagrams for the production.}
  \label{fig:production}
} 
This has required the generalization of the matrix element
corrections in \cite {Norrbin:2000uu} to the case of a massive photon:
\ba
|M|^2_{f\bar{f}\rightarrow F_v \bar{F_v}} & = & 
 \left( 1 - 4 r_1 \right)^{3/2} ~, \\
\frac{|M|^2_{f\bar{f}\rightarrow F_v \bar{F_v}\gamma_v}}% 
{|M|^2_{f\bar{f}\rightarrow F_v \bar{F_v}}} & = &  
(r_3 + 2 r_1) (-1 + 4 r_1) \left( \frac{1}{(1 - x_1)^2} +
\frac{1}{(1 - x_2)^2} \right) \nonumber \\
& + & \frac{-1 + 8 r_1 - x_2}{1 - x_1} 
+ \frac{-1 + 8 r_1 - x_1}{1 - x_2} \nonumber \\ 
& + & \frac{2(1 - 6 r_1 + 8 r_1^2 + 4 r_3 r_1)}{(1 - x_1)(1 - x_2)}
+ 2 ~. 
\ea 
Here $r_1 = r_2 = \overline{m}^2/ m_0^2$ and 
$r_3 = m^2_{\gamma_v} / m_0^2$. (Expressions for $r_1 \neq r_2$ have 
also been obtained but, by the preceding trick, are not needed.)  
Coupling constants have been omitted, as discussed before for the 
shower. Furthermore, to simplify calculations, the process is taken 
to proceed via the exchange of a scalar particle instead of a spin 1 
gauge boson. The $|M|^2_{f\bar{f}\rightarrow F_v \bar{F_v}\gamma_v}$ spin 
information, relevant for decay angular distributions, will be lost 
this way. Effects are known to be minor for the ME correction ratio 
\cite {Norrbin:2000uu}. As an illustration, the above expression 
reduces to $(x_1^2 + x_2^2)/((1-x_1)(1-x_2)) + 2$  for 
$r_1 = r_2 = r_3 = 0$, where the first term is the familiar 
expression for $e^+ e^- \to \gamma^*/Z^* \to q \bar q$, and the 
second finite term comes in addition for a spin 0 exchanged particle. 

\subsubsection{Matrix element for radiation in decay}

\FIGURE[ht]{ \epsfig{file=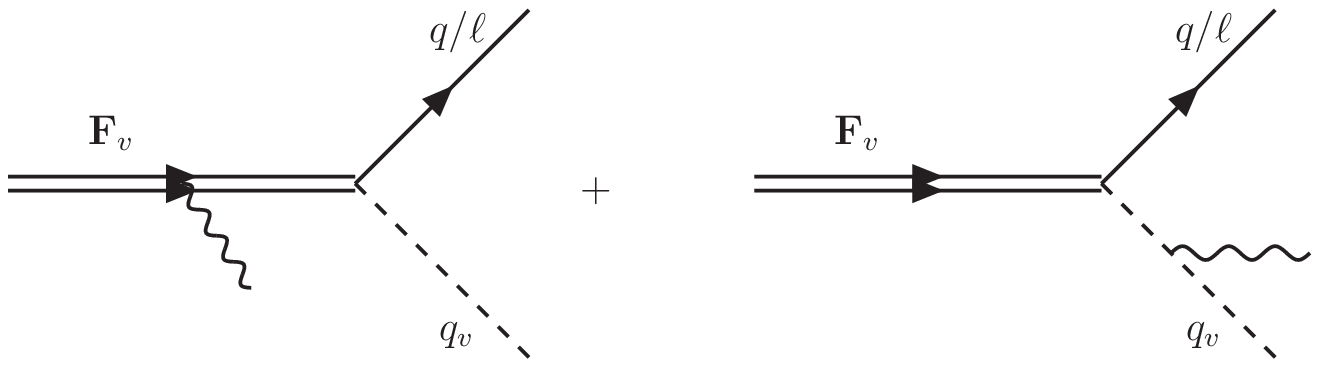,width=0.7\textwidth}
  \caption{The Feynman diagrams for the $F_v\rightarrow qv\hspace{4pt}
    q/\ell $ decay.}
  \label{fig:decay}
} 

The matrix elements corresponding to $F_v\rightarrow q_v f \gamma_v$ are
\ba
|M|^2_{F_v\rightarrow q_v f} &=& (1 - r_1 + r_2 + 2 q_2)
\sqrt{(1 - r_1 - r_2)^2 - 4 r_1 r_2} \\
\frac{|M|^2_{F_v\rightarrow q_v f \gamma_v}}{|M|^2_{F_v\rightarrow q_v f}}
&=& \frac{(r_3/2 + 2r_1^2 + r_2 r_3/2 + q_2 r_3 - 2r_1
  - r_1 r_3/2 - 2 r_1 r_2 - 4 r_1 q_2)}{(1 + r_2 - r_1 - x_2)^2}\nonumber\\
&+& \frac{(-2 + 2r_2^2 + 2r_1^2 + 2r_2 r_3 - 4q_2 + 2 q_2 r_3
  + 4q_2 r_2 - 4r_1 r_2 - 4r_1 q_2)}{(1+r_2-r_1-x_2)(r_3-x_3)} \nonumber\\
&+& \frac{(-2 - r_3/2 - 2r_2 - 4q_2 + 2r_1)}{(1+r_2-r_1-x_2)}\nonumber\\
&+& \frac{ (-2 - r_3 - 2r_2 - r_2 r_3 - 4q_2 - 2q_2 r_3
  + 2r_1 + r_1 r_3 )} {(r_3 - x_3)^2}\nonumber\\
&+& \frac{( -1 - r_3 - r_2 - 4q_2 + r_1 - x_2 )}{(r_3 - x_3)} + 1 ~.  
\ea
where $r_1 = m_{q_v}^2/ m_F^2$,  $r_2 = m_f^2/ m_F^2$,  
$r_3 = m^2_{\gamma_v} / m_F^2$ and $q_2 = m_f/ m_F = \sqrt{r_2}$.
The calculation has been done for the specific choice of $F_v$ and $f$ 
being fermions, and $q_v$ a scalar, but again the result should be 
representative also for other spin choices. 

\subsection{Hidden sector hadronization}
\label{sec:fragmentation}

If the $G$ group is the unbroken $SU(N)$, the gauge boson $g_v$ is
massless and the partons are confined. The picture therefore is
closely similar to that of QCD, and we will use exactly the same 
framework to describe hadronization physics as in QCD: the Lund 
string model \cite{Andersson:1983ia}. 

For the hidden sector, the model is most easily illustrated for the 
production of a back-to-back $q_v \bar q_v$ pair, with the perturbative
emission of additional $g_v$s neglected for now. In that case,  
as the partons move apart, the physical picture is that of a 
$v$-colour flux tube being stretched between the $q_v$ and the 
$\bar q_v$. If the tube is assumed to be uniform along its length,
this automatically leads to a confinement picture with a linearly
rising potential, $V(r) = \kappa r$.
 
In order to obtain a Lorentz covariant and causal description of the 
energy flow due to this linear confinement, the most straightforward 
approach is to use the dynamics of the massless relativistic string 
with no transverse degrees of freedom. The mathematical, one-dimensional 
string can be thought of as parameterizing the position of the axis of 
a cylindrically symmetric flux tube.

Now consider the simple $q_v \bar q_v$ two-parton event further.
As the $q_v$ and $\bar q_v$ move apart from the creation vertex, 
the potential energy stored in the string increases, and the string 
may break by the production of a new $q_v' \bar q_v'$ pair, so that 
the system splits into two colour singlet systems $q_v \bar q_v'$ and 
$q_v' \bar q_v$. If the invariant mass of either of these systems is 
large enough, further breaks may occur, and so on until only  
$v$-mesons remain. A system with $n$ primary $v$-mesons thus requires
$n-1$ breaks $q_{v,i}\bar q_{v,i}$ to produce a chain of $v$-mesons 
$q_v\bar q_{v,1}$, $q_{v,1} \bar q_{v,2}$, $q_{v,2}\bar q_{v,3}$, 
\ldots, $q_{v,n-1} \bar q_v$ stretching from the $q_v$ end to the 
$\bar q_v$ one.

The flavour of each $q_{v,i}\bar q_{v,i}$ is supposed to be a random 
choice among the $N_{\mrm{flav}}$ different flavours. Since all are
taken to have the same mass, for now, they are also produced at the 
same rate. This thus also goes for the $N_{\mrm{flav}}^2$ different 
$v$-meson flavour combinations. If the $q_v$ are fermions then both 
pseudoscalar and vector $v$-mesons can be produced, $\pi_v$ and 
$\rho_v$. Again disregarding possible effects of a mass splitting, 
simple spin counting predicts a relative production rate  
$\pi_v : \rho_v = 1 : 3$. 

The possibility of higher excited states is disregarded, as is known 
to offer a good approximation for the QCD case. Also $v$-baryon 
production is left out, which is a 10\% effect in QCD. For a 
generic $SU(N)$ group a $v$-baryon needs to consist of $N$ $v$-quarks. 
This should lead to exceedingly tiny rates for $N > 3$, while $N = 2$ 
could offer a more robust $v$-baryon production rate. 

The space--time picture of the string motion can be mapped onto 
a corresponding energy--momentum picture by noting that the constant 
string tension implies that the $v$-quarks lose a constant amount of 
energy per distance traveled. The different breaks are space-like 
separated, but two adjacent breaks are constrained by the fact that 
the string piece created by them has to be on the mass shell for the 
$v$-meson being produced. The space-like separation implies that the
fragmentation process can be traced in any order, e.g.\ from one of 
the endpoints inwards, while the constraint implies that there is 
only one kinematical degree of freedom for each new $v$-meson.
Typically it is chosen to be $z$, the light-cone momentum fraction
that the new $v$-meson takes from whatever is left in the system 
after previously produced $v$-meson have been subtracted off. 

By symmetry arguments one arrives at the Lund-Bowler shape of the 
$z$ probability distribution \cite{Bowler:1981sb}
\begin{equation}
f(z) \propto \frac{1}{z^{1 + b m_{q_v}^2}} \, (1 - z)^a \, 
\exp\left( - \frac{b m_{m_v}^2}{z} \right) ~,
\label{eq:bowler}
\end{equation}
where $m_{m_v} \approx 2 m_{q_v}$ is the mass of the produced $v$-meson.
The equation contains two free parameters, $a$ and $b$. Roughly 
speaking, these regulate the average rapidity spacing  of the 
$v$-mesons, and the size of the fluctuations around this average.
While $a$ is dimensionless, $b$ is not, which means that it becomes
necessary to adjust $b$ as $m_{q_v}$ is changed. For instance, assume 
that the $q_v$ mass is related to the strong-interaction scale 
$\Lambda_v$. Then, if $\Lambda_v$, $m_{q_v}$, $m_{m_v}$ and 
the collision energy are scaled up by a common factor, we would want 
to retain the same rapidity distribution of produced $v$-mesons.
This is achieved by rewriting $b m_{m_v}^2 = 
(b m_{q_v}^2) (m_{m_v}^2/m_{q_v}^2) = b' (m_{m_v}^2/m_{q_v}^2)$,
where now $b'$ can be assumed constant.  

In additional to fluctuations in the longitudinal fragmentation,
it is assumed that each new $q_v' \bar q_v'$ pair produced when the 
string breaks also carries an opposite and compensating transverse
momentum component. The $\pT$ of the $q_{v,i-1}\bar q_{v,i}$ meson is 
then given by the vector sum of its two constituent $\pT$ values.
The pair $\pT$ naturally arises in a tunneling production process, 
which also leads to a Gaussian $\pT$ distribution. The width $\sigma$ 
of this Gaussian again should scale like $\Lambda_v$, so we rewrite 
as $\sigma = (\sigma / m_{q_v}) m_{q_v} = \sigma' m_{q_v}$. 
When the $v$-mesons are allowed to acquire a $\pT$ it should be 
noted that the $m_{m_v}^2$ in eq.~(\ref{eq:bowler}) must be replaced 
by $m_{\perp m_v}^2 = m_{m_v}^2 + \pT^2$. 

In lack of further knowledge, it is convenient to assign $b'$ and 
$\sigma'$ values by analogy with standard QCD. To be more specific, 
we have in mind something like the $s$ quark, with a bare mass 
of the same order as $\Lambda$. For heavy quarks, like $c$ and $b$
in QCD, tunneling is suppressed, and the framework would have to be
further modified. To assess uncertainties in a scenario, it would
make sense to vary $b'$ and $\sigma'$ values over some range, say
a factor of two in either direction.

So far, the emission of $g_v$s has been neglected. When it is included, 
more complicated string topologies can arise. Like in QCD, the 
complexity is reduced by using the planar or large-$N_C$ limit 
\cite{'tHooft:1973jz}. In it a $v$-gluon is assigned an incoherent sum 
of a ($v$-)colour charge and a different anticolour one. In a branching 
$q_v \to q_v g_v$ the initial $q_v$ colour is taken away by the $g_v$
and a new colour-anticolour pair is stretched between the final $q_v$
and $g_v$. Similarly $g_v \to g_v g_v$ is associated with the creation
of a new colour. That way partons nearby in the shower evolution also 
come to be colour-connected. This leads to a picture of a single string,
consisting of several separate string pieces, stretching from one 
$q_v$ end to the $g_v$ it shares one colour with, on to the next 
colour-related $g_v$, and so on until the $\bar{q_v}$ string end is 
reached. Several separate string pieces could have formed, had 
perturbative branchings $g_v \rightarrow q_v \bar{q_v}$ been included,  
but, as in QCD, $g_v \rightarrow q_v \bar{q_v}$ should be rare both 
in relation to the more singular $g_v \rightarrow g_v g_v$
and in absolute terms.  

The motion of a string with several gluon kinks can be quite 
complicated, but it is possible to extend the fragmentation framework
of a single straight string also to the more complex topologies
\cite{Sjostrand:1984ic}. Basically the string will break up along its
length by the production of new $q_v' \bar q_v'$ pairs, with two 
adjacent breaks correlated in such a way that the $v$-meson produced
between them is on the mass shell. Sometimes the two breaks will be on 
either side of a $g_v$ string corner.

One of the key virtues of the string fragmentation approach is that it
is collinear and infrared safe. That is, the emission of a gluon 
disturbs the overall string motion and fragmentation 
vanishingly little in the small-angle/energy limit. Therefore the 
choice of lower cut-off scale for parton showers is not crucial: 
letting the shower evolve to smaller and smaller scales just adds 
smaller and smaller wrinkles on the string, which still maintains 
the same overall shape. 

The complete $v$-string fragmentation scenario contains a set of 
further technical details that are not described here. The key point, 
however, is that essentially all of the concepts of normal string
fragmentation framework can be taken over unchanged. The one new aspect
is what to do when the invariant mass of the hidden-valley system is 
too large to produce one single on-shell $v$-meson and too small to 
give two of them. As already explained, then the emission of 
$v$-glueballs is used to balance energy-momentum. 

\subsection{Decays back into the SM sector}

Disregarding the trivial direct decay $F_v\rightarrow f q_v$, the main 
decay modes back into the SM are through $\gamma_v$ kinetic mixing or
$Z^\prime$ decay. 
For $G = \not\hspace{-3pt}U(1)$ the $\gamma_v$ therefore can decay 
to SM particles with the same branching ratios as a photon of 
corresponding mass, i.e.\ $\propto e_f^2 N_c$, with 
$N_c = 1$ for leptons. For $G = SU(N)$ only the flavour-diagonal 
mesons can decay, either with a $\gamma_v$ 
or a $Z'$. (The former would imply that $G = SU(N) \times U(1)$,
which would require some further extensions relative to the scenarios
studied here.) A $\rho_v$ meson, with spin 1, could have the same
branching ratios as above, or slightly modified depending on the
$Z'$ couplings. A $\pi_v$ meson, with spin 0, would acquire an 
extra helicity factor $m_f^2$ that would favor the heaviest 
fermions kinematically allowed. Should the $\pi_v$ be scalar rather
than pseudoscalar there would also be a further threshold suppression,
in addition to the phase space one. 

The decay back into the standard model would be accompanied by normal
QED and QCD radiation, where relevant. Quarks and gluons would further
hadronize, as described by the normal Lund string model. That model is 
not carefully set up to handle different exclusive states if the 
$\gamma_v$ or $\rho_v/\pi_v$ mass is very low, of the order 1 or 2 GeV, 
but should be good enough as a starting point. For studies that 
zoom in on one specific mass, more carefully constructed decay tables 
could be used instead.

\section{Analysis of the different scenarios}
\label{sec:analysis}

The tools described above allow us to simulate several different
setups. We concentrate on the phenomenology of the six scenarios
listed in Table~\ref{tab:scenarios}.  Three different production mechanisms
are involved: $s$-channel  pair production via kinetic mixing with the
light $\gamma_v$ (KM$_{\gamma_v}$), $s$-channel  pair production
mediated by a $Z^\prime$  (M$_{Z^\prime}$) and $s$-channel pair
production via SM gauge bosons (SM) and the $F_v$ particles. For each
of these production mechanisms an Abelian setup and a non-Abelian one
are considered, labeled by A and NA respectively. Note that the
Abelian/non-Abelian group we refer to in the following analyses
correspond to the $G$ gauge group, not to the production
mechanisms. In the Abelian case $G=\brk$, while in the non-Abelian
case $G=SU(3)$.

\TABLE[h]{
\begin{tabular}{|c|c|c|c|c|} \hline \hspace{2pt} & production &
radiation & hadronization & decay to SM\\ \hline AM$_{Z^\prime}$ &
$e^+ e^- \rightarrow Z^\prime \rightarrow q_v \bar{q}_v$ &
$q_v\rightarrow q_v \gamma_v$ & --- & $\gamma_v\rightarrow$ SM \\
NAM$_{Z^\prime}$  & $e^+ e^- \rightarrow Z^\prime \rightarrow q_v
\bar{q}_v$ & $q_v \rightarrow q_v g_v$, $g_v \rightarrow g_v g_v$ &
$q_v\bar{q_v}\sim \pi_v/\rho_v$ & $\pi_v/\rho_v\rightarrow$ SM\\
KMA$_{\gamma_v}$  & $e^+ e^- \rightarrow \gamma_v \rightarrow q_v
\bar{q}_v$ & $q_v\rightarrow q_v \gamma_v$ & --- &
$\gamma_v\rightarrow$ SM\\ KMNA$_{\gamma_v}$ & $e^+ e^- \rightarrow
\gamma_v \rightarrow q_v  \bar{q}_v$ & $q_v \rightarrow q_v g_v$, $g_v
\rightarrow g_v g_v$ & $q_v\bar{q_v}\sim \pi_v/\rho_v$ &
$\pi_v/\rho_v\rightarrow$ SM\\ SMA & $e^+ e^- \rightarrow \gamma^*
\rightarrow E_v \bar{E}_v$ & $q_v\rightarrow q_v \gamma_v$ & --- &
$\gamma_v\rightarrow$ SM\\ SMNA & $e^+ e^- \rightarrow \gamma^*
\rightarrow E_v \bar{E}_v$ & $q_v \rightarrow q_v g_v$, $g_v
\rightarrow g_v g_v$ & $q_v\bar{q_v}\sim \pi_v/\rho_v$ &
$\pi_v/\rho_v\rightarrow$ SM\\ \hline
\end{tabular}
\label{tab:scenarios}
\caption{The six scenarios studied.}  }

The phenomenology of the six scenarios is a function of the pair
production cross section, which will in general depend upon the
specific  model realization of each setup. In particular,  for the
KM scenarios, the cross section will depend upon the size of the
kinetic mixing parameter $\epsilon$, while for the  $Z^\prime$ mediate
ones on the mass of the $Z^\prime$ and on its couplings  to the SM
particles and to the $v$-quarks.  The analysis is performed on
per-event distributions, so as to factor out this model dependence.
Assuming the same number of events are produced, the phenomenology 
of the setups will
depend upon a different number of parameters.  For the KMA$_{\gamma_v}$ and the 
AM$_{Z^\prime}$ one
must fix the $q_v$ masses, the $\gamma_v$ mass and the $\brk$ coupling
constant $\alpha_v$, while for the SMA production one must also  fix
the $F_v$ masses. In the corresponding KMNA$_{\gamma_v}$, AM$_{Z^\prime}$
 and SMNA one must fix the
meson masses, but these will be connected to the $q_v$ masses chosen,
and furthermore the $g_v$ remains massless. We select a scenario in 
which $m_{q_v}\sim \Lambda$, so that $m_{\pi_v}\sim m_{\rho_v}$.
This in turn ensures (as already described in section~\ref{sec:decay}) 
 that meson decay into SM lepton is not supressed.
For simplicity, in the
following analysis we assume only one mass for all $v$-quark flavours,
and only one common $\pi_v/\rho_v$ mass $m_{\pi_v/\rho_v}=2m_{q_v}$.  
One additional
simplification in the following analysis, is that for the SM cases we
assume the pair production of one single $E_v$ belonging to the standard 
model doublet with  no consideration for anomaly cancellation issues. In the
non-Abelian case we have assumed simple proportions $1 : 3$ for
$\pi_v : \rho_v$ production from fragmentation,  which comes from spin
counting when the $q_v$ has spin $1/2$. The branching ratios of the 
decays to standard model particles are fixed by the kinetic mixing 
mechanism.

We concentrate on the phenomenology of the six setups at an $e^+e^-$
collider with center-of-mass (CM) energy of $800$ GeV.  A similar study
for $pp$ colliders like the LHC is also  possible, and obviously more
relevant in the near future, but makes it less transparent to
compare and understand the properties of the models. Bremsstrahlung
corrections have been included, and we shall see that these can give a
non-negligible effect, whereas the machine-specific beamstrahlung
has not. All of the figures in this section are based on a Monte Carlo
statistics of 10000 events. 

As a consequence of the $e^+e^-$ collider choice, the events have a
spherical symmetry rather than a cylindrical one, i.e. are described
in terms of particle energy and $(\theta, \phi)$ variables rather than
in terms of $E_T$ and $(\eta, \phi)$.  The jet clustering algorithms
are thus determined by the spherical topology and we primarily use the
\textsc{Pythia} built-in \ttt{ClusterJet} Jade algorithm
\cite{Bartel:1985ey, Bartel:1986ua}. The Jade algorithm is geared
towards clustering objects nearby in mass, and so for clustering a
variable number of fixed-mass $\gamma_v/\pi_v/\rho_v$ systems it
is more relevant than clustering e.g.\ in transverse momenta or angles. 

\subsection{Basic distributions}
\label{sec:basic_distr} 

The phenomenology of the six different setups
is discussed in detail in dedicated subsections. We here would like to
discuss the general features of the secluded sector signals
and to introduce the observables we focus  on.

\FIGURE[t]{
\begin{minipage}[b]{0.4\linewidth}
\epsfig{file=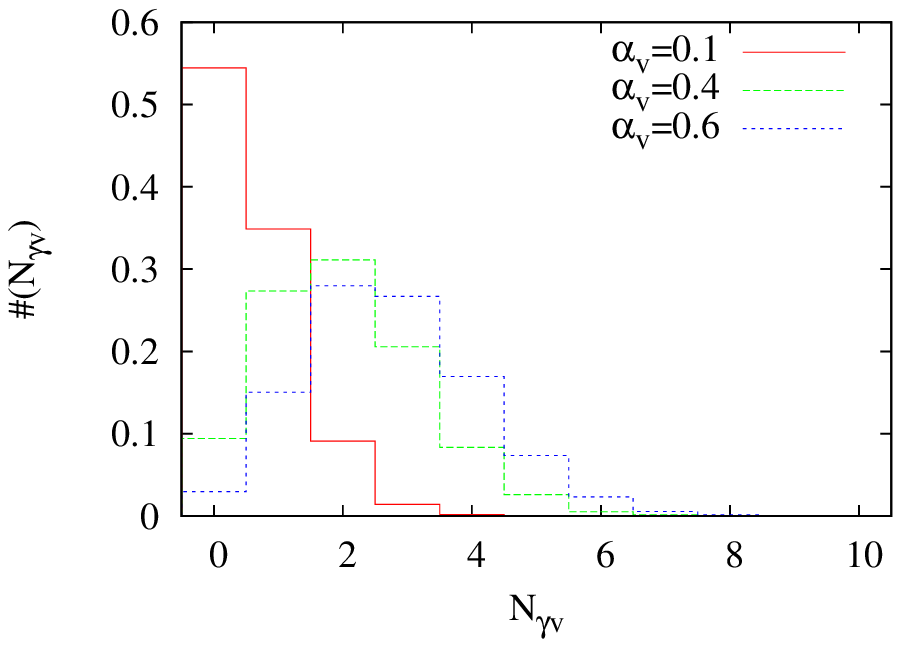,width=1.2\linewidth,angle=0}
\end{minipage}
\hspace{1cm}
\begin{minipage}[b]{0.4\linewidth}
\epsfig{file=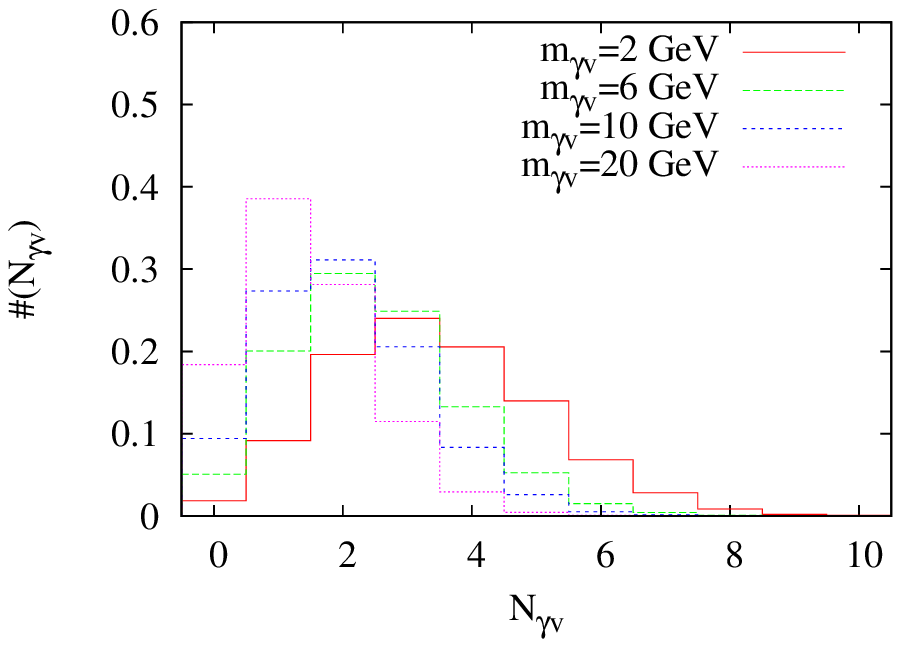,width=1.2\linewidth,angle=0}
\end{minipage}
\caption{{\itshape AM$_{Z^\prime}$}: the number of $\gamma_v$ gauge
bosons emitted per event. On the left we highlight the $\alpha_v$
dependence, while on the right the $m_{\gamma_v}$ dependence. On the left side
$m_{q_v}=50$ GeV, $m_{\gamma_v}=10$ GeV and $\alpha_v=0.1, 0.4, 0.6$, while on the right side
$m_{\gamma_v}=2, 6, 10, 20$ GeV and the coupling is fixed at $\alpha_v=0.4$.}
\label{fig:KMAz_Ngv} }

One of the benefits of a Monte Carlo simulation is that one may look
behind the scene, to study also the distributions of the invisible
secluded sector particles. These can then be compared with the SM
particle distributions to determine which  features are governed by
the the secluded sector dynamics, and which come from the decays to
the SM.  In this spirit, Fig.~\ref{fig:KMAz_Ngv} shows the number of 
$\gamma_v$ gauge bosons emitted per event in the AM$_{Z^\prime}$ case. 
On the left we highlight the $\alpha_v$ dependence, on the right 
the $m_{\gamma_v}$ dependence. Not unexpectedly, the number of
$\gamma_v$ increases almost linearly with $\alpha_v$, up to saturation
effects from energy--momentum conservation.

\FIGURE[t]{
\begin{minipage}[b]{0.45\linewidth}
\epsfig{file=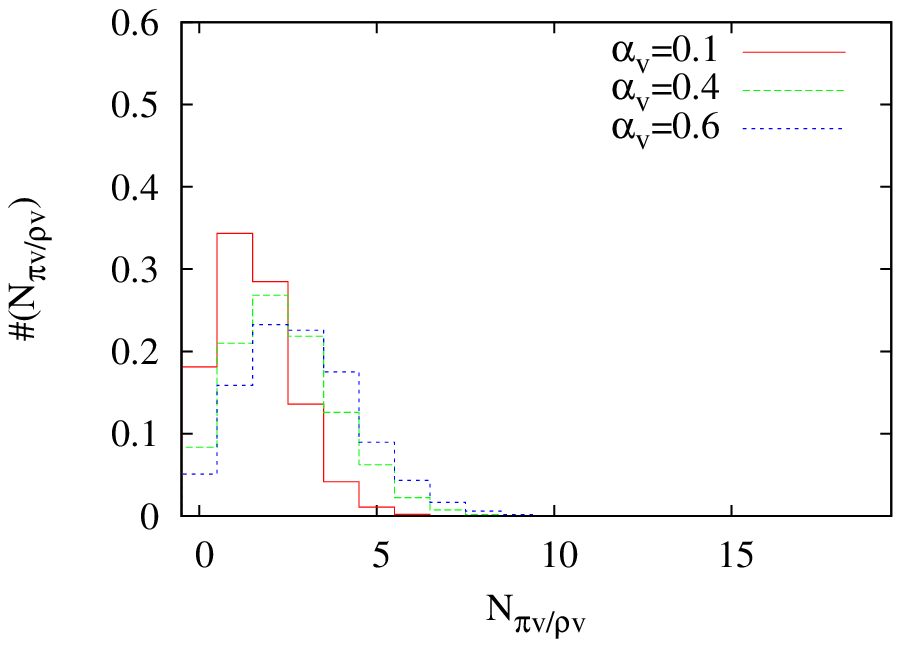,scale=0.7,angle=0}
\end{minipage}
\begin{minipage}[b]{0.45\linewidth}
\epsfig{file=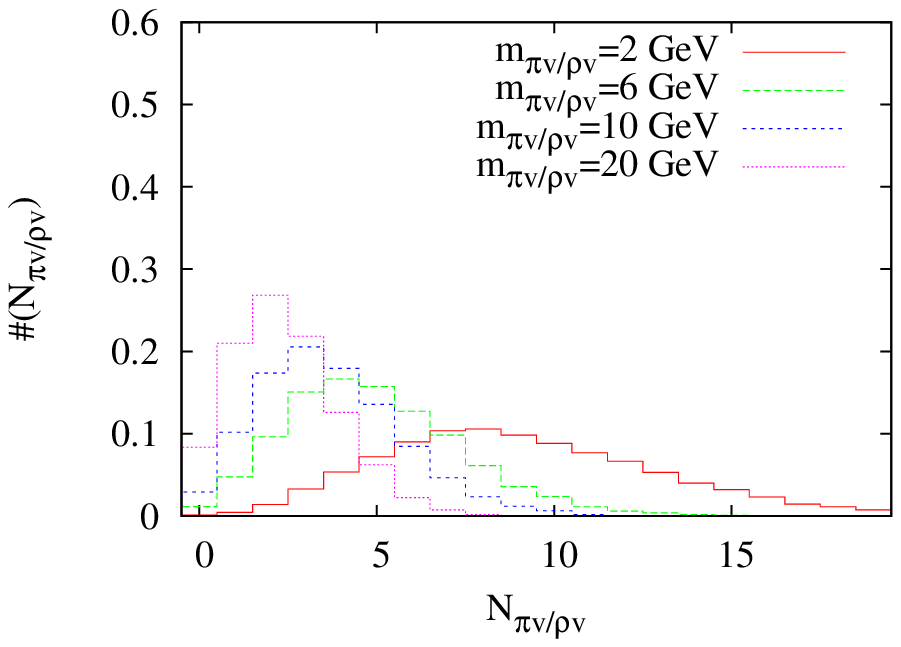,scale=0.7,angle=0}
\end{minipage}
\caption{{\itshape NAM$_{Z^\prime}$}: the number of flavour diagonal $\pi_v/\rho_v$ mesons emitted per event. On the left we emphasize the $\alpha_v$ dependence, while on the right the $m_{\pi_v/\rho_v}$ dependence. On the left the meson mass is fixed at $m_{\pi_v/\rho_v}=10$ GeV and the coupling varies among $\alpha_v=0.1, 0.4, 0.6$, while on the right side $m_{\pi_v/\rho_v}=2, 6, 10, 20$ GeV (which in turn implies $m_{q_v}=1,3,5,10$ GeV), the coupling is fixed at $\alpha_v=0.4$ and the number of flavours is $N_\mrm{flav}=4$. 
}
\label{fig:KMNAz_Nmes} }

Compare this distribution with the corresponding non-Abelian
AM$_{Z^\prime}$ case in Fig.~\ref{fig:KMNAz_Nmes} for the flavour
diagonal $\pi_v/\rho_v$. Again the number of $g_v$ grows with 
$\alpha_v$ \footnote{Note that, in this case, the emission rate
$q_v \to q_v g_v$ is proportional to $C_F\alpha_v$, with 
$C_F = (N^2 - 1)/(2 N)$, and $g_v \to g_v g_v$ to $N\alpha_v$.}, 
but the number of $v$-mesons does not primarily reflect this 
$\alpha_v$ dependence. Instead the number of $v$-mesons produced 
by string fragmentation rather reflects the masses of the $v$-quarks 
(and thereby of the mesons) and the fragmentation parameters, see
Fig.~\ref{fig:KMNAz_Nmes} right plot. Specifically, even with
$\alpha_v$ set to zero for the perturbative evolution, there would 
still be non-perturbative production of $v$-mesons from the single 
string piece stretched directly from the $q_v$ to the $\bar{q}_v$.
With $\alpha_v$ nonzero the string is stretched via a number of 
intermediate $g_v$ gluons that form transverse kinks along the string,
and this gives a larger multiplicity during the hadronization.

Comparing the number of $\gamma_v$ in Fig.~\ref{fig:KMAz_Ngv}  
with the corresponding distributions in the other Abelian setups, 
 KMA$_{\gamma_v}$ and SMA, in Fig.~\ref{fig:KMAg_Ngv} and  
Fig.~\ref{fig:SMA_Ngv} respectively, the two KMA$_{\gamma_v}$ and 
AM$_{Z^\prime}$ setups produce 
similar distributions, while the SMA produces much fewer $\gamma_v$.
The SMA difference is due to the more complicated kinematics, where
the electrons from the $E_v\rightarrow e q_v$ decays take away a large 
fraction of energy and momentum that then cannot be used for  
$\gamma_v$ emissions.  

The average charged multiplicity of an event, 
Fig.~\ref{fig:KMz_Np}, will be directly proportional to the number 
of $\gamma_v/\pi_v/\rho_v$ produced. The trends from above are thus 
reproduced, that the non-Abelian multiplicity varies only mildly
with $\alpha_v$, while the variation is more pronounced in the 
Abelian case. The constant of proportionality 
depends on the $\gamma_v/\pi_v/\rho_v$ mass, with more massive states
obviously producing more charged particles per state. This offsets 
the corresponding reduction in production rate of more massive
$\gamma_v/\pi_v/\rho_v$, other conditions being the same.
Similarly the number of jets should be proportional to the number of
$\gamma_v/\pi_v/\rho_v$ emitted, see further Sec.~\ref{sec:KMz}.

\FIGURE[t]{
\begin{minipage}[b]{0.45\linewidth}
\epsfig{file=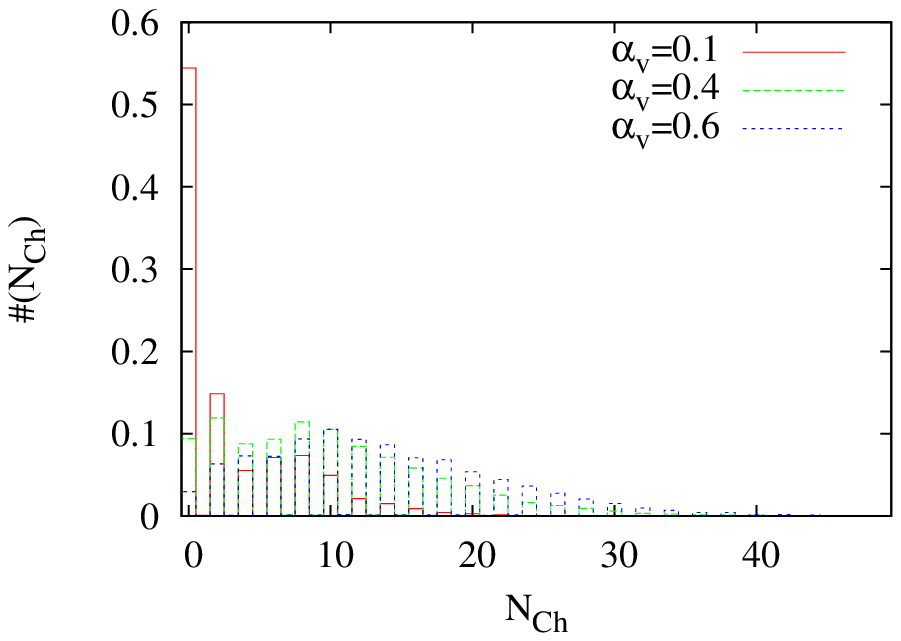,scale=0.75,angle=0}
\end{minipage} \hspace{0.5cm}
\begin{minipage}[b]{0.45\linewidth}
\epsfig{file=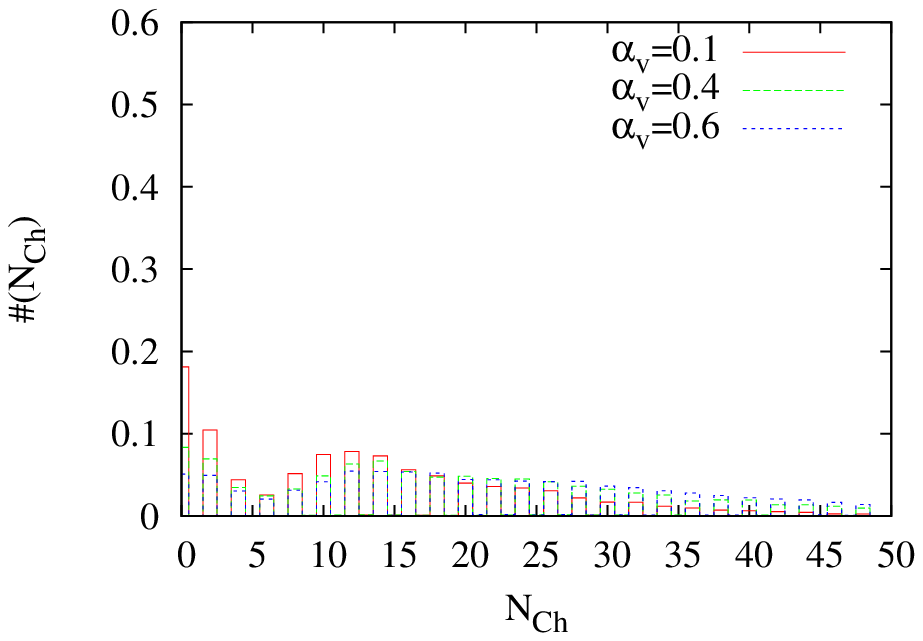,scale=0.75,angle=0}
\end{minipage}
\caption{{\itshape AM$_{Z^\prime}$, NAM$_{Z^\prime}$}: $\alpha_v$ dependence
 of the overall number of charged particles emitted per event in the Abelian (left) and non-Abelian (right) case. For the Abelian cases  $m_{q_v}=50$ GeV and $m_{\gamma_v}=10$ GeV, for the non-Abelian cases $m_{q_v}=5$,  $m_{\pi_v/\rho_v}=10$ GeV.}
\label{fig:KMz_Np} }

Without an understanding of the $\gamma_v/\pi_v/\rho_v$ mass spectra,
the mix of effects would make an $\alpha_v$ determination nontrivial,
especially in the non-Abelian case. Even with a mass fixed, e.g.\ by 
a peak in the lepton pair mass spectrum, other model parameters will 
enter the game. One such parameter is the number $N_\mrm{flav}$ of 
$q_v$ flavours.
Since only $1/N_\mrm{flav}$ of the $\pi_v/\rho_v$ would decay back into 
the SM the visible energy is reduced accordingly. With all $q_v$ having 
the same mass, the relation 
$\langle E_{\mrm{visible}} \rangle/E_{\mrm{cm}} = 1/N_\mrm{flav}$
works fine to determine $N_\mrm{flav}$, but deviations should be expected 
for a more sophisticated mass spectrum . Furthermore the
$\pi_v : \rho_v$ mix, with different branching ratios for the two,
needs to be considered. If the $\pi_v$ fraction is large, the number
of heavy leptons and hadrons produced may increase substantially, see
\cite{Strassler:2006im}.
  
\FIGURE[t]{
\begin{minipage}[b]{0.45\linewidth}
\epsfig{file=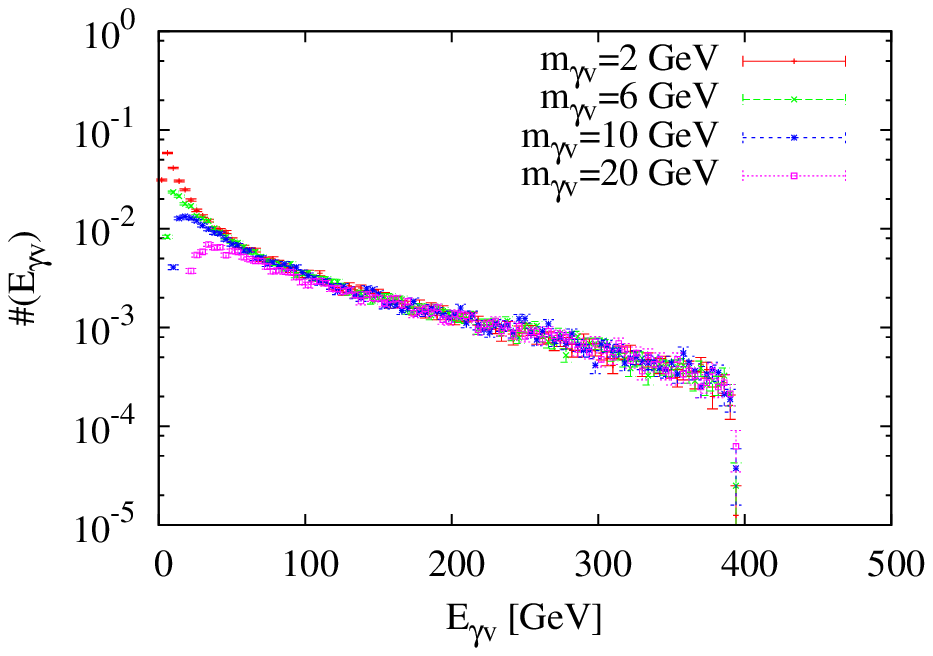,scale=0.7,angle=0}
\epsfig{file=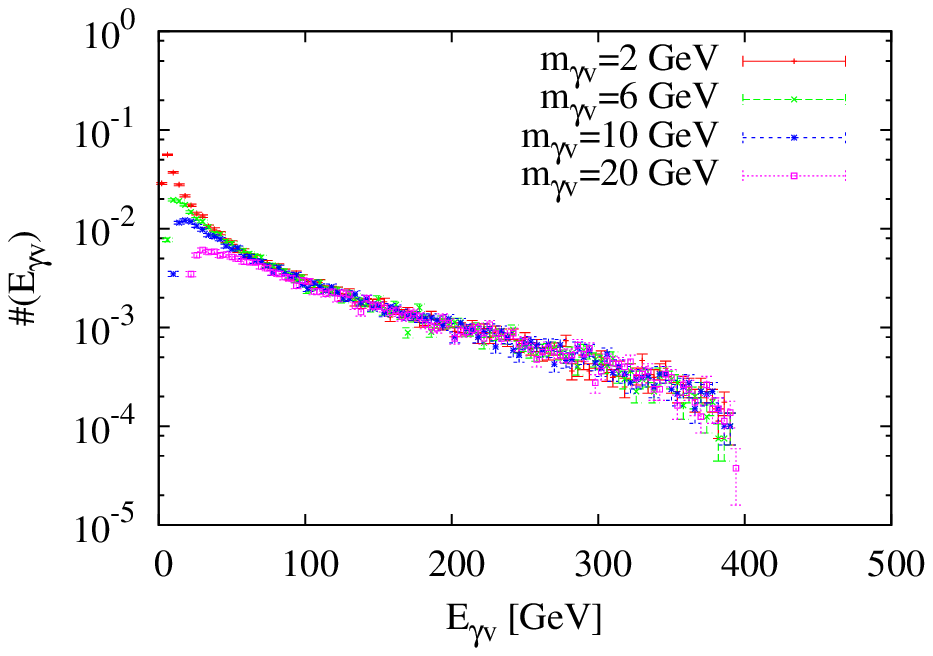,scale=0.7,angle=0}
\end{minipage}
\begin{minipage}[b]{0.45\linewidth}
\epsfig{file=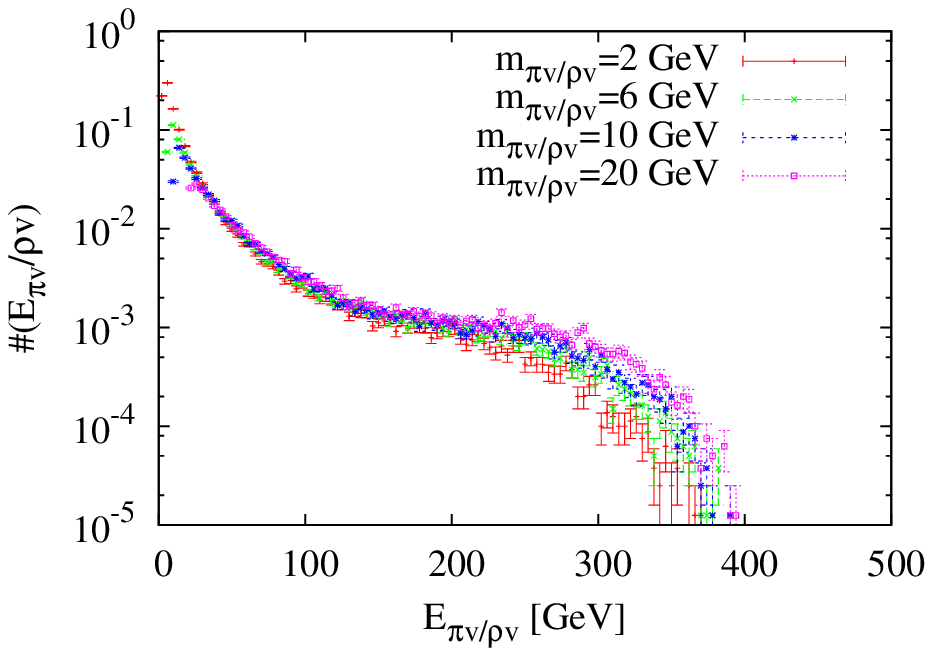,scale=0.7,angle=0}
\epsfig{file=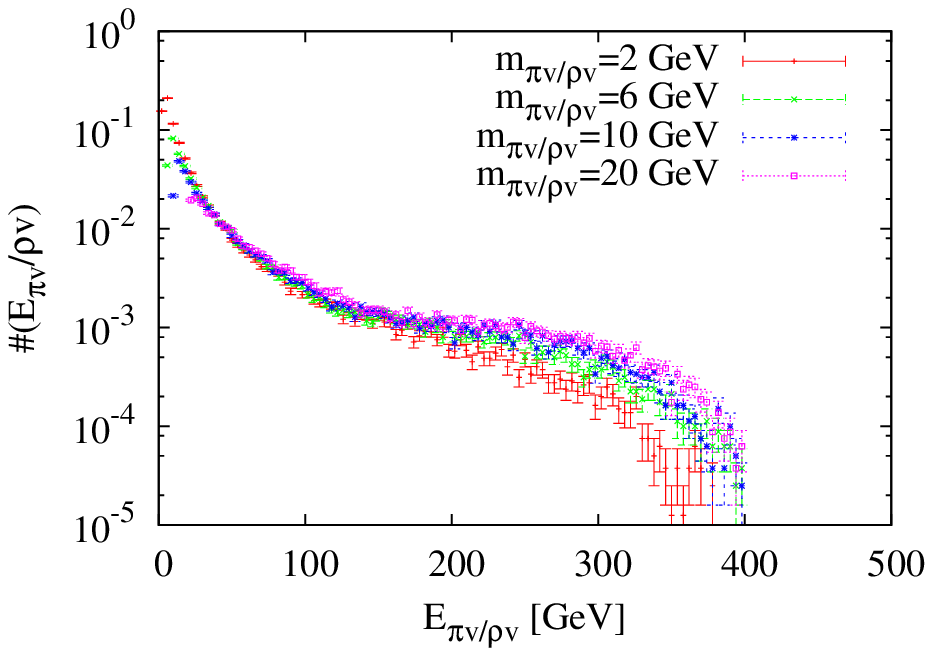,scale=0.7,angle=0}
\end{minipage}
\caption{{\itshape M$_{Z^\prime}$} vs {\itshape KM$_{\gamma_v}$}: the energy spectrum of the $\gamma_v$ and diagonal $\pi_v/\rho_v$ emitted per event. On top: left side shows the energy distribution for AM$_{Z^\prime}$, the right side shows the corresponding one for NAM$_{Z^\prime}$. 
Bottom: left side shows the energy distribution for KMA$_{\gamma_v}$, right side shows KMNA$_{\gamma_v}$. For the abelian cases $mq_v=50$ GeV, for the non abelian cases $m_{q_v}=m_{\gamma_v}/2$. In all four cases $\alpha_v=0.4$.}
\label{fig:KMz_KMg_Egv/Emes} }

In Fig.~\ref{fig:KMz_KMg_Egv/Emes} we show the energy spectra of the
hidden sector $\gamma_v$ and $\rho_v/\pi_v$. Note the difference
between the NAM$_{Z^\prime}$ setup and the KMNA$\gamma_v$ one. This
is due to the difference in the amount of initial-state radiation in
the two cases, as discussed in Sec.~\ref{sec:KMg} and shown in
Fig.~\ref{fig:KMNAg_eISR}.

\FIGURE[t]{
\begin{minipage}[b]{0.45\linewidth}
\epsfig{file=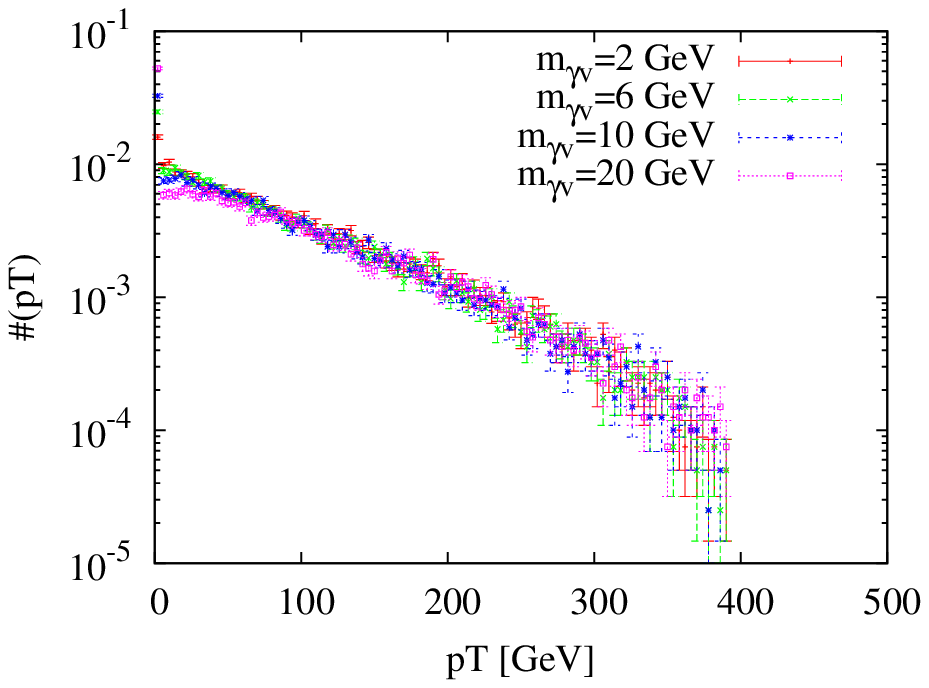,scale=0.71,angle=0}
\epsfig{file=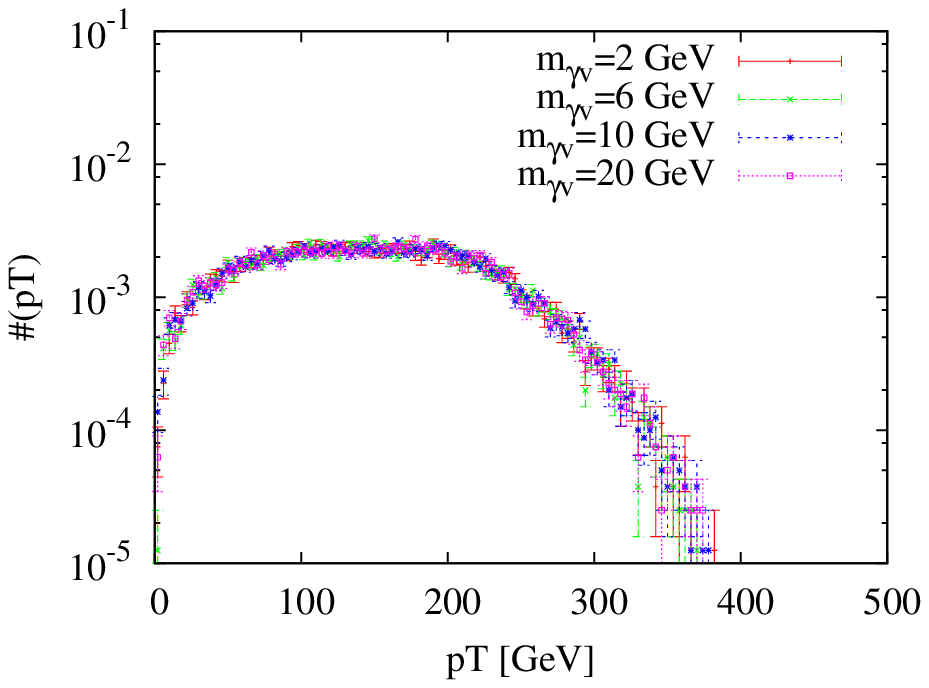,scale=0.71,angle=0}
\end{minipage}
\hspace{0.5cm}
\begin{minipage}[b]{0.45\linewidth}
\epsfig{file=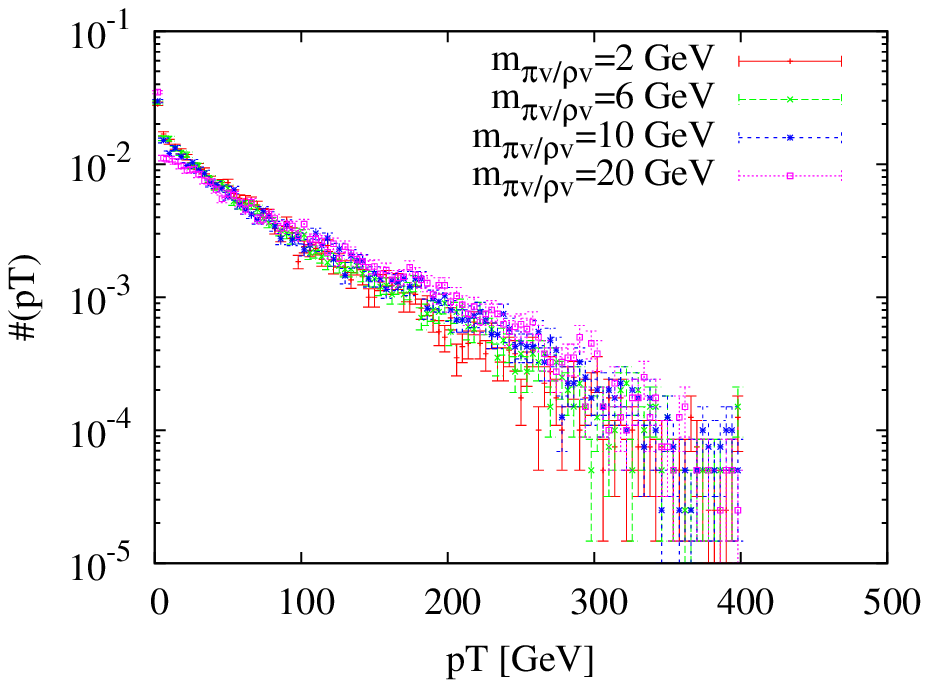,scale=0.71,angle=0}
\epsfig{file=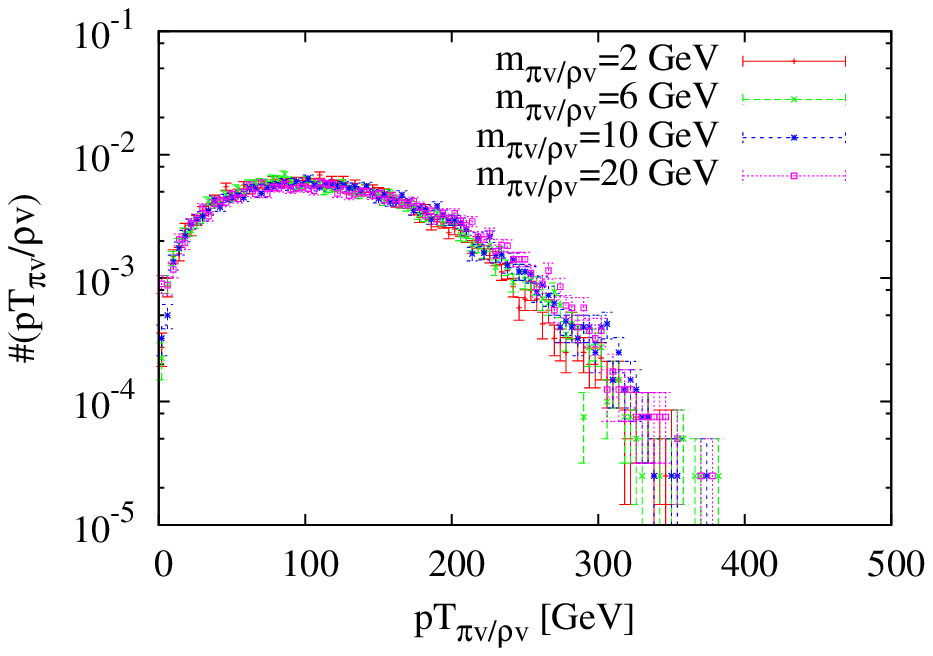,scale=0.71,angle=0}
\end{minipage}
\caption{{\itshape KMA$_{\gamma_v}$, KMNA$_{\gamma_v}$, SMA, SMNA}: the 
$\npT$ spectrum in each event. For the Abelian cases (left) the $\npT$ 
is due to the $q_v$ escaping detection. For the non-Abelian cases (right) 
it is due to the $v$-flavoured mesons not decaying into SM particles. In 
the Abelian cases $m_{q_v}=50$ GeV, while in the non-Abelian cases (right) 
$m_{q_v}=1,3,5,10$.  For all plots $\alpha_v=0.4$.}
\label{fig:KMg_SM_pTmiss}}

The energy and momenta of the $v$-sector particles not decaying back
into the SM is the prime source of the missing $\pT$ distributions,
\footnote{with some extra effects from neutrinos e.g.\ in $b$, $c$ 
and $\tau$ decays, included in the plots but here not considered on
their own.} see Fig.~\ref{fig:KMg_SM_pTmiss}.
In each of the six setups there is only
one source of missing energy. In the Abelian ones it is the $q_v$s 
that escape detection, while in the non-Abelian ones it is the stable 
non-diagonal $v$-mesons. For KMA$_{\gamma_v}$ the falling Abelian $\npT$ spectra are
easily understood from the bremsstrahlung nature of the $\gamma_v$
emissions. The spike at $\npT = 0$ comes from events without any 
emissions at all, where all the energy is carried away by the 
invisible $q_v$s, and would hardly be selected by a detector trigger.
(ISR photons might be used as a trigger in this case, but with
irreducible backgrounds e.g.\ from $Z^0 \to \nu \bar{\nu}$ it is not 
likely.). In the KMNA case the momentum of non-diagonal $v$-mesons does 
not leak back, this again allows a falling slope and a spike at 
$\npT = 0$, for events in which equal amount of energy in the non-diagonal 
mesons radiated from either side of the $q_v \bar{q}_v$ system. 
For the SMA and SMNA scenarios, on the other hand, the
starting point is the pT imbalance that comes from the $e^+$ and
$e^-$ from the $E_v$ and $\bar{E}_v$ decays, which have no reason to
 balance each other. So even without $\gamma_v$ emission,
or diagonal $\pi_v/\rho_v$, there will will be a $\pT$ imbalance. 

 In the
SMNA case, though the spectrum is shifted towards lower 
missing $\npT = 0$ because on average a higher number of mesons
are radiated, so it is less likely to have an event with the two
leptons back-to-back. There could also be $\npT=0$ cases in which 
all the mesons are flavour diagonal and all the energy-momenta 
decays back into the SM, but these events are very rare. 

In the Abelian case, the missing $\pT$ distribution is directly
connected to the mass parameter values $m_{q_v}$ and, in the SMA case,
to the $m_{E_v}$. The value of $m_{q_v}$ in the KM/${Z^\prime}$
mediated cases may be extracted from the kinematic limit given by the
``shoulder'' of the distribution. In the SM-mediated case, where two
different fermion mass scales are involved, one can extract a
relationship for the relative size of the two from lepton energy 
distributions such as the one in Fig.~\ref{fig:SM_eLeptons}, 
see \cite{Carloni:2010tw} for details. The distribution that 
directly pinpoints the mass of the particle decaying back into the 
SM, though, is the invariant mass of the lepton pairs
produced, and that of the hadronic jets. We will discuss these
distributions in the sections dedicated to each scenario. 

\subsection{AM$_{Z^\prime}$ and NAM$_{Z^\prime}$}
\label{sec:KMz} 

In discussing the phenomenology of the different scenarios we will 
describe the $v$-sector particle distributions first, then the 
visible particle distributions followed by the jet distributions.

The number of particles of $\gamma_v$ photons emitted in the
AM$_{Z^\prime}$ and NAM$_{Z^\prime}$ scenarios was described in
Fig.~\ref{fig:KMAz_Ngv} and Fig.~\ref{fig:KMNAz_Nmes} in the previous
section. The difference between the Abelian and non-Abelian 
dependence on the $\alpha_v$ and $m_{\gamma_v/\pi_v/\rho_v}$
parameters has already been highlighted in the same
Sec.~\ref{sec:basic_distr}, as well as the $\gamma_v/\pi_v/\rho_v$ 
energy distributions, the charged multiplicity and the $\npT$ spectrum.
The difference between KMA$_{\gamma_v}$ and SMA was also discussed.

%Lisa! You need to explain the following section!
The difference between the Abelian and non-Abelian $\npT$ distribution
in Fig.~\ref{fig:KMg_SM_pTmiss} is more subtle. In the Abelian case an 
event has maximum $\pT$
inbalance when one of the $q_v/\bar{q}_v$ produced emits 
a collinear $\gamma_v$ which takes most of the $q_v$ ($\bar{q}_v$) 
momentum  while the other $v$-quark has no emission and goes undetected. 
The more $\gamma_v$ are emitted, the less likely it is that the undetected 
$\bar{q}_v$ will have maximal energy. This remains true for all the 
$m_{\gamma_v}$ contemplated (except in the low-$\pT$ region). 
In the non-Abelian case, to have large $\pT$ inbalance the event must 
 produce few energetic mesons back-to-back and have the mostly 
flavoured mesons at one end and mostly flavour neutral mesons at the other 
end.  When the meson mass is lower, there is a higher probability 
of the string producing a large number of mesons and the likelihood 
of having large $\npT$ falls rapidly. When the meson masses are higher and 
fewer $\pi_v/\rho_v$ are produced the high $\npT$ distribution falls off 
less rapidly. 
 
\FIGURE[t]{
\begin{minipage}[b]{0.45\linewidth}
\epsfig{file=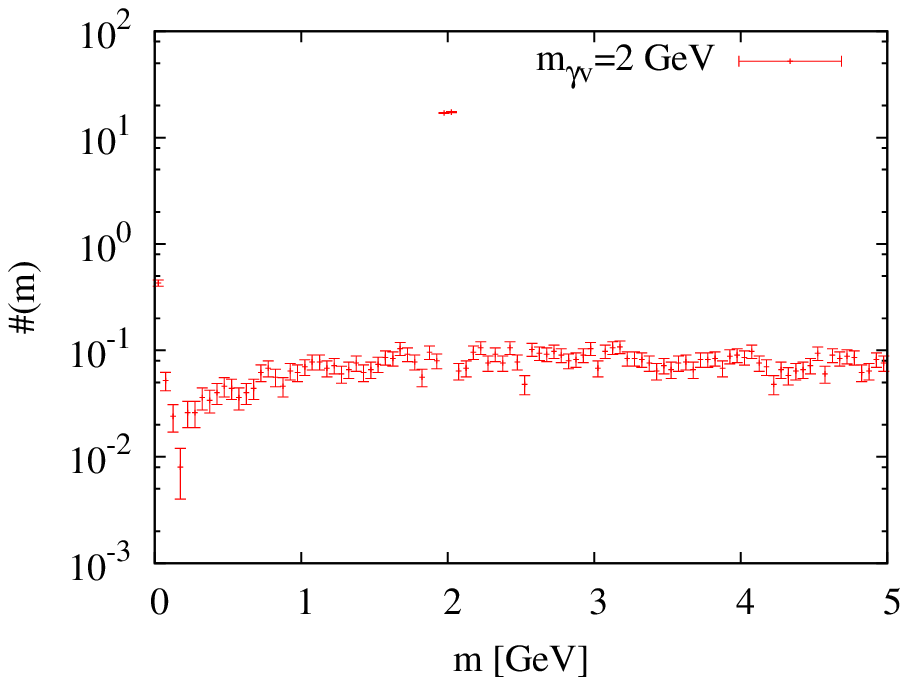,scale=0.7,angle=0}
\end{minipage}
\begin{minipage}[b]{0.45\linewidth}
\epsfig{file=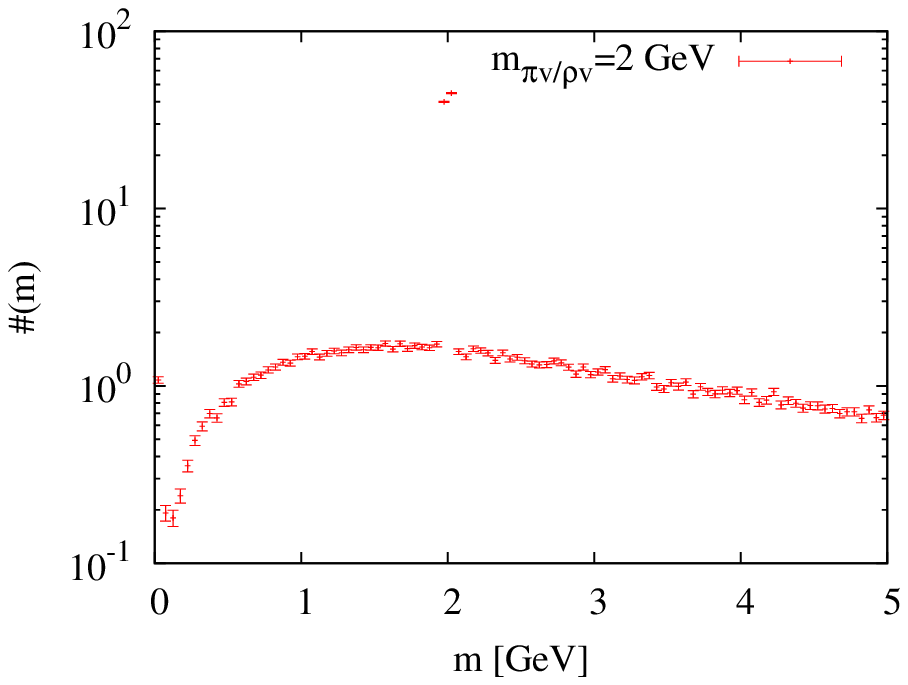,scale=0.7,angle=0}
\end{minipage}
\caption{{\itshape AM$_{Z^\prime}$} and {\itshape NAM$_{Z^\prime}$}. The distribution of the invariant mass of the lepton pairs. 
Note the peak at 2 GeV, in both cases corresponding to the mass to be reconstructed $m_{\gamma_v}=m_{\rho_v/\pi_v}$. On the left, in the Abelian case, $m_{q_v}=50$ GeV, while on the right, in the non-Abelian case, $m_{q_v}=1$ GeV. The coupling is fixed at $\alpha_v=0.4$.}
\label{fig:KMz_mLeptons} }

The $\gamma_v/\pi_v/\rho_v$ mass can be extracted from the
lepton pair invariant mass, where it shows up as a well-defined 
spike, Fig.~\ref{fig:KMz_mLeptons}. 
(The additional spike near zero mass is mainly related to 
Dalitz decay $\pi^0 \to e^+ e^- \gamma$.) Once the 
mass is known, the remaining hadrons and photons may be clustered
using the Jade algorithm, with $m_{\gamma_v/\pi_v/\rho_v}$ as the joining
scale. The corresponding number of jets  and invariant mass
distribution for the hadronic jets is given in
Fig.~\ref{fig:KMAz_Nj}.  The jet invariant mass distribution clearly
shows the peaks connected to the invariant mass of the $\gamma_v$. 
The background comes from several sources. The spike at zero mass
is mainly related to ISR photons; although we assume no detection
within 50 mrad of the beam directions, some isolated photons do show 
up above this angle and form jets on their own. When kinematically
possible, $\tau$ decays and $c$ and $b$ decays will also occur. 
These contain neutrino products that reduce the visible mass, thus
contributing to a continuum below the mass peak. Finally, 
misidentifications among partly overlapping systems leads to tails
on both sides of the peak.

\FIGURE[t]{
\begin{minipage}[b]{0.45\linewidth}
\epsfig{file=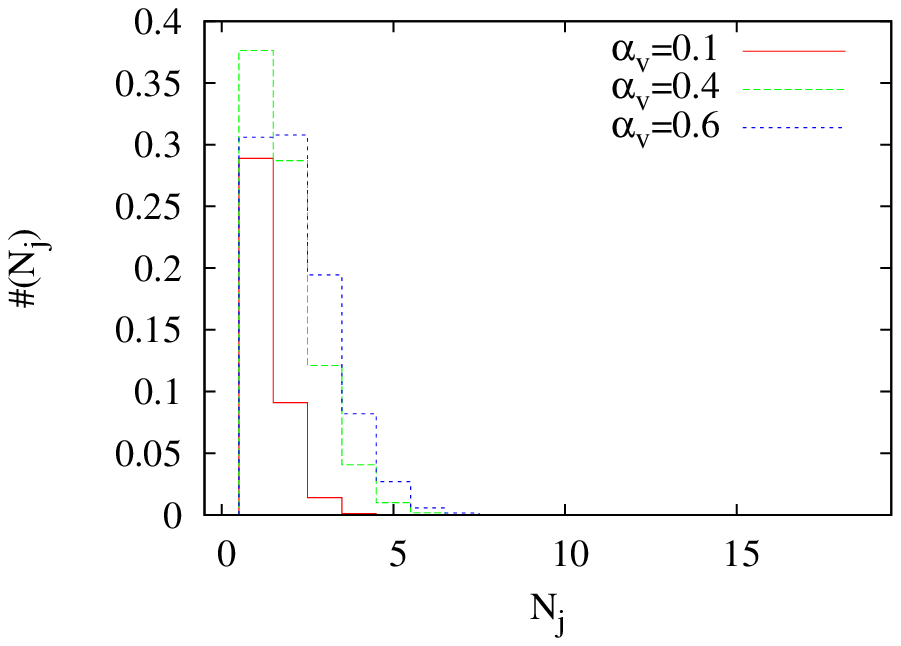,scale=0.7,angle=0}
\end{minipage}
\begin{minipage}[b]{0.45\linewidth}
\epsfig{file=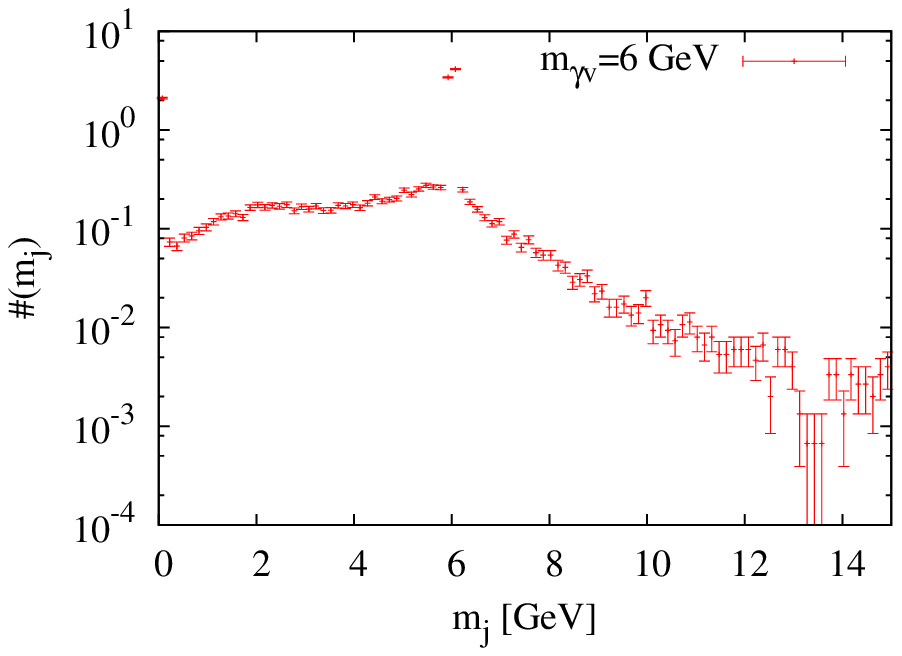,scale=0.7,angle=0}
\end{minipage}
\caption{ {\itshape AM$_{Z^\prime}$}: the number of jets per event and the distribution of the invariant mass of the  of the jets.  In both plots the $q_v$ mass is $m_{q_v}=50$ GeV. The left side shows the $\alpha_v$ dependence for $m_{\gamma_v}=10$ GeV, while the right side shows the jet invariant mass distribution for $\alpha_v=0.4$ and $m_{\gamma_v}=6$ GeV. The jet reconstruction algorithm is Jade, with $m_\mrm{cut}$ corresponding to the $\gamma_v$ mass. Note the peak at $6$ GeV. 
}
\label{fig:KMAz_Nj} }

An efficient clustering algorithm should maintain the ratio between the
average number of $\gamma_v$ particles produced and the number of jets
found, as is confirmed by comparing the plots on
left side of Figs.~\ref{fig:KMAz_Ngv} and \ref{fig:KMAz_Nj}.  
In this particular exercise we have relied on the extraction of 
the relevant mass scale from the lepton pair invariant mass distribution. 
This should be guaranteed by the presence of leptons in all six scenarios. 
Specifically, if the only way to decay back into the SM is via kinetic mixing, 
the 
$\gamma_v\rightarrow SM$ branching ratios are fixed by the off-shell
$\gamma^*$ branching ratios. In the non-Abelian case the absence of a
spin 1 $v$-meson would reduce the rate of $e^+e^-$ and $\mu^+\mu^-$
pairs by helicity suppression. However, also in such scenarios, 
a simple trial-and-error approach with a range of jet clustering
scales would suffice to reveal a convincing jet mass peak.

More information about the event and the model parameters can be
extracted from the angular distributions in
Sec.~\ref{sec:angular_eShapes}.

\subsection{KMA$_{\gamma_v}$ and KMNA$_{\gamma_v}$}
\label{sec:KMg} 

The $Z^\prime$ mediated and $\gamma_v$ mediated setups
are effectively very similar, once the difference in coupling
constants is factored out. The phenomenology can appear rather
different, however, because of the initial state radiation from 
the electron/positron beams. To view this, recall that the 
photon bremsstrahlung spectrum is spiked at small energy fractions,
$\propto \mrm{d}z_{\gamma}/z_{\gamma}$, and that therefore the 
electron-inside-electron PDF roughly goes like $\mrm{d}z_e/(1 - z_e)$,
with $z_e = 1 - z_{\gamma}$. In the case of a $\gamma$ or light 
$\gamma_v$ propagator, behaving like $1/\hat{s} = 1/(z_es)$, this 
combines to give a $\mrm{d}z_e/(z_e(1 - z_e)) 
= \mrm{d}z_{\gamma}/((1 - z_{\gamma}) z_{\gamma})$ spectrum.
The complete description includes the emission of multiple photons 
off both incoming beams, but the key features above described are readily 
visible in Fig.~\ref{fig:KMNAg_eISR}. Specifically, the spike at  
$E_{\mrm{ISR}} = 400$ GeV corresponds to the emission of an energetic
photon on one side only, while the non-negligible tail above that
requires hard emissions on both sides.

\FIGURE[!h]{
\begin{minipage}[b]{\linewidth}
\epsfig{file=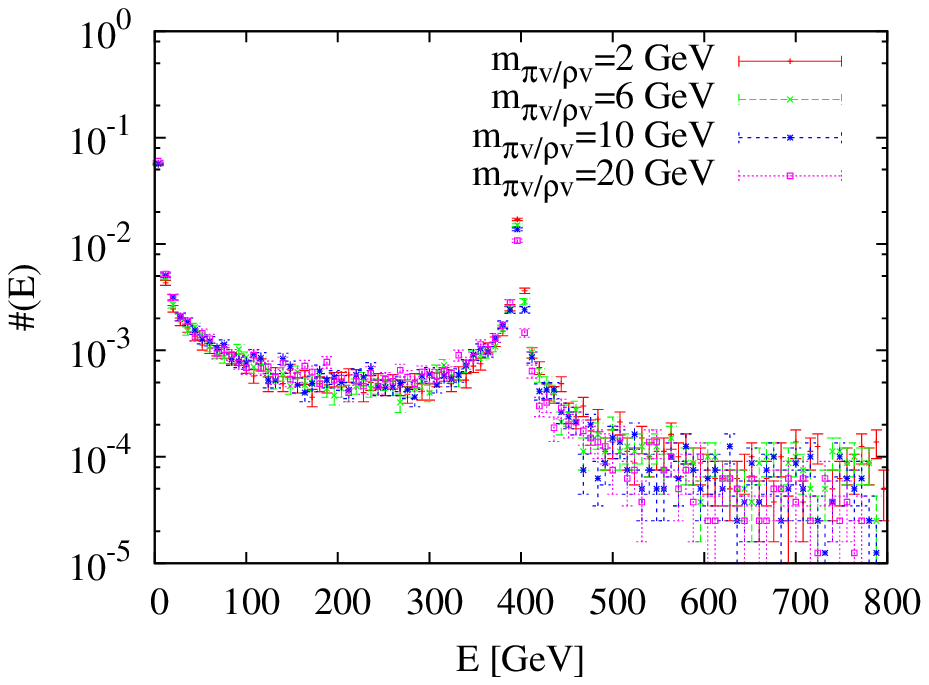,scale=0.7,angle=0}
\end{minipage}
\caption{{\itshape KMNA$_{\gamma_v}$}: The total energy radiated in
ISR photons. Note the spike around 400 GeV: a large fraction of the events  
 will have a reduced $\hat{s}$ due IS photon emission. The $q_v$ mass is fixed by $m_{q_v}=m_{\pi/\rho_v}/2$.}
\label{fig:KMNAg_eISR} }

\FIGURE[t]{
\begin{minipage}[b]{0.45\linewidth}
\epsfig{file=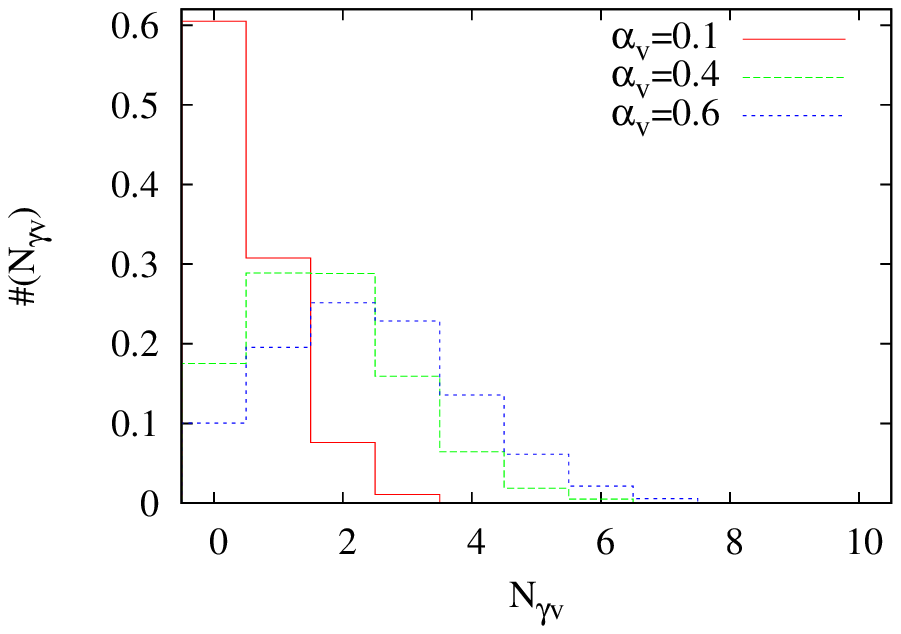,scale=0.7,angle=0}
\end{minipage}
\begin{minipage}[b]{0.45\linewidth}
\epsfig{file=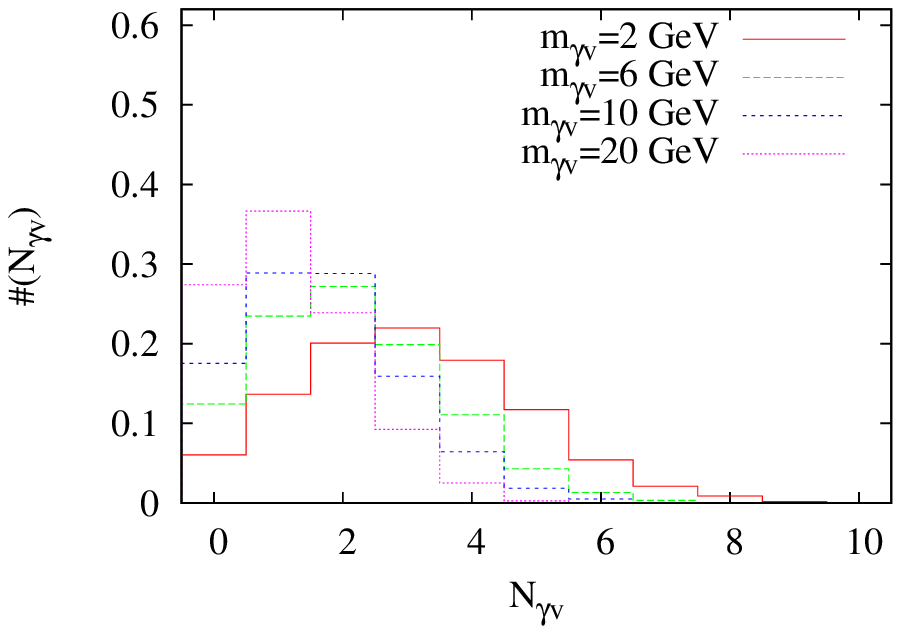,scale=0.7,angle=0}
\end{minipage}
\caption{{\itshape KMA$_{\gamma_v}$}: the number of $\gamma_v$ gauge bosons emitted per event. On the left we emphasize the $\alpha_v$ dependence, while on the right the $m_{\gamma_v}$ dependence. In both plots  the $q_v$ mass is $m_{q_v}=10$ GeV.  On the left side $m_{\gamma_v}=10$ GeV and $\alpha_v=0.1, 0.4, 0.6$, while on the right side $m_{\gamma_v}=2, 6, 10, 20$ GeV and the coupling is fixed at $\alpha_v=0.4$.
}
\label{fig:KMAg_Ngv} }

\FIGURE[t]{
\begin{minipage}[b]{0.45\linewidth}
\epsfig{file=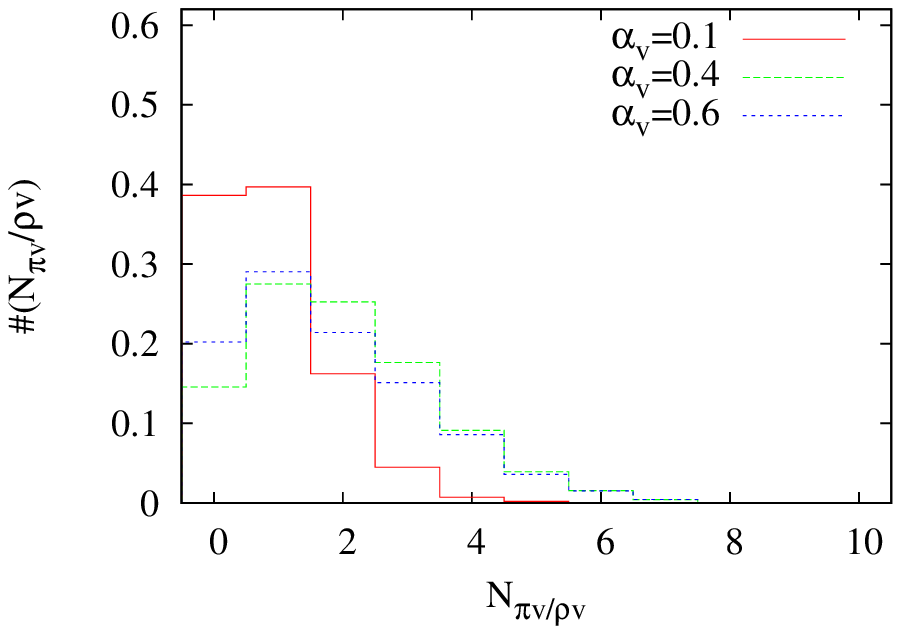,scale=0.7,angle=0}
\end{minipage}
\begin{minipage}[b]{0.45\linewidth}
\epsfig{file=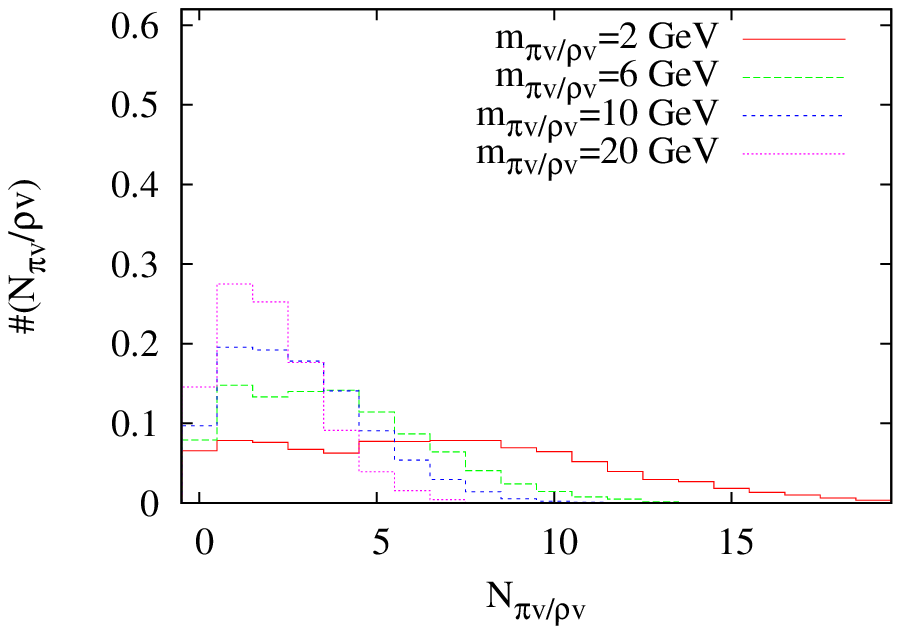,scale=0.7,angle=0}
\end{minipage}
\caption{{\itshape KMNA$_{\gamma_v}$}: the number of flavour diagonal $\pi_v/\rho_v$ gauge bosons emitted per event. On the left we emphasize the $\alpha_v$ dependence, while on the right the $m_{\pi_v/\rho_v}$ dependence. On the left side $m_{\pi_v/\rho_v}=10$ GeV and $\alpha_v=0.1, 0.4, 0.6$, while on the right side $m_{\pi_v/\rho_v}=2, 6, 10, 20$ GeV and the coupling is fixed at $\alpha_v=0.4$.}
\label{fig:KMNAg_Nmes}
}

Note that while the number of $v$-particles produced are
similar for KMA$_{\gamma_v}$ and AM$_{Z^\prime}$ in the Abelian
case, Fig.~\ref{fig:KMAz_Ngv} vs. Fig.~\ref{fig:KMAg_Ngv}, 
they are different in the non-Abelian case, 
Fig.~\ref{fig:KMNAz_Nmes} vs. Fig.~\ref{fig:KMNAg_Nmes}.
Specifically, the higher average multiplicity in the 
non-Abelian case leads to a double spike in the distribution,
a normal one from events with little ISR and an extra 
low-multiplicity one from events with much ISR.

\subsection{SMA and SMNA}

\FIGURE[t]{
\begin{minipage}[b]{0.45\linewidth}
\epsfig{file=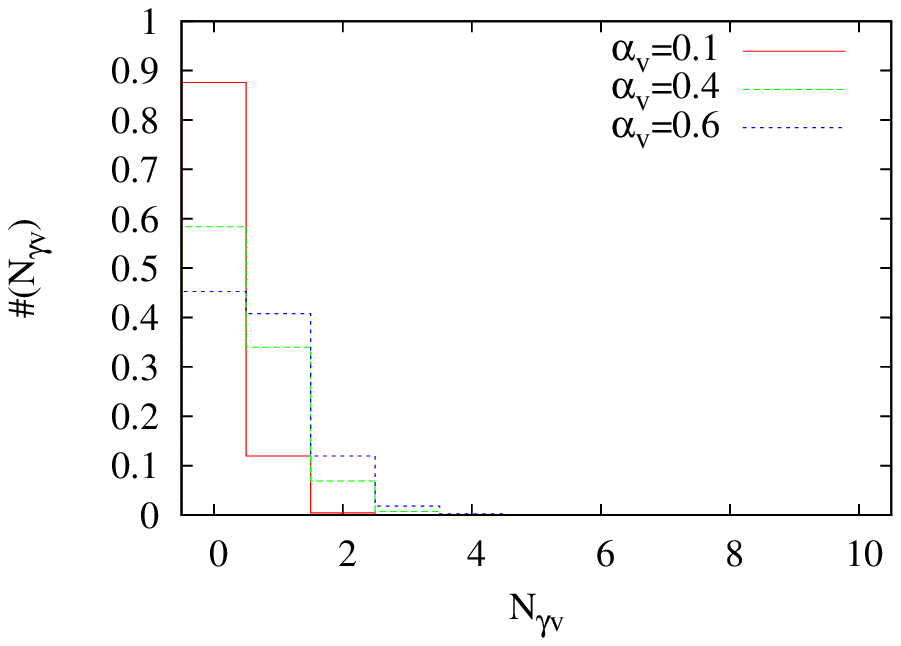,scale=0.7,angle=0}
\end{minipage}
\begin{minipage}[b]{0.45\linewidth}
\epsfig{file=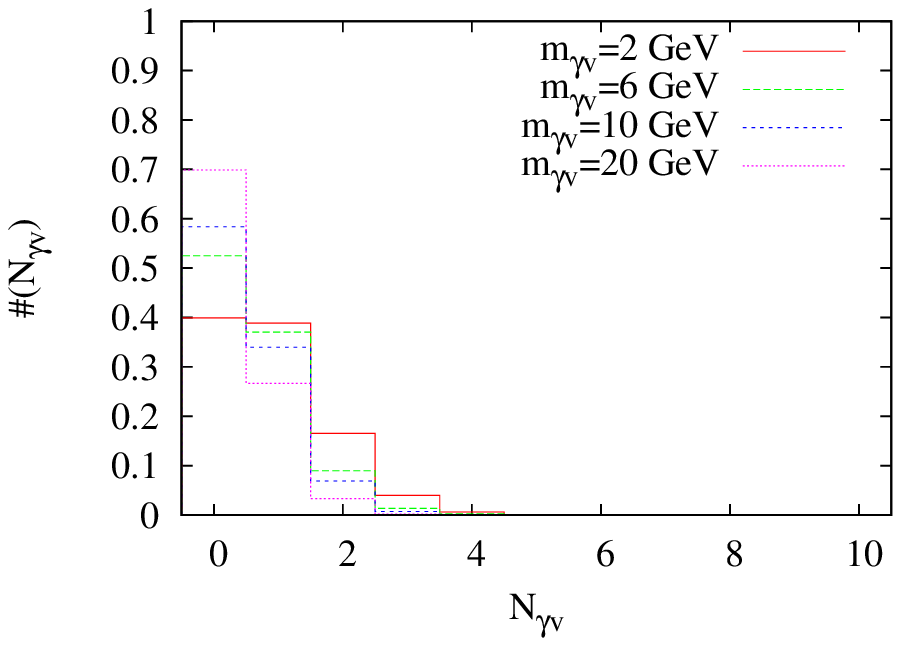,scale=0.7,angle=0}
\end{minipage}
\caption{{\itshape SMA}: the number of $\gamma_v$ gauge bosons emitted
per event. On the left we emphasize the $\alpha_v$ dependence, while
on the right the $m_{\gamma_v}$ dependence. In both plots $m_{E_v}=$
250 GeV and  $m_{q_v}=50$ GeV.  On the left side $m_{\gamma_v}=10$ GeV and
$\alpha_v=0.1, 0.4, 0.6$, while on the right side $m_{\gamma_v}=2, 6, 10,
20$ GeV and the coupling is fixed at $\alpha_v=0.4$. }
\label{fig:SMA_Ngv} }

The number of $v$-particles produced in the standard model mediated
scenarios is shown in Fig.~\ref{fig:SMA_Ngv}, and the energy of the 
$\gamma_v$ photons and of the $v$-mesons in 
Fig.~\ref{fig:SMA_SMNA_Egv/Emes}. The kinetic boundary
$E^\mrm{max}$ is different in the two cases, owing to the choice of 
$m_{q_v} = 50$~GeV in the Abelian case. This reduces the energy
available for $\gamma_v$ emissions. The  $\npT$ distribution was 
shown in Fig.~\ref{fig:KMg_SM_pTmiss} and has already been discussed.

\FIGURE[t]{
\begin{minipage}[b]{0.45\linewidth}
\epsfig{file=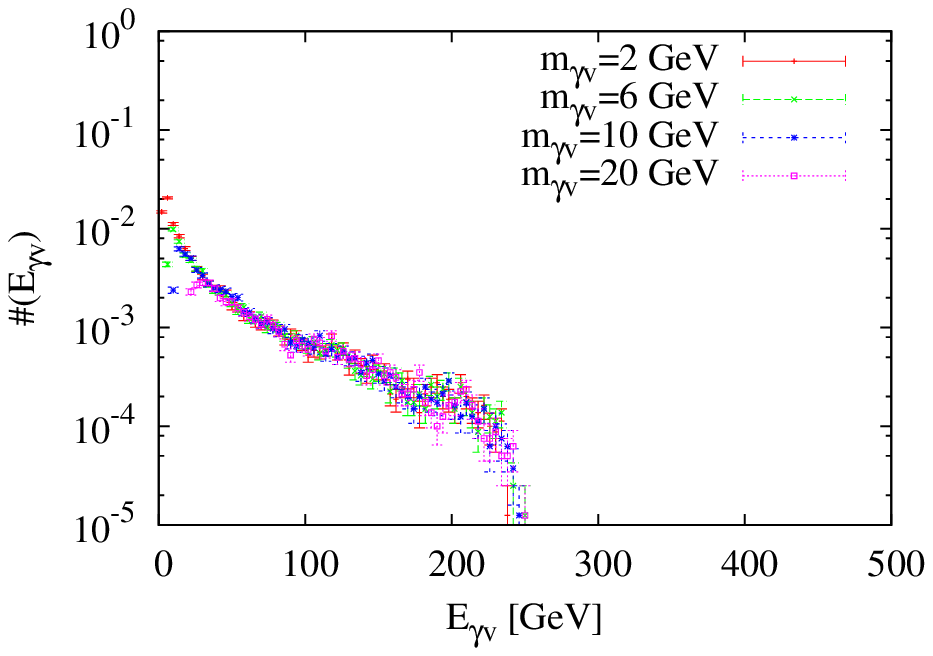,scale=0.7,angle=0}
\end{minipage}
\begin{minipage}[b]{0.45\linewidth}
\epsfig{file=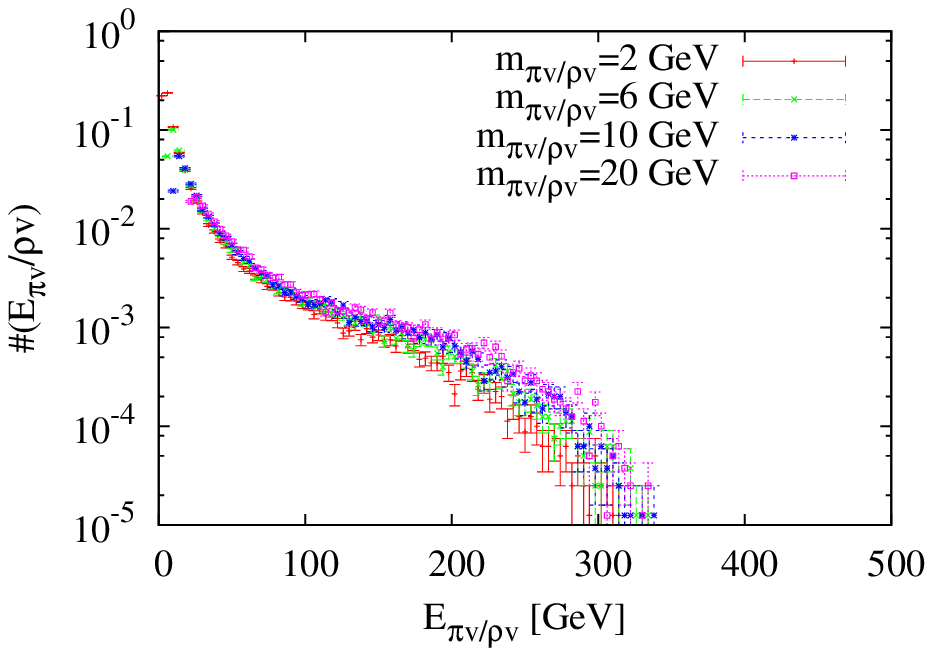,scale=0.7,angle=0}
\end{minipage}
\caption{{\itshape SMA} vs {\itshape SMNA}: the energy spectra of the $\gamma_v$ and of the diagonal $\pi_v/\rho_v$ emitted per event. The left side shows the energy distribution for SMA $m_{E_v}=250$, GeV $m_{q_v}=50$ GeV, the right side shows the corresponding distribution for SMNA (in this case $m_{q_v}=1,3,5,10$ GeV corresponds to half of the $\pi_v/\rho_v$ mass.).}
\label{fig:SMA_SMNA_Egv/Emes} }

The most important distribution to pinpoint the masses of the model 
is the lepton energy spectrum, Fig.~\ref{fig:SM_eLeptons}. In this 
case leptons may come from both the kinetic mixing decays of the 
$\gamma_v$ and from the decays of the $E_v$ into $e q_v$. 
The energy spectra are very different in the two cases. The leptons
coming from the $E_v$ decays tend be highly energetic, while the rest
are less so. A reasonable first approximation
is to associate the highest energy electron and positron with the
two $E_v$ decays, and the rest with the $\gamma_v/\pi_v/\rho_v$ 
ones. The curve in Fig.~\ref{fig:SM_eLeptons} represents the sum of the steeply falling spectrum associated to the leptons coming from $\gamma_v$ decay,  and a flat spectrum associated to the leptons from $E_v\rightarrow e q_v$ decay.  
The upper and lower shoulders of the former energy distributions 
then give a relationship between the $E_v$ and the $q_v$ masses 
\cite{Carloni:2010tw}. 

\FIGURE[t]{
\begin{minipage}[b]{0.45\linewidth}
\epsfig{file=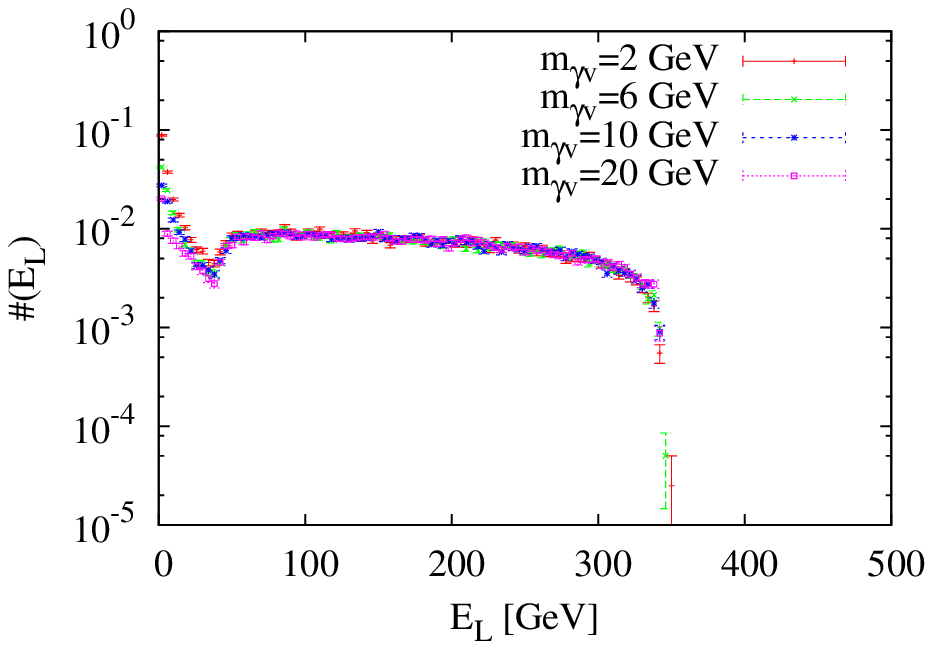,scale=0.7,angle=0}
\end{minipage}
\begin{minipage}[b]{0.45\linewidth}
\epsfig{file=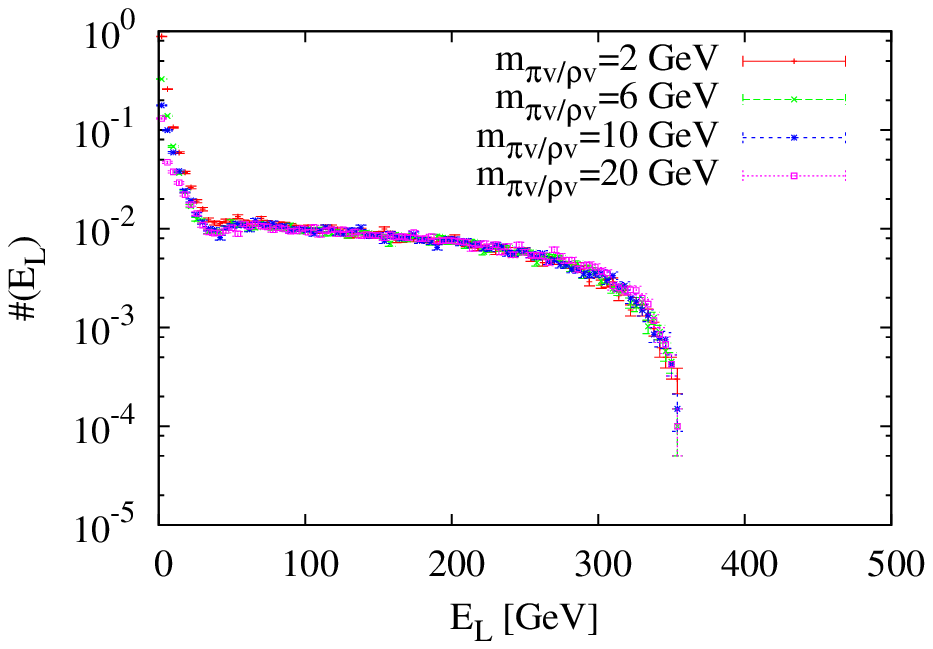,scale=0.7,angle=0}
\end{minipage}
\caption{{\itshape SMA} vs {\itshape SMNA}: the lepton energy spectra in the Abelian (left) case and in the non-Abelian (right) case. In the Abelian case  GeV $m_{q_v}=50$ GeV, while in the non-Abelian case $m_{q_v}=1,3,5,10$ GeV. In both cases $m_{E_v}=250$ GeV.}
\label{fig:SM_eLeptons} }

\subsection{Angular distributions and event shapes} 
\label{sec:angular_eShapes}

The distribution of the production cross-section as a function of the angle 
between the jets and the beam axis has a characteristic dependence on the 
spin of the pair-produced particles. This fact may be used to identify the 
$q_v$ spin.

\FIGURE[t]{
\begin{minipage}[b]{0.45\linewidth}
\epsfig{file=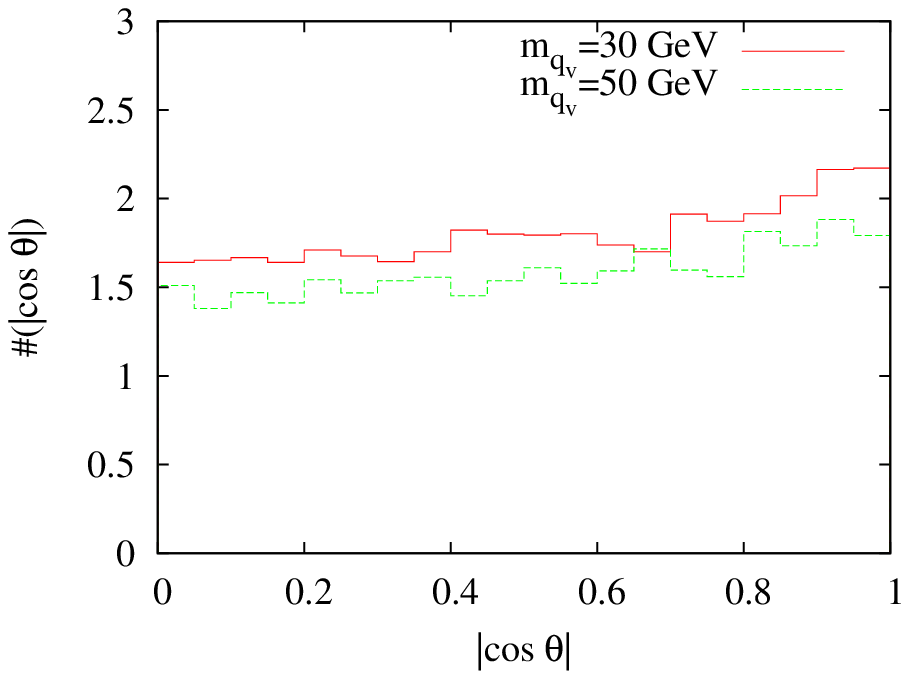,scale=0.7,angle=0}
\end{minipage}
\begin{minipage}[b]{0.45\linewidth}
\epsfig{file=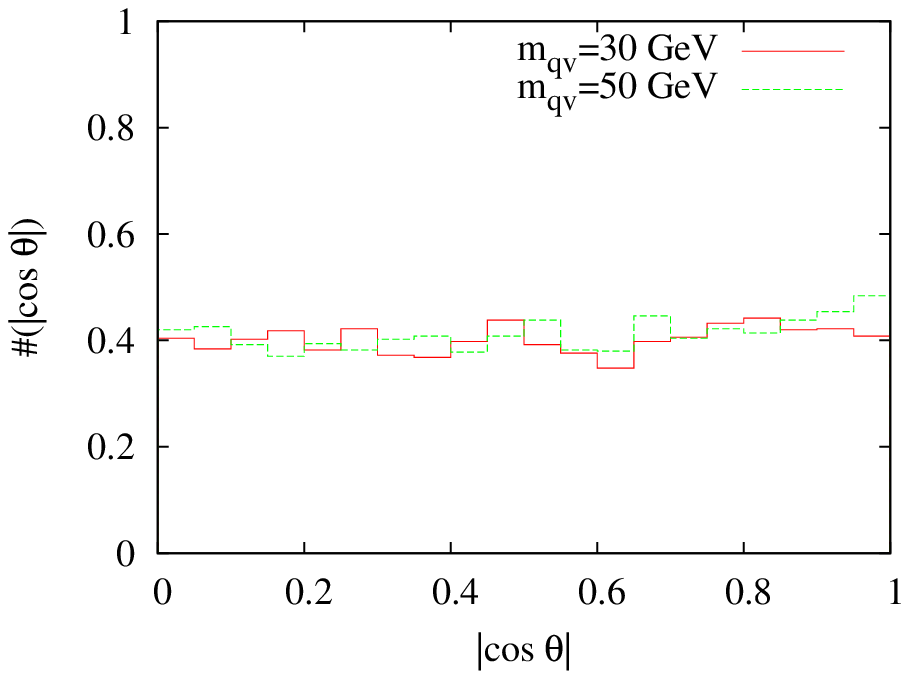,scale=0.7,angle=0}
\end{minipage}
\caption{{\itshape AM$_{Z^\prime}$} and {\itshape SMA}: the
distribution of $\cos\theta_i$, where $\theta_i$ is the angle
between the $i^\mrm{th}$ jet and the beam axis, for $m_{q_v}=30$ GeV and for $m_{q_v}=50$ GeV. A lower cut on the
invariant mass $m_j >m_{\gamma_v/\pi_v/\rho_v}/2$ was applied in this
case. In both cases $m_{\gamma_v}=10$ GeV and the $v$-coupling is fixed 
to $\alpha_v=0.4$, $m_{q_v}=10$ GeV.}
\label{fig:thetaJ}
}

In Fig.~\ref{fig:thetaJ} one may observe the $\cos\theta_{i}$
distribution, where $\theta_{i}$ is the polar angle between the
$i^\mrm{th}$ jet and the beam axis, for $m_{\gamma_v}=10$~GeV. 
Only jets with a reconstructed mass $m_j > m_{\gamma_v}/2$
are shown, since lower-mass jets are strongly contaminated by ISR 
photons above the $\theta_\mrm{cut} = 50$~mrad cut. The
production cross-section for $e^+ e^-\rightarrow q_v\bar{q}_v$, with
$q_v$ a massless spin 1/2 fermion, is proportional to $1+\cos^2\theta$. 
In the AM$_{Z^\prime}$ case one must allow for corrections due to 
the $\theta_\mrm{cut}$, to the $q_v$ being massive  and to the $\gamma_v$ 
radiation; typically this leads to a somewhat flatter distribution. 
In the SMA case the isotropic decays $E_v\rightarrow e q_v$ flattens 
whatever original $e^+e^-\rightarrow E_v \bar{E}_v$ distribution, 
unless the $e q_v$ decay products are highly boosted. In our case a 
small ISR contamination is still visible close to $\cos\theta = 1$, 
but otherwise the distribution is flat. 

\FIGURE[t]{
\begin{minipage}[b]{0.45\linewidth}
\epsfig{file=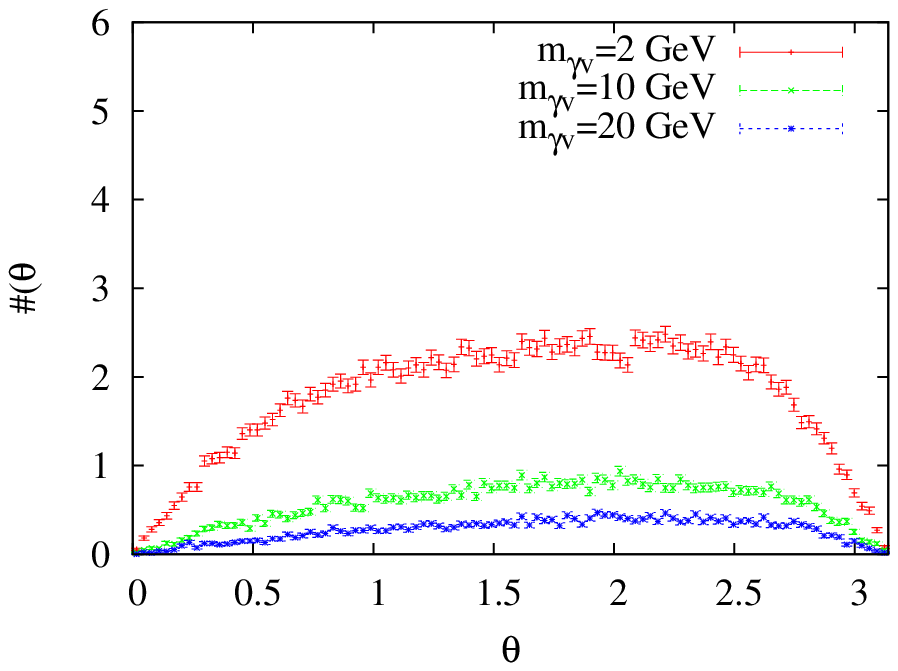,scale=0.7,angle=0}
\end{minipage}
\begin{minipage}[b]{0.45\linewidth}
\epsfig{file=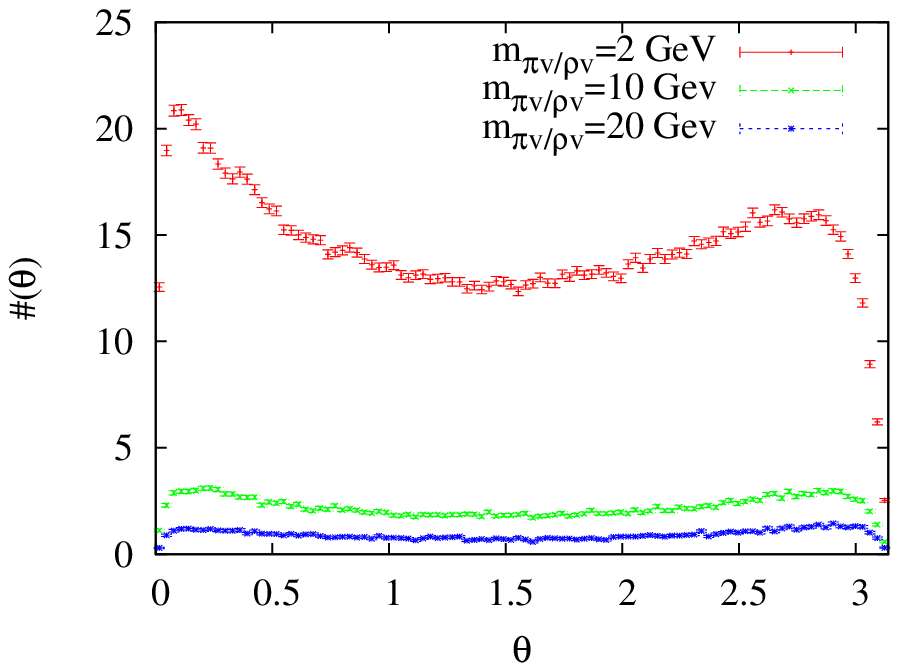,scale=0.7,angle=0}
\end{minipage}
\caption{{\itshape AM$_Z^\prime$} and {\itshape NAM$_Z^\prime$}: the
distribution of $\theta_{ij}$, the angle between the $\gamma_v$s in the
Abelian case (left), or the angle between the diagonal mesons in the
non-Abelian case (right). The distributions are shown as functions of 
the mass $m_{\gamma_v}=m_{\pi_v/\rho_v}$. The coupling constant is
$\alpha_v=0.4$ in both cases. Note that the number of $\gamma_v$ and
diagonal $\pi_v/\rho_v$ is different in the two cases.}
\label{fig:KM_thetaij_v}
}
 
The distribution of opening angles in pairs of $v$-particles, 
$\theta_{ij}$, should give some insight whether the secluded sector 
$G$ is an Abelian or a confining non-Abelian group. In an Abelian event 
the $q_v \bar{q_v}$ quarks define a dipole emission axis. To first
approximation the $v$-gammas are emitted independently, i.e.\ with a flat 
distribution in the $\phi$ angle around the dipole axis, and uniformly in
rapidity along this axis.  In the non-Abelian case the emissions occur along 
a chain of dipoles, that is reconfigured by each new emission, since the $g_v$s 
carry $v$-colour charge. This implies a different underlying correlation 
structure, but it is unclear what happens with this correlation on
the way through the $v$-hadronization process and the decays back into the
standard sector, and how best to search for it.
  
In Fig.~\ref{fig:KM_thetaij_v} we show the $\theta_{ij}$ distribution
of $v$-particle pairs, 
for the Abelian AM$_{Z^\prime}$ and non-Abelian NAM$_{Z^\prime}$ cases. Note how 
the $\theta_{ij}~0$ and $\theta_{ij}~\pi$ angles are preferred in non-Abelian case. 
The comparison is somewhat misleading, however, in the sense that we compare 
scenarios with the same $m_{\gamma_v} = m_{\pi_v/\rho_v}$ and the same 
$\alpha_v$, but with different numbers of $v$-particles per event (see
Fig.~\ref{fig:KMAz_Ngv}) and different $v$-particle energy
distributions. In Sec.~\ref{sec:compare} we will discuss further these
distributions under more similar conditions.
   
In Fig.~\ref{fig:KM_thetaJij} we show the reconstructed jet-jet 
$\cos\theta_{ij}$ distributions corresponding to the $v$-particle 
distributions in Fig.~\ref{fig:KM_thetaij_v}. Note how the relative 
difference between the Abelian and non-Abelian scenarios is maintained.

\FIGURE[t]{
\begin{minipage}[b]{0.45\linewidth}
\epsfig{file=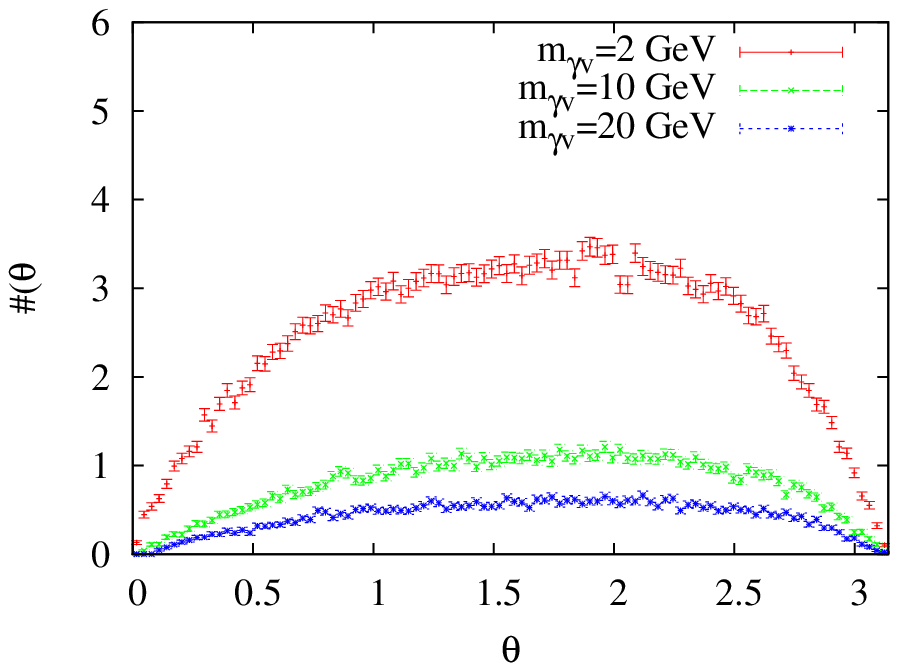,scale=0.7,angle=0}
\end{minipage}
\begin{minipage}[b]{0.45\linewidth}
\epsfig{file=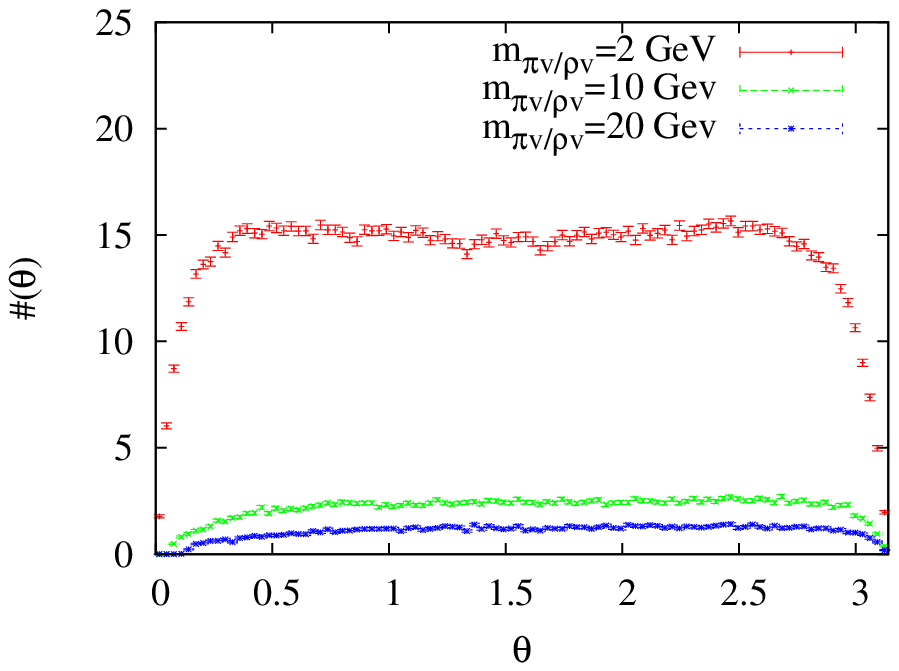,scale=0.7,angle=0}
\end{minipage}
\caption{{\itshape AM$_Z^\prime$} and {\itshape NAM$_Z^\prime$}: the
distribution of $\theta_{ij}$, the angle between the jets in the Abelian
case (left) and in the non-Abelian case (right). The distributions are
shown as functions of the mass $m_{\gamma_v}=m_{\pi_v/\rho_v}$. The
$v$-coupling is fixed to $\alpha_v=0.4$. }
\label{fig:KM_thetaJij}
} 

In order to characterize the shape of the events one may also use 
thrust and (the linearized version of) sphericity 
\cite{Ellis:1991qj, Brandt:1964sa, Bjorken:1969wi}.  
These indicate whether an event is more pencil-like, $T$=1 and $S=0$,
or more spherical, $T=1/2$ and $S=1$. Sphericity and thrust 
are primarily intended for events analyzed in their own rest frame, 
while the visible systems we study have a net momentum that is 
compensated by the stable secluded-section particles, plus ISR photons 
going down the beam pipe and neutrinos. Since we are interested in 
the properties of the visible system itself, not in its net motion,
the analysis is performed in the rest frame of this visible system. 
In addition, for the SMA scenarios, a further distortion occurs
by the kinematics of the $E_v\rightarrow e q_v$ decays, and by the
presence of the resulting $e^{\pm}$ in the final state. To this end,
the highest-energy electron and positron are excluded from the 
definition of the visible system. 

\FIGURE[h]{
\begin{minipage}[b]{0.4\linewidth}
\epsfig{file=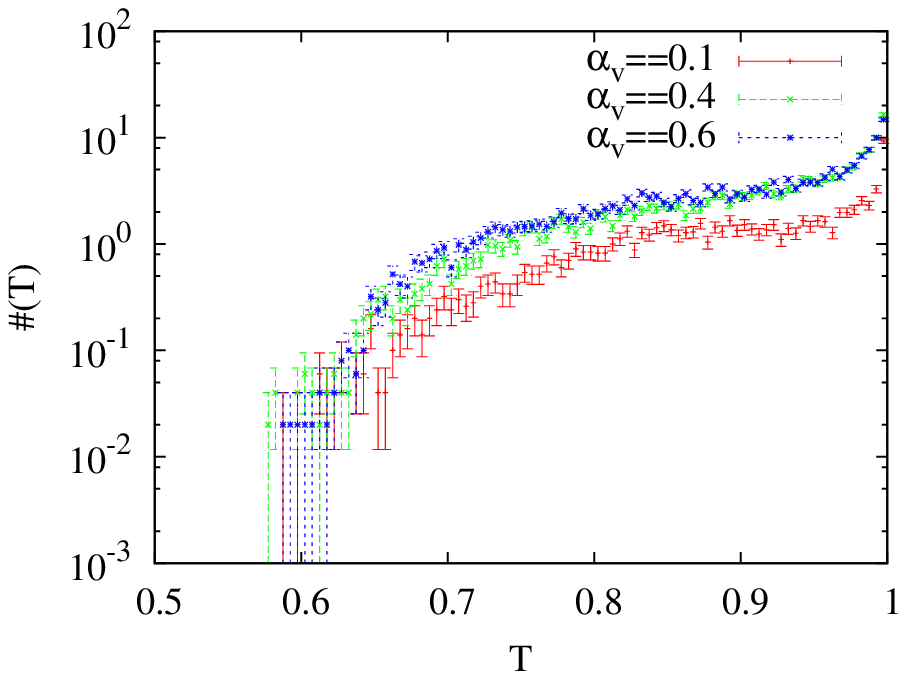,scale=0.7, angle=0}
\epsfig{file=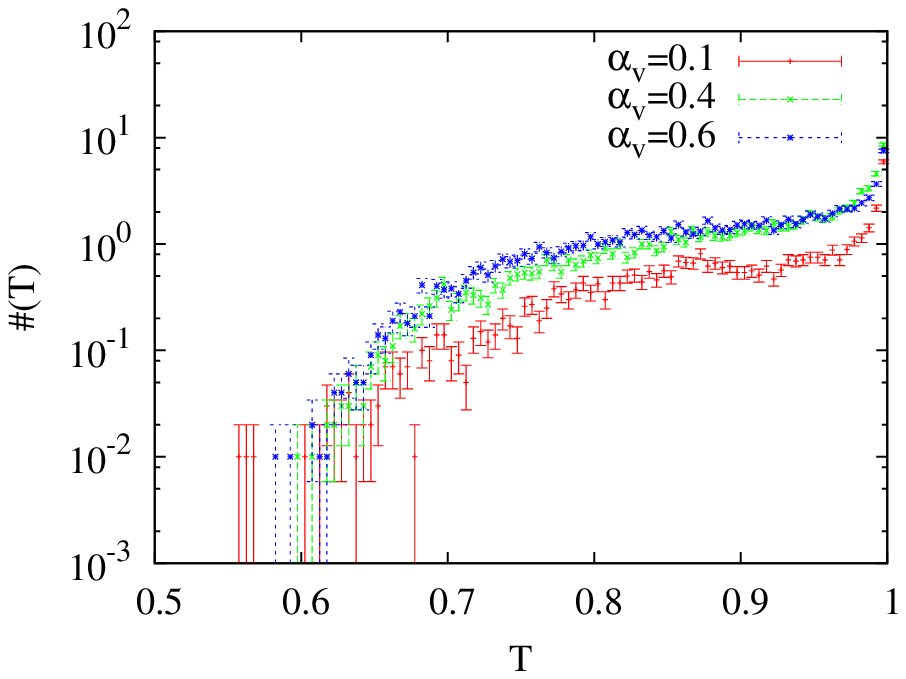,scale=0.7, angle=0}
\epsfig{file=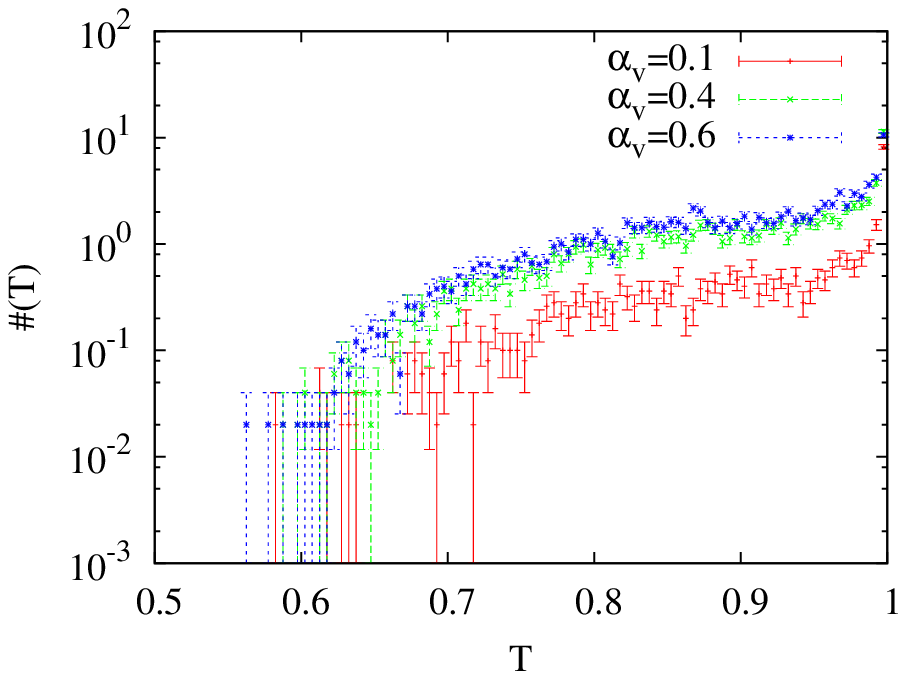,scale=0.7, angle=0}
\end{minipage} \hspace{0.5cm}
\begin{minipage}[b]{0.4\linewidth}
\epsfig{file=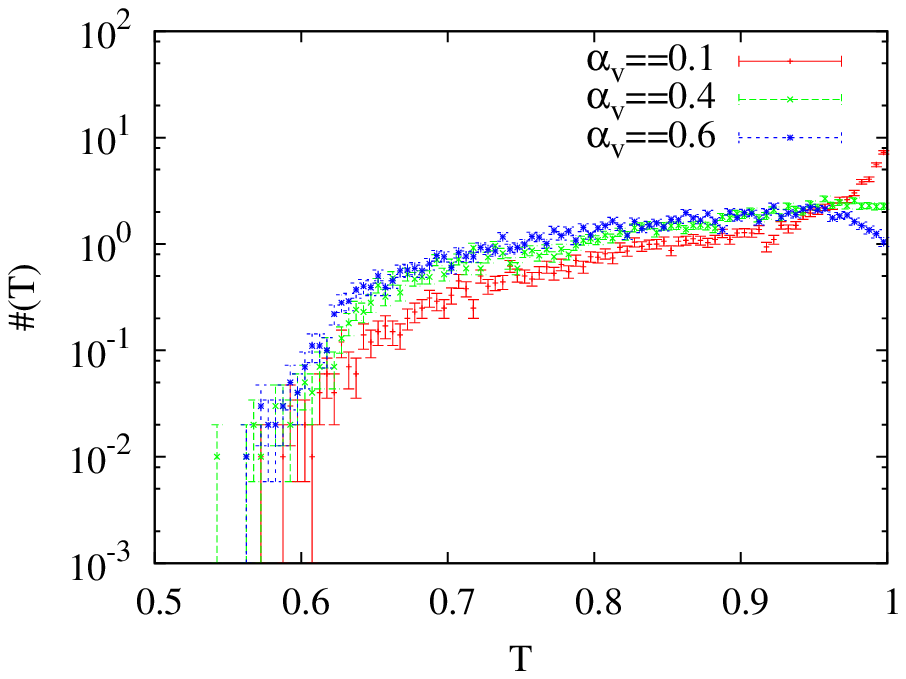,scale=0.7, angle=0}
\epsfig{file=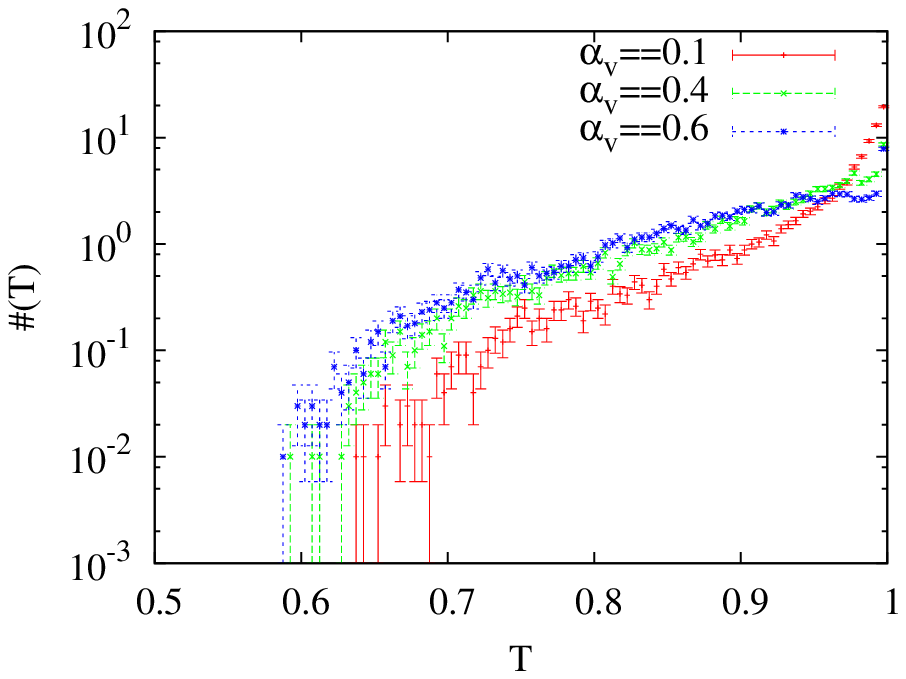,scale=0.7, angle=0}
\epsfig{file=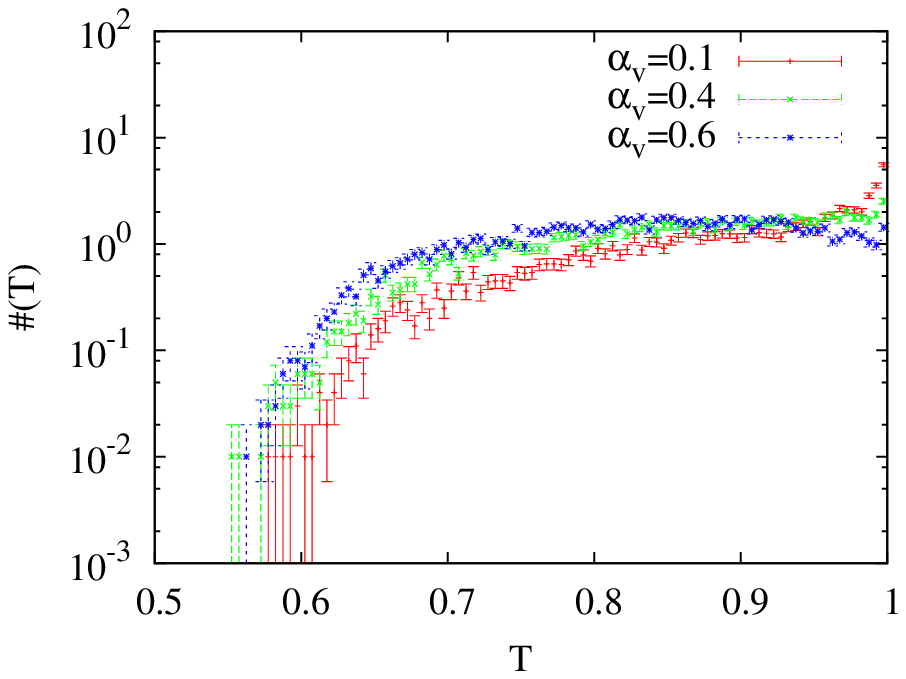,scale=0.7, angle=0}
\end{minipage}
\caption{The plots show the thrust distribution for the six scenarios
as a function of the $\alpha_v$. These correspond to, from top to
bottom, M$_{Z^\prime}$, KM$\gamma_v$ and SM production. The plots on the
left show the Abelian case while the plot on the right show the
non-Abelian cases. For all Abelian plots, the parameters are set to
$m_{q_v}=50$ GeV and $m_{\gamma_v}=10$ GeV.  For the SMA case $m_{E_v}=250$
GeV is set as well. For the non-Abelian cases, the mass of the mesons
is fixed to $m_{\gamma_v/\pi_v/\rho_v}$} 
\label{fig:thrust} }

In Fig.~\ref{fig:thrust} we show the thrust distributions in the six 
scenarios, and in Fig.~\ref{fig:spheri} the sphericity distributions.  
 As one may have predicted, the
events become less pencil-like as the coupling $\alpha_v$ grows. In addition 
 the SM mediated events are less likely to be
pencil-like than the KM or $Z^\prime$ mediated ones. 
Note that events with smaller $\alpha_v$, in which nothing is radiated are 
not analyzed.

\FIGURE[t]{
\begin{minipage}[b]{0.4\linewidth}
\epsfig{file=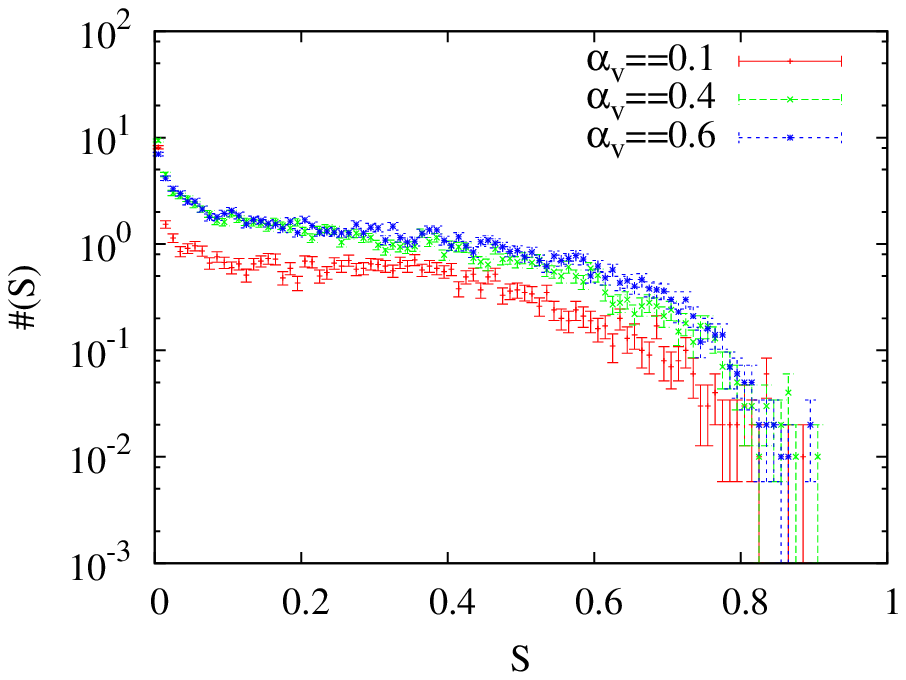,scale=0.7, angle=0}
\epsfig{file=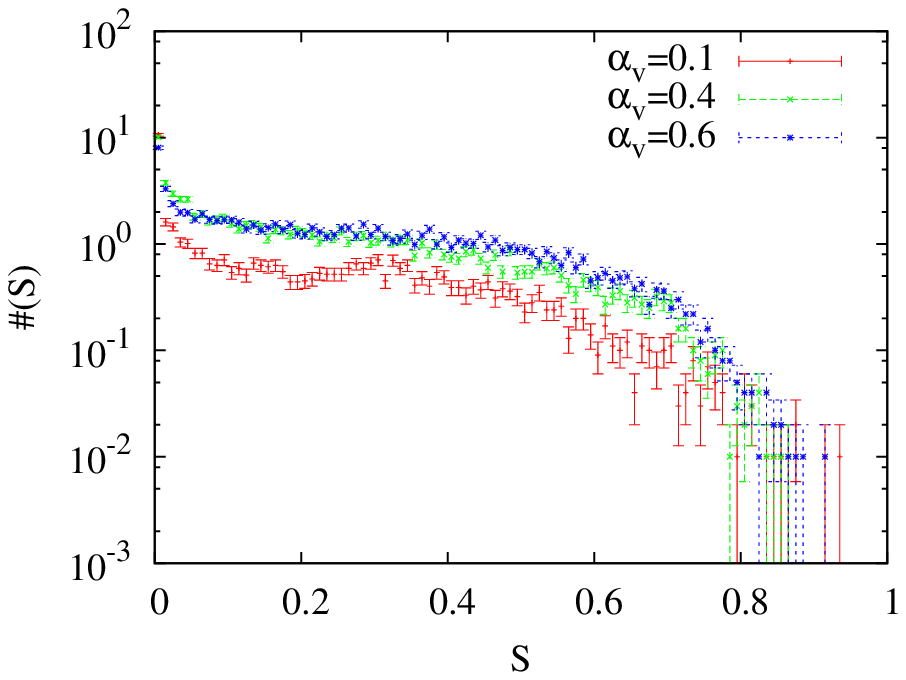,scale=0.7, angle=0}
\epsfig{file=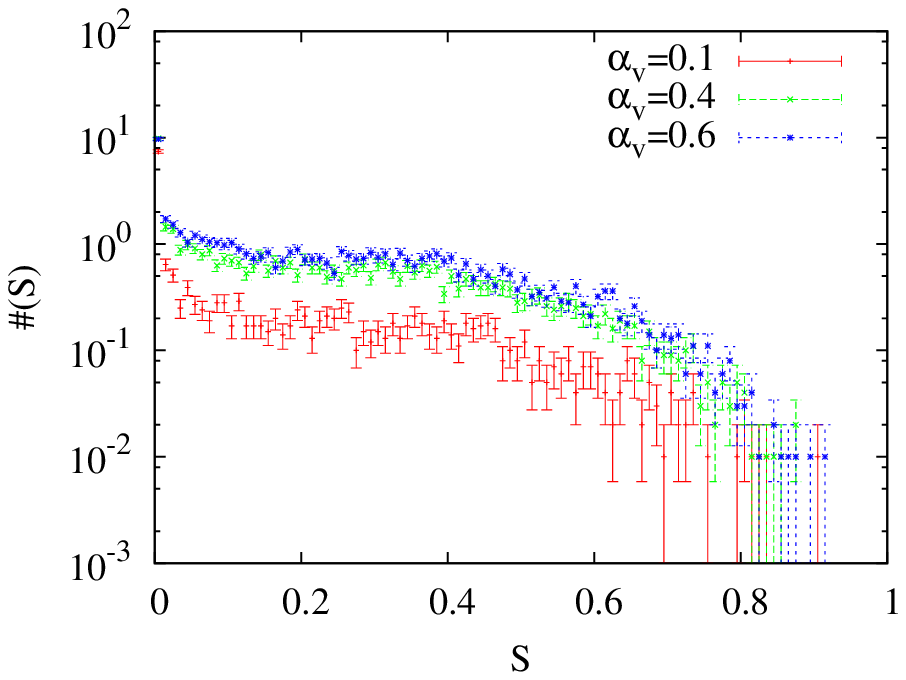,scale=0.7,
angle=0}
\end{minipage} \hspace{0.5cm}
\begin{minipage}[b]{0.4\linewidth}
\epsfig{file=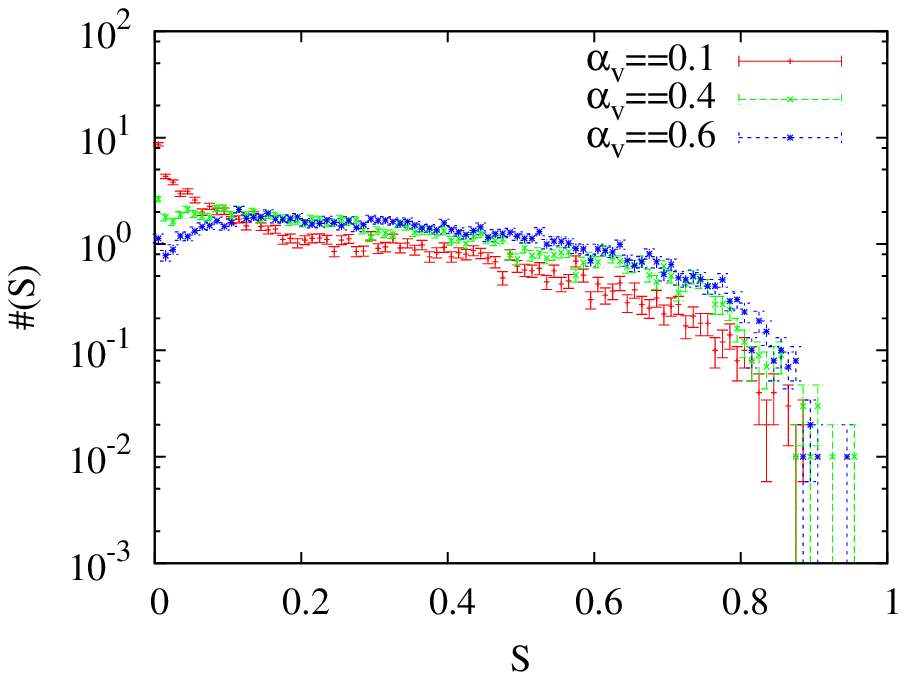,scale=0.7, angle=0}
\epsfig{file=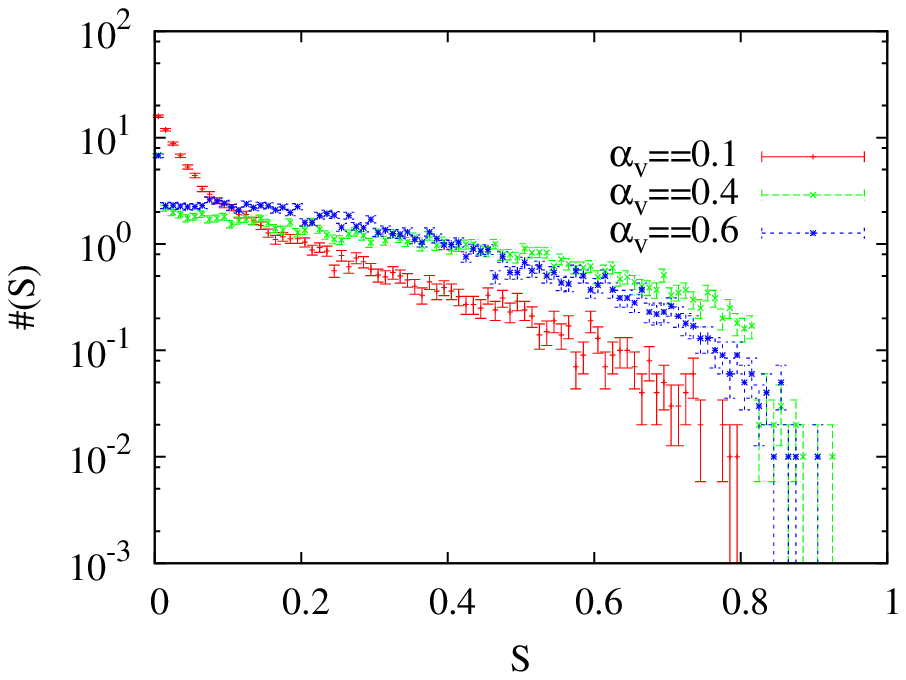,scale=0.7, angle=0}
\epsfig{file=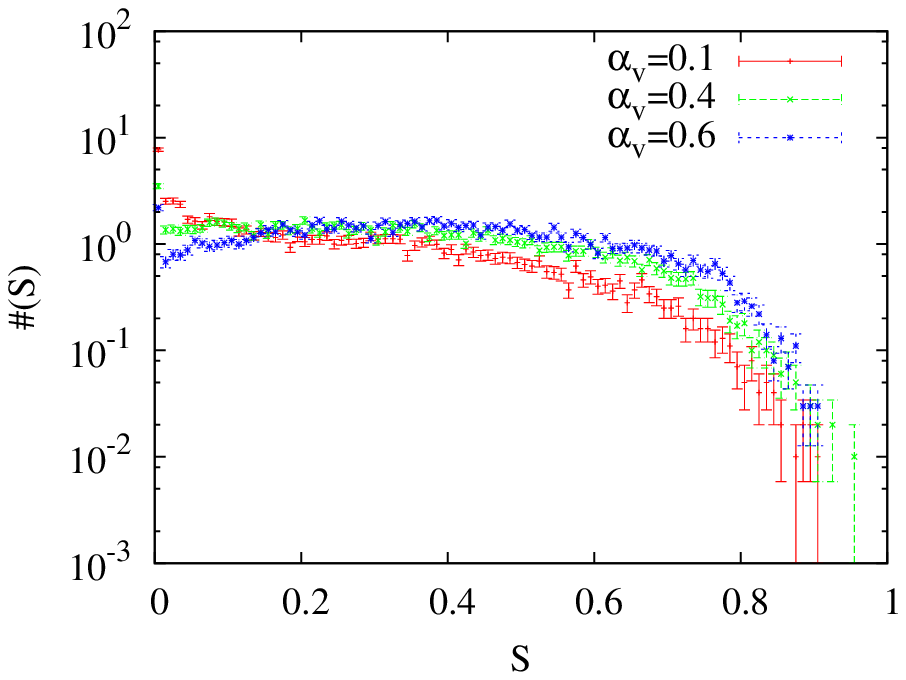,scale=0.7, angle=0}
\end{minipage}

\caption{The plots show the (linearized) sphericity distribution for the 
six scenarios as a function of the $\alpha_v$. These correspond to, from
top to bottom, KM$_{Z^\prime}$, KM$\gamma_v$ and SM production. The plots
on the left show the Abelian case while the plot on the right show the
non-Abelian cases. For all Abelian plots, the parameters are set to
$m_{q_v}=50$ GeV and $m_{\gamma_v}=10$ GeV.  For the SMA case $m_{E_v}=250$
GeV is set as well. For the non-Abelian cases, the mass of the mesons
is fixed to  $m_{\pi_v/\rho_v}=m_{\gamma_v}$.} 
\label{fig:spheri} }

\section{Analysis: comparing $\not \hspace{-3pt} U(1)$ and $SU(N)$}
\label{sec:compare}

In this section we begin to address the issue of discriminating between 
Abelian and non-Abelian scenarios in cases in which smoking-gun discriminating 
signals are absent. To this end we consider the most challenging scenario, 
in which $\gamma_v$ and $\pi_v,\rho_v$ have the same mass, and the same average 
number of $v$-particles leak back into the SM sector, carrying the
same average amount of energy. We also consider the same production 
mechanism, to reduce model dependence and to isolate the effects of 
the hidden sector dynamics.

\FIGURE[t]{
\begin{minipage}[b]{0.4\linewidth}
\epsfig{file=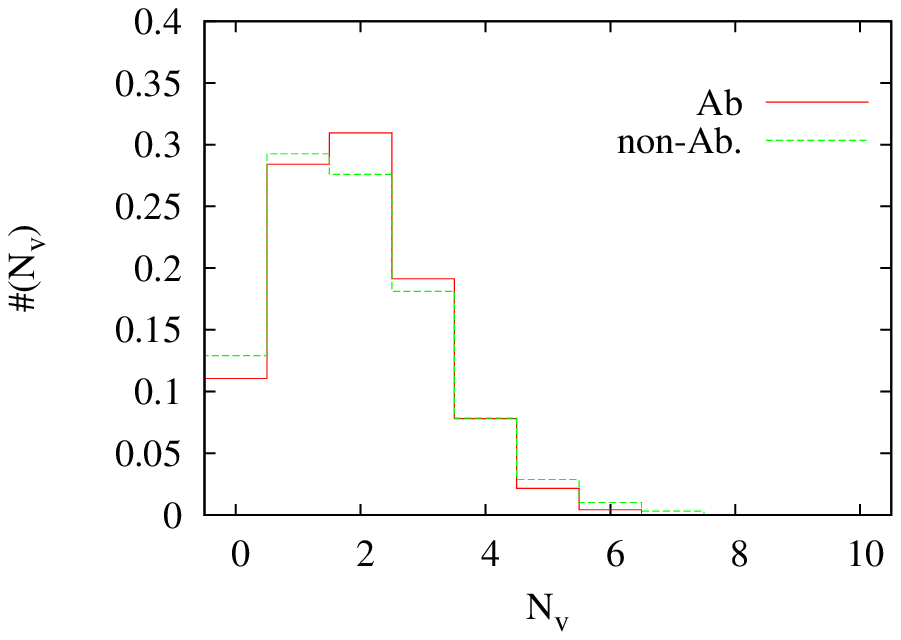,scale=0.7, angle=0}
\epsfig{file=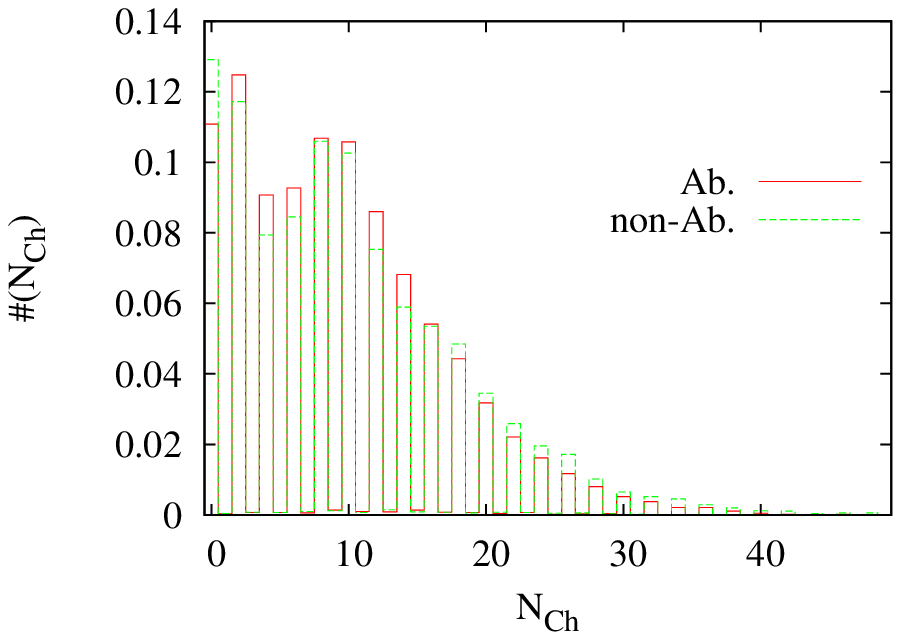,scale=0.7, angle=0}
\end{minipage} \hspace{0.5cm}
\begin{minipage}[b]{0.4\linewidth}
\epsfig{file=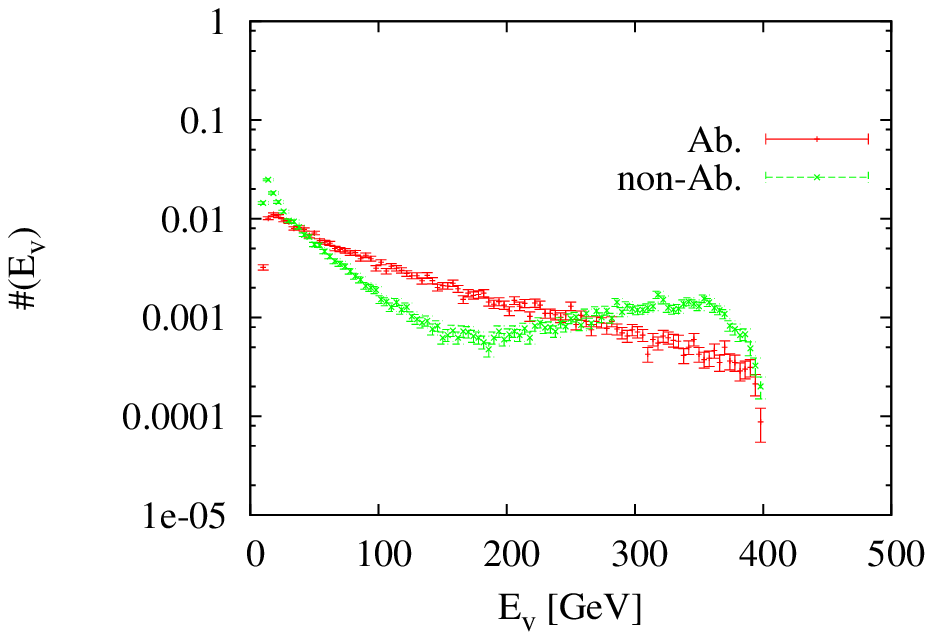,scale=0.7, angle=0}
\epsfig{file=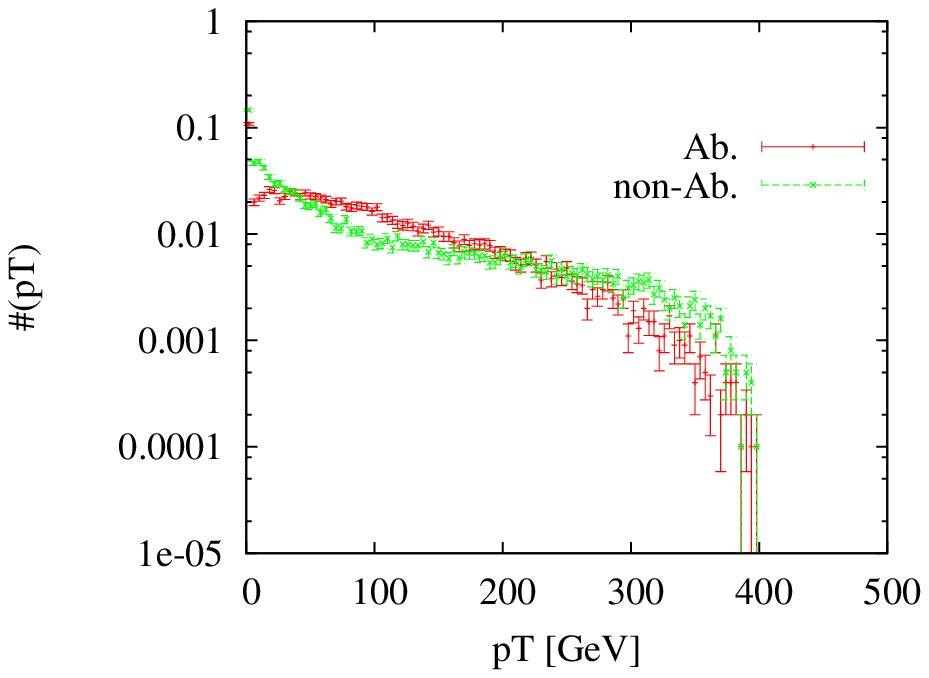,scale=0.7, angle=0}
\end{minipage}
\caption{The plots show the comparison between the Abelian and the 
non-Abelian setups: the number of $\gamma_v$s (Abelian) or diagonal 
$\pi_v$s$/\rho_v$s (non-Abelian) (top left), the $v$-particle energy 
distribution (top right), the number of SM charged particles produced, 
and the missing $\pT$. The scenarios are chosen to yield similar 
distributions.} 
\label{fig:compare_Ngv} }

So how could the Abelian and non-Abelian scenarios come to have so 
similar properties?  First off, of course, 
$m_{\gamma_v} = m_{\pi_v} = m_{\rho_v}$ needs to be assumed.
Thereafter the value of $N_{\mrm{flav}}$ in the non-Abelian model 
specifies that exactly an average fraction $1/N_{\mrm{flav}}$ of the
full energy leaks back into the visible sector. In the Abelian model,
for a given $m_{q_v}$, $\alpha_v$ is the only free parameter that could
be fixed to give that average energy. For this parameter set, the number
and energy spectrum of $\gamma_v$s are predicted entirely by
the perturbative cascade. These distributions now need to be roughly
reproduced by the non-Abelian model, which first of all means the same
average number of $v$-particles decaying back into the SM. While
$m_{q_v} =  m_{\pi_v}/2$ is fixed in this case, there is freedom in the
choices of $\alpha_v$ and non-perturbative fragmentation parameters.
Recall that the number of $v$-particles will not vanish in the 
$\alpha_v \to 0$ limit for the non-Abelian model, unlike the Abelian one.
Actually it turns out to be slightly difficult to reduce the non-Abelian
multiplicity down to the level set by the Abelian scenario. With an 
$\alpha_v$ comparable to that of a QCD cascade at a corresponding 
energy/mass ratio, the longitudinal fragmentation function needs to be
made harder by decreasing $a$ and increasing $b'$ relative to the QCD
values.  

Using such a strategy, a matching pair of scenarios have been constructed,
an AM$_Z^\prime$ model with $m_{q_v}=20$~GeV, $m_{\gamma_v}=10$~GeV and 
$\alpha_v=0.3$, and a NAM$_Z^\prime$ with $m_{q_v}=5$~GeV, $m_{\pi_v/\rho_v}=10$~GeV,
$N_{\mrm{flav}}=4$, $\alpha_v=0.15$, $a=0.12$ and $b'=2$. This gives fair 
agreement, as can be seen in Fig.~\ref{fig:compare_Ngv}.

\FIGURE[t]{
\begin{minipage}[b]{0.4\linewidth}
\epsfig{file=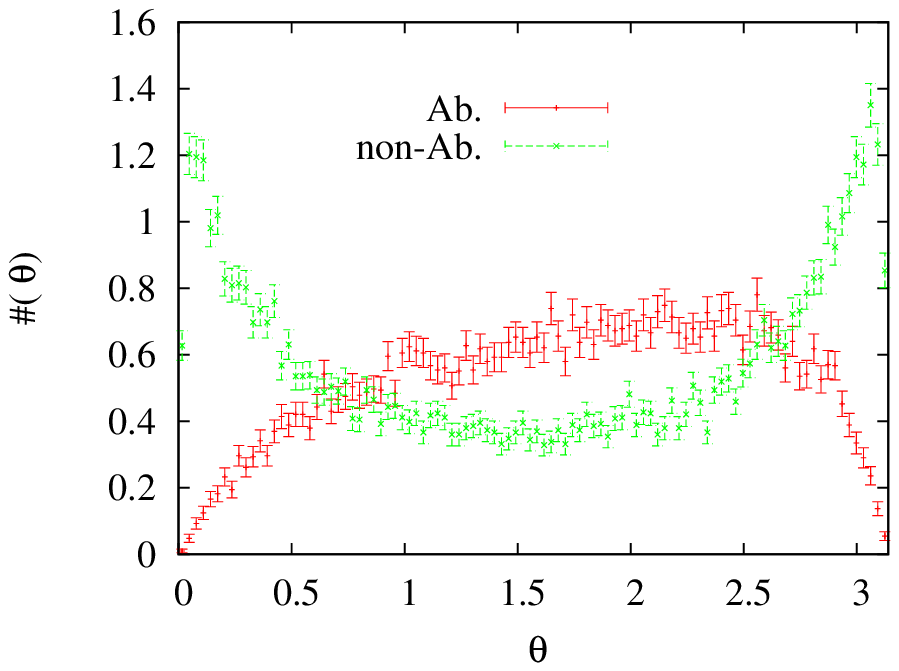,scale=0.7, angle=0}
\end{minipage} \hspace{0.5cm}
\begin{minipage}[b]{0.4\linewidth}
\epsfig{file=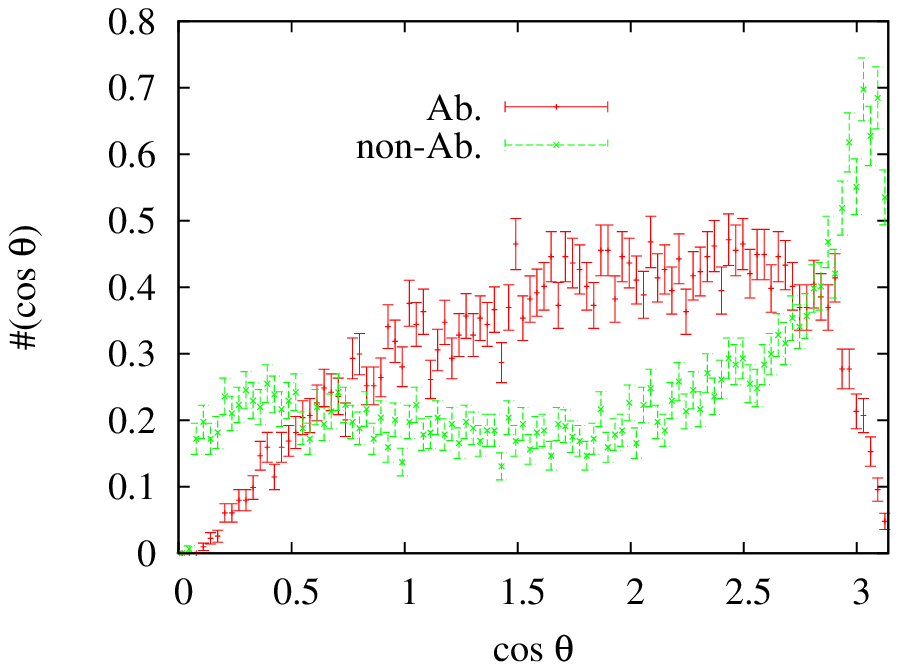,scale=0.7, angle=0}
\end{minipage}
\caption{The plots show the comparison between the Abelian and the non
Abelian setups. On the left is the $\cos\theta_{ij}$ between the
$v$-particles which can decay back into the SM. On the right is the
corresponding $\cos\theta_{ij}$ between the jets in the detector.} 
\label{fig:compare_angles} }

%\FIGURE[t]{
%\begin{minipage}[b]{0.4\linewidth}
%\epsfig{file=U1vsSU4_sphericity_compare_small.eps,scale=0.7, angle=0}
%\end{minipage} \hspace{0.5cm}
%\begin{minipage}[b]{0.4\linewidth}
%\epsfig{file=U1vsSU4_thrust_compare_small.eps,scale=0.7, angle=0}
%\end{minipage}
%\caption{The plots show the comparison between the Abelian and the non
%Abelian setups. On the left is sphericity and on the right is thrust.} 
%\label{fig:compare_spheri} }

We can now compare the angular distributions in the two cases. In
Fig.~\ref{fig:compare_angles} (left) one may observe how in the
non-Abelian case the $\theta_{ij}$ distribution of the angle 
between the visible $v$-particles is much more peaked near 0 and $\pi$ 
than in the Abelian one. The plot on the right shows how this 
characteristic is maintained in
jet distributions. 
%The event shapes confirm that the non-Abelian
%events are more pencil-like than the Abelian ones, see
%Fig.~\ref{fig:compare_spheri}.
 
%\FIGURE[t]{
%\begin{minipage}[b]{0.4\linewidth}
%\epsfig{file=U1vsSU4_ThV_compare_large.eps,scale=0.7, angle=0}
%\end{minipage} \hspace{0.5cm}
%\begin{minipage}[b]{0.4\linewidth}
%\epsfig{file=U1vsSU4_ThJJ_compare_large.eps,scale=0.7, angle=0}
%\end{minipage}
%\caption{The plots show the comparison between the Abelian and the non
%Abelian setups. On the left is the $\cos\theta_{ij}$ between the
%$v$-particles which can decay back into the SM. On the right is the
%corresponding $\cos\theta_{ij}$ between the jets in the detector.} 
%\label{fig:compare_angles_large} }

We have repeated the study for some different model parameters and found sumilar results. 
In the second comparison the AM$_Z^\prime$ model has $m_{q_v}=50$~GeV,
$m_{\gamma_v}=10$~GeV and $\alpha_v=0.43$, while NAM$_Z^\prime$ has
$m_{q_v}=5$~GeV, $m_{\pi_v/rho_v}=10$~GeV, $N_{\mrm{flav}}=4$, $\alpha_v=0.18$,
$a=0.2$ and $b'=2$. 

%The angular distributions in 
%Fig.~\ref{fig:compare_angles_large} and the event shapes in 
%Fig.~\ref{fig:compare_spheri_large} shows the same picture as before:
%the non-Abelian scenario is more likely to give a back-to-back,
%pencil-like structure.

It is tempting to ascribe the observed differences to 
different radiation/hadronization patterns in the two scenarios,
ultimately deriving from the different dipole emission topologies
already discussed. One attractive possibility is that the relative 
lack of peaking at $\cos\theta_{ij} = \pm 1$ for the Abelian scenario
is a consequence of the dead cone effect, i.e.\ the suppression of
emissions parallel to a massive radiating particle. A back-to-back 
$q_v\bar{q}_v$ pair undergoing non-Abelian hadronization would have 
no corresponding suppression for $v$-hadron formation along this axis.  
However, considering of the flexibility that exists in the 
tuning of non-perturbative hadronization parameters, and the differences 
observed in the $v$-particle energy spectrum, we will not now be as 
bold as to exclude the possibility of a closer match.
If such a match required straining the non-perturbative model
to behave rather differently from QCD extrapolations, however, then 
such tunes would not be particularly credible.

\section{Summary and Outlook}
\label{sec:conclusions}

In this article we have compared six different scenarios of a 
generic secluded-sector character. These are either kinetic mixing 
with a light $\gamma_v$, or $Z^\prime$, or new $F_v$ particles, and 
in each case either with a broken $U(1)$ or an unbroken $SU(N)$.
The $F_v$ particles are charged both under the Standard Model groups
and the new secluded-sector groups, and so are guaranteed a 
significant production rate, whenever kinematically possible,
whereas the rate via $\gamma_v$ or $Z^\prime$ depends on a number of 
parameters such as the $\gamma / \gamma_v$ mixing parameters or the 
$Z^\prime$ coupling structure and mass. In this article we have 
completely disregarded such rate issues and instead studied the 
properties of the different models on an event-by-event basis. 
 
In order to do so, we have developed a new flexible framework that
implements hadronization in the hidden sector. Similar modeling in
the past have relied on simple rescaling of QCD, whereas
here we set up hidden-sector string fragmentation as a completely
separate framework, though sharing the same underlying space--time
structure of the hadronization process. We have also expanded on 
our previous work with parton showers in the hidden sector, possibly
interleaved with radiation in the visible sector, by allowing for the
emission of massive $\gamma_v$s, when the $U(1)$ group is broken.
In order to obtain the correct behaviour, both in the soft/collinear
limits and for hard emissions, the shower is matched to first-order
matrix elements we have calculated for the massive $\gamma_v$ cases.
Much of the framework presented here could be applied also to other
related scenarios, although there are limits. For instance, 
implicitly it has been assumed that the $q_v$ masses are not too
dissimilar from the confinement scale $\Lambda$ of the new $SU(N)$
group --- hadronization would look rather different in the limit
$m_{q_v} \gg \Lambda$. 

In the scenarios where $F_v$s are produced and promptly decay like
$F_v \to f q_v$, the presence of the fermions $f$ in the final state
is a distinguishing factor, and the $f$ energy spectrum can be
used to extract information on mass scales in the secluded sector.
At first glance, the production mediated by a $\gamma_v$ or a 
$Z^\prime$ would seem to be more similar. The  different location 
of the propagator mass peaks leads to quite significant patterns
of initial-state photon radiation, however, that would be easily
observed. 

The challenge, thus, is to distinguish an Abelian and  a non-Abelian 
scenario interactions in the secluded sector. In certain cases
that would be straightforward, e.g. if there is only one $q_v$
species, so that all energy decays back into the standard sector
in the non-Abelian models. 
It is possible to fix parameters in the non-Abelian cases so that
only some fraction decays, so that the level of activity matches
Abelian $\alpha_v$ one.

Our first investigations here point to differences  
showing up in event properties related to the overall structure of 
the energy and particle flow. More elaborate tunings possibly 
might bring the models closer together, but one would hope that
data still would favour ``what comes naturally'' in either of the 
models. 

There are also more handles than the ones we have used. We have
shown that not only lepton pairs but also jets are amenable to 
mass peak identification, which would allow to divide events
into several subsystems and analyze their relative location,
e.g.\ searching for coherence effects. Should lifetimes be
long enough to induce displaced vertices, not only would that 
be a spectacular signal, but it would also be a boon to such
analysis efforts.
 
The most obvious next step would be to study these models for 
consequences at the LHC. The task can be split into three parts: 
cross sections, triggers and model-specific event properties.
The cross sections are so intimately related to the choice of
masses and couplings that it will be impossible to exclude the 
possibility of a secluded sector, only to separate excluded
and not-excluded regions of parameter space, in close analogy
with SUSY. The obvious trigger would be $\npT$, but we have seen
that this would not work for non-Abelian scenarios with one 
$q_v$ flavour. It would then need to be supplemented by the 
presence of (multiple) lepton pairs of some fixed invariant mass  
and, if we are lucky, displaced vertices. The final step would 
be to understand whether the more busy environment in hadronic 
events would still allow to distinguish Abelian and non-Abelian
models --- the separation between the three production scenarios
we have considered here should still be straightforward. Chances 
are that this will bring us full circle to the cross sections 
issue, since more sophisticated analyses will require a decent 
event rate to start out from.

\acknowledgments

We would like to thank Matt Strassler, Katryn Zurek, Peter Skands and Bob McElrath for
helpful suggestions. 
Work supported in part by Marie-Curie Early Stage Training program "HEP-EST''
(contract number MEST-CT-2005-019626), by the Marie-Curie MCnet program, and by the
Swedish Research Council (contract numbers 621-2010-3326 and 621-2008-4219)

\appendix
\label{appendix}
\appendix
\section{Scenario selection and setup}

We here present some information relevant to get going with secluded-sector
event generation in \textsc{Pythia~8}. Basic knowledge of the program
is assumed \cite{Sjostrand:2007gs}.
 
\TABLE[t]{
\begin{tabular}{|c|c|c|l|}
\hline
name & \ttt{name}      & identity       & comment \\ 
\hline
$D_v$ & \ttt{Dv}       & \ttt{4900001}  & partner to the $d$ quark\\
$U_v$ & \ttt{Uv}       & \ttt{4900002}  & partner to the $u$ quark\\
$S_v$ & \ttt{Sv}       & \ttt{4900003}  & partner to the $s$ quark\\
$C_v$ & \ttt{Cv}       & \ttt{4900004}  & partner to the $c$ quark\\
$B_v$ & \ttt{Bv}       & \ttt{4900005}  & partner to the $b$ quark\\
$T_v$ & \ttt{Tv}       & \ttt{4900006}  & partner to the $t$ quark\\
$E_v$ & \ttt{Ev}       & \ttt{4900011}  & partner to the $e$ lepton\\
$\nu_{E_v}$ & \ttt{nuEv}  & \ttt{4900012}  & partner to the $\nu_e$ neutrino\\
$M_v$ & \ttt{MUv}      & \ttt{4900013}  & partner to the $\mu$ lepton\\
$\nu_{M_v}$ & \ttt{nuMUv} & \ttt{4900014}  & partner to the $\nu_\mu$ neutrino\\
$T_v$ & \ttt{TAUv}     & \ttt{4900015}  & partner to the $\tau$ lepton\\
$\nu_{T_v}$ & \ttt{nuTAUv} & \ttt{4900016} & partner to the $\nu_\tau$ neutrino\\
$g_v$ & \ttt{gv}   & \ttt{4900021}  & the $v$-gluon in an $SU(N)$ scenario\\
$\gamma_v$ & \ttt{gammav} & \ttt{4900022}  
   & the $v$-photon in a $U(1)$ scenario \\
$Z', Z_v$ & \ttt{Zv} & \ttt{4900023}    
   & massive gauge boson linking SM- and $v$-sectors\\
$q_v$ & \ttt{qv} & \ttt{4900101}  & matter particles purely in  $v$-sector\\
$\pi_v^{\mrm{diag}}$ & \ttt{pivDiag}  & \ttt{4900111}  
   & flavour-diagonal spin 0 $v$-meson\\
$\rho_v^{\mrm{diag}}$ & \ttt{rhovDiag} & \ttt{4900113}  
   & flavour-diagonal spin 1 $v$-meson\\\
$\pi_v^{\mrm{up}}$ & \ttt{pivUp}    & \ttt{4900211}  
   & flavour-nondiagonal spin 0 $v$-meson\\\
$\rho_v^{\mrm{up}}$ & \ttt{rhovUp}   & \ttt{4900213}  
   & flavour-nondiagonal spin 1 $v$-meson\\\
      & \ttt{ggv} & \ttt{4900991}  & glueball made of $v$-gluons\\
\hline
\end{tabular}
\label{tab:Fv_codes}
\caption{The allowed new particles in valley scenarios. Names are gives
as used in this text and as in \textsc{Pythia~8} event listings.
The identity code is an integer identifier, in the spirit of the PDG 
codes, but is not part of the current PDG standard \cite{Nakamura:2010zzi}.}
}

The $v$-particle content is summarized in Tab.~\ref{tab:Fv_codes}. Needless
to say, not all of them would be relevant for each specific scenario. 
Internally further copies of $q_v$ may be used, up to code 4900108, 
but these do not appear in the event record.
Properties of the particles can be set to modify the scenarios, notably 
mass (\ttt{m0}); only the $g_v$ must remain massless. If $F_v \to f q_v$
is allowed, masses must be chosen so that the decay is kinematically 
possible. The $\pi_v$ and $\rho_v$ masses should be set at around twice 
the $q_v$ one. For unstable particles the width (\ttt{mWidth}) and 
allowed mass range (\ttt{mMin} and \ttt{mMax}) can be set. To generate
displaced vertices the $c \tau$ value must be set (\ttt{tau0}). 
Spin choices are described later.

Several particles by default are set stable, so it is necessary to switch 
on their decay (\ttt{mayDecay}). For $\gamma_v$, $\pi_v^{\mrm{diag}}$ and
$\rho_v^{\mrm{diag}}$ no decay channels are on by default, since that set 
depends so strongly on the mass scale selected. The simple way, of 
switching on everything (\ttt{onMode = on}) works in principle, but 
e.g. a 10 GeV $\gamma_v$ would then be above the $b \bar b$ threshold
but below the $B \bar B$ one, and so generate a trail of (harmless) 
error messages. Also the branching ratios of the decay channels may need 
to be adjusted, based on the scenario. The $Z_v$ ones are mainly 
place-fillers, to give one example. 
 
\TABLE[t]{
\begin{tabular}{|l|l|l|}
\hline
code & flag & process \\
\hline
4901 & \ttt{HiddenValley:gg2DvDvbar} & $g g \to D_v \bar D_v$\\
4902 & \ttt{HiddenValley:gg2UvUvbar} & $g g \to U_v \bar U_v$\\
4903 & \ttt{HiddenValley:gg2SvSvbar} & $g g \to S_v \bar S_v$\\
4904 & \ttt{HiddenValley:gg2CvCvbar} & $g g \to C_v \bar C_v$\\
4905 & \ttt{HiddenValley:gg2BvBvbar} & $g g \to B_v \bar B_v$\\
4906 & \ttt{HiddenValley:gg2TvTvbar} & $g g \to T_v \bar T_v$\\
4911 & \ttt{HiddenValley:qqbar2DvDvbar} & $q \bar q \to g^* \to D_v \bar D_v$\\
4912 & \ttt{HiddenValley:qqbar2UvUvbar} & $q \bar q \to g^* \to U_v \bar U_v$\\
4913 & \ttt{HiddenValley:qqbar2SvSvbar} & $q \bar q \to g^* \to S_v \bar S_v$\\
4914 & \ttt{HiddenValley:qqbar2CvCvbar} & $q \bar q \to g^* \to C_v \bar C_v$\\
4915 & \ttt{HiddenValley:qqbar2BvBvbar} & $q \bar q \to g^* \to B_v \bar B_v$\\
4916 & \ttt{HiddenValley:qqbar2TvTvbar} & $q \bar q \to g^* \to T_v \bar T_v$\\
4921 & \ttt{HiddenValley:ffbar2DvDvbar} 
     & $f \bar f \to \gamma^* \to D_v \bar D_v$\\
4922 & \ttt{HiddenValley:ffbar2UvUvbar} 
     & $f \bar f \to \gamma^* \to U_v \bar U_v$\\
4923 & \ttt{HiddenValley:ffbar2SvSvbar} 
     & $f \bar f \to \gamma^* \to S_v \bar S_v$\\
4924 & \ttt{HiddenValley:ffbar2CvCvbar} 
     & $f \bar f \to \gamma^* \to C_v \bar C_v$\\
4925 & \ttt{HiddenValley:ffbar2BvBvbar} 
     & $f \bar f \to \gamma^* \to B_v \bar B_v$\\
4926 & \ttt{HiddenValley:ffbar2TvTvbar} 
     & $f \bar f \to \gamma^* \to T_v \bar T_v$\\
4931 & \ttt{HiddenValley:ffbar2EvEvbar} 
     & $f \bar f \to \gamma^* \to E_v \bar E_v$\\
4932 & \ttt{HiddenValley:ffbar2nuEvnuEvbar} 
     & $f \bar f \to \gamma^* \to \nu_{E_v} \bar \nu_{E_v}$\\
4933 & \ttt{HiddenValley:ffbar2MUvMUvbar} 
     & $f \bar f \to \gamma^* \to M_v \bar M_v$\\
4934 & \ttt{HiddenValley:ffbar2nuMUvnuMUvbar} 
     & $f \bar f \to \gamma^* \to \nu_{M_v} \bar \nu_{M_v}$\\
4935 & \ttt{HiddenValley:ffbar2TAUvTAUvbar} 
     & $f \bar f \to \gamma^* \to T_v \bar T_v$\\
4936 & \ttt{HiddenValley:ffbar2nuTAUvnuTAUvbar} 
     & $f \bar f \to \gamma^* \to \nu_{T_v} \bar \nu_{T_v}$\\
4941 & \ttt{HiddenValley:ffbar2Zv} 
     & $f \bar f \to Z_v^* (\to q_v \bar q_v)$\\
\hline
\end{tabular}
\label{tab:processes}
\caption{Allowed processes that can be switched on individually.}
}

The list of processes is shown in Tab.~\ref{tab:processes}.
It would be possible to switch on all of them with 
\ttt{HiddenValley:all = on}, but normally that would imply a mix of
different scenarios that does not appear plausible. Many processes
should also be viewed in the context of the other choices made.
 
\TABLE[t]{
\begin{tabular}{|l|l|l|}
\hline
parameter & def. & meaning \\
\hline
\multicolumn{3}{|c|}{Scenario} \\ \hline
\ttt{HiddenValley:Ngauge} & 3 & 1 for $U(1)$, $N$ for $SU(N)$ \\
\ttt{HiddenValley:spinFv} & 1 & 0, 1 or 2 for $F_v$ spin 0, $1/2$ and 1\\
\ttt{HiddenValley:spinqv} & 0 & $q_v$ spin 0 or 1 when $s_{F_v} = 1/2$\\
\ttt{HiddenValley:kappa} & 1. & $F_v$ anomalous magnetic dipole moment\\
\ttt{HiddenValley:doKinMix} & off & allow kinetic mixing\\
\ttt{HiddenValley:kinMix} & 1. & strength of kinetic mixing, if on\\
\hline
\multicolumn{3}{|c|}{Showers in secluded sector} \\ \hline
\ttt{HiddenValley:FSR} & off & allow final-state radiation\\
\ttt{HiddenValley:alphaFSR} & 0.1 & constant coupling strength\\
\ttt{HiddenValley:pTminFSR} & 0.4 & lower cutoff of shower evolution\\
\hline
\multicolumn{3}{|c|}{Hadronization in secluded sector} \\ \hline
\ttt{HiddenValley:fragment} & off & allow hadronization\\
\ttt{HiddenValley:nFlav} & 1 & $N_{\mathrm{flav}}$, number of distinct 
                               $q_v$ species\\
\ttt{HiddenValley:probVector} & 0.75 & fraction of spin-1 $v$-mesons \\
\ttt{HiddenValley:aLund} & 0.3 & $a$ parameter in eq.~(\ref{eq:bowtwo})\\
\ttt{HiddenValley:bmqv2} & 0.8 & $b' = bm_{q_v}^2$ parameter in 
                                 eq.~(\ref{eq:bowtwo})\\
\ttt{HiddenValley:rFactqv} & 1.0 & $r$ parameter in eq.~(\ref{eq:bowtwo})\\
\ttt{HiddenValley:sigmamqv} & 0.5 & $\sigma'$, 
                            such that $\sigma = \sigma' m_{q_v}$\\
\hline
\end{tabular}
\label{tab:parameters}
\caption{The parameters that can be set to select the model to study,
with default values and some expanations.}
}

Finally, the list of relevant model parameters is shown in 
Tab.~\ref{tab:parameters}. On top is the choice between a $U(1)$ and
an $SU(N)$ scenario. The $F_v$ and $q_v$ spins must be selected in a 
coordinated fashion, to be consistent with $F_v \to f q_v$ decays.
The choice of $F_v$ spin directly affects the process (differential) 
cross sections. If $F_v$ has spin 1 also the choice of an anomalous
magnetic moment would have an influence.

The kinetic mixing switch allows to reuse the $\gamma^*$-mediated 
processes in a completely different context than originally foreseen,
in which the $F_v$ have no Standard Model coupling but are produced
by $\gamma - \gamma_v$ mixing. Actually this redefines the $F_v$ to
be equivalent with what we normally call $q_v$. Thus a process like 
$f \bar f \to \gamma^* \to E_v \bar E_v$ becomes
$f \bar f \to \gamma^* \to \gamma^*_v \to q_v \bar q_v$.
To complete this transformation you need to set the $E_v$ stable 
(\ttt{mayDecay = false}), uncharged (\ttt{chargeType = 0}) and 
invisible (\ttt{isVisible = false}).

The shower parameters should be self-explanatory. The lower cutoff
scale can be raised in proportion to the characteristic mass scales, 
but ought to be no more than $m_{q_v}/2$, say. A lower cutoff means 
longer execution time without any significant change of event
properties.

The hadronization parameters have also been discussed before, 
except for $r$, which is providing slightly more flexibility
to the Lund--Bowler fragmetation function than in 
eq.~(\ref{eq:bowler})
\begin{equation}
f(z) \propto \frac{1}{z^{1 + r b'}} \, (1 - z)^a \, 
\exp\left( - \frac{b' m_{m_v}^2}{z \, m_{m_q}^2} \right) ~.
\label{eq:bowtwo}
\end{equation}
where $r$ could be set anywhere between 0 and 1.
The dimensionless $\sigma'$ parameter is normalized so that the 
$q_v$ of each new pair produced in the hadronization has a 
$\langle \pT^2 \rangle = (\sigma' m_{q_v})^2$.  

Behind the scenes, the \ttt{HiddenValleyFragmentation} class can
reuse most of the standard \ttt{StringFragmentation} and 
\ttt{MiniStringFragmentation} machineries. Specifically,
already for the Standard Model hadronization, the selection of
flavour, $z$ and $\pT$ is relegated to three ``helper'' classes.
The three new classes \ttt{HVStringFlav}, \ttt{HVStringZ} and
\ttt{HVStringPT} derive from their respective SM equivalent,
and cleanly replace these three aspects while keeping the rest of
the handling of complex string topologies. Specifically, it would 
be straightforward to expand towards a richer flavour structure 
in the secluded sector. Note, however, that it is important to
select $v$-quark ``constituent'' masses that reasonably match
the intended $v$-meson mass spectrum, since such relations are
assumed in parts of the code. Furthermore, with new $q_v$ 
defined with separate particle data, one must disable the few lines 
in \ttt{HiddenValleyFragmentation::init(\ldots)} that now duplicate 
the one $q_v$ into several identical copies.

\end{document}